\documentclass{aa}  
\usepackage{txfonts}
\usepackage{verbatim}
\usepackage{graphicx}
\usepackage{xcolor}
\usepackage{booktabs}
\usepackage[colorlinks=true,linkcolor=blue,urlcolor=blue,citecolor=blue]{hyperref}
\usepackage{xspace}
\usepackage{pdflscape}
\usepackage{afterpage}
\usepackage{subcaption}
\usepackage{placeins}
\usepackage{longtable}
\usepackage{threeparttablex}
\usepackage{lscape}       
\newcommand{\msun}{{\rm M}_{\odot}}

\newcommand{\eg}{e.g.\@\xspace}
\newcommand{\cf}{c.f.\@\xspace}
\newcommand{\ie}{i.e.\@\xspace}

\begin{document}
\title{Revisiting the electromagnetic perspective on the black hole mass distribution}

\titlerunning{Revisiting the electromagnetic perspective on the black hole mass distribution}

\author{
    Aina S. Nambena\inst{\ref{ITP}, \ref{HITS}}\fnmsep\thanks{nambena@thphys.uni-heidelberg.de}
    \and Fabian R. N. Schneider\inst{\ref{HITS}, \ref{ZAH}}\fnmsep\thanks{fabian.schneider@h-its.org}
    \and Eva Laplace\inst{\ref{HITS}, \ref{KUL}, \ref{KUL2}, \ref{AMS}}
    \and Jorge Casares\inst{\ref{IAC}, \ref{DALL}}
    \and Teo Mu\~noz-Darias \inst{\ref{IAC}, \ref{DALL}}
    \and Manuel A. P. Torres \inst{\ref{IAC}, \ref{DALL}}
    \and Gast\'on J. Escobar\inst{\ref{IAC}, \ref{DALL}}
}
\authorrunning{Nambena et al.}

\institute{Institut für Theoretische Physik, Ruprecht-Karls-Universität Heidelberg, Philosophenweg 16, 69120 Heidelberg, Germany\label{ITP}
    \and Heidelberger Institut f{\"u}r Theoretische Studien, Schloss-Wolfsbrunnenweg 35, 69118 Heidelberg, Germany\label{HITS}
    \and Zentrum f{\"u}r Astronomie der Universit{\"a}t Heidelberg, Astronomisches Rechen-Institut, M{\"o}nchhofstr.\ 12-14, 69120 Heidelberg, Germany\label{ZAH},
    \and Institute of Astronomy, KU Leuven, Celestijnenlaan 200D, 3001 Leuven, Belgium\label{KUL},
    \and Leuven Gravity Institute, KU Leuven, Celestijnenlaan 200D, box 2415, 3001 Leuven, Belgium\label{KUL2}
    \and Anton Pannekoek Institute of Astronomy, Science Park 904, University of Amsterdam, 1098XH Amsterdam, The Netherlands\label{AMS}
    \and Instituto de Astrof\'\i{}sica de Canarias, E-38205 La Laguna, Tenerife, Spain\label{IAC}
    \and Departamento de Astrof\'isica, Universidad de La Laguna, E-38206 La Laguna, Tenerife, Spain\label{DALL}
    }

\date{Received xxx}

\abstract{
Gravitational wave (GW) detections have opened a new window into our Universe, offering unprecedented insights into stellar-mass black holes (BHs), while electromagnetic (EM) searches provide a complementary perspective. Previous EM studies suggested a mass gap between the most massive neutron stars (NSs) and the least massive BHs, based on 20 BH mass measurements. This potential gap does not appear in GW observations. Here, we compile an updated sample of 40 BHs with known masses from X-ray binaries, inert BH binary systems, and one microlensing event, doubling the previous EM sample. Using a hierarchical Bayesian framework, we fit four parametric models of increasing flexibility to constrain the observed BH mass distribution. Across all models, we find a dominant component peaking at $5\text{--}10\,\msun$, encompassing about half of the systems. This peak is well described by a Gaussian with $\mu= 8.2_{-0.5}^{+0.5}\,\msun$ and $\sigma=2.6_{-0.4}^{+0.5}\,\msun$. Below $5\,\msun$, the BH mass distribution declines smoothly, and there is no evidence for a sharp NS-BH mass gap. Instead, we observe a dearth of BHs, populated by several systems, in agreement with GW observations. At higher masses, the distribution decreases gradually, and hints at a secondary high-mass component dominated by BHs from stars formed at sub-solar metallicities; however, the current sample is too small to constrain its shape. Simple functional forms, such as power-law and Gaussian functions, have difficulty reproducing the main BH component as well as the low- and high-mass components at the same time. BH mass measurements from EM observations are challenging, and the total number of BHs from EM observations is still rather small, rendering any conclusions inherently uncertain. As our analysis only considers present-day and not birth masses of BHs, future studies need to disentangle metallicity, spin, and evolutionary channels to constrain the birth mass distribution of BHs.
}

\keywords{Stars: black holes --
            binaries: general --
            methods: statistical --
             X-rays: binaries
           }

\maketitle
\nolinenumbers

\section{\label{sec:introduction}Introduction}

The advent of gravitational wave (GW) astronomy has revolutionized our understanding of stellar-mass black holes (BHs), as the confident detection of hundreds of compact binary mergers allowed for detailed statistical analyses of their masses and other properties \citep[\eg][]{theligoscientificcollaboration2025a, theligoscientificcollaboration2026b}. For instance, features in the chirp mass distribution, such as a clear peak at $8\,\msun$, followed by a prominent dearth at $10\,\msun$ with subsequent rise to higher masses and further peaks at $13\text{--}14\,\msun$ and $27\,\msun$, offer valuable insights into the formation channels of BHs in binary systems \citep[\eg][]{tiwari2021a, tiwari2022a, tiwari2025a, abbott2023a, schneider2023a, willcox2025a}. In addition to GW discoveries, electromagnetic (EM) observations continue to identify new BHs in binary systems. These include X-ray binaries, where the BH accretes from a stellar companion \citep[\eg][]{miller-jones2021a, binder2021a, casares2014b}; X-ray quiet or inert BHs, identified via spectroscopic or astrometric measurements \citep[\eg][]{giesers2019a, el-badry2023a, panuzzo2024a, wang2024b}; and gravitational lensing events with BHs \citep{wyrzykowski2020a, sahu2022a, sahu2025a}. Together, these complementary methods expand the sample of known stellar-mass BHs and provide an independent window into the demographics of BH populations \citep[see][for a recent review]{bambi2025a}. Indeed, if GW BHs typically form from low-metallicity stars \citep[\eg][]{rinaldi2024a}, EM-observed BHs often stem from solar-like metallicity stars, and probe an earlier stage in the evolution of binary stars that ultimately may end up in a binary BH merger, motivating our focus on EM-observed populations.

From a sample of $15\text{--}20$ BHs in X-ray binary systems, \citet{farr2011a} and \citet{kreidberg2012a} applied hierarchical Bayesian techniques to infer the observed BH mass distribution. One of their main findings was evidence for a mass gap at $2\text{--}5\,\msun$, between neutron stars (NSs) and BHs, confirming earlier works by \citet{bailyn1998a} and \citet{ozel2010a}. However, the robustness of this gap has since been questioned, particularly given the likely BH-NS merger GW230529, with a BH of mass $2.5\text{--}4.5\,\msun$ \citep{abac2024a}, and compact objects in non-interacting binaries possibly falling into this gap \citep{barr2024a}\footnote{Also 2MASS\,J05215658+4359220 \citep{thompson2019a} and V723~Mon \citep{jayasinghe2021a} were considered having compact objects in the NS-BH mass gap but are now revealed to be stellar, non-compact objects \citep{el-badry2022a, bianchi2024a, kochanek2025a}.}, motivating a re-examination of the BH mass distribution.

Beyond demographics, BHs and their masses hold crucial information on the evolutionary history of their progenitor star, providing insights into the late evolution of massive stars beyond core helium burning and the complex core collapse mechanism. Hence, the BH mass distribution offers indirect constraints on stellar physics, including nuclear burning stages, neutrino transport, and the conditions for successful or failed supernova explosions, direct BH formation or partial fallback events \citep[\eg][]{sukhbold2016a, limongi2017a, woosley2020a, ertl2020a, heger2023a, schneider2023a, laplace2025a}. In other words, new observations from binary mergers detected via GW or EM emissions hold promise to shed light on physical processes in massive stars, and ultimately on the formation of stellar-mass BHs.

Many BH progenitors -- especially those in GW events and EM binaries -- likely experienced binary mass transfer phases, stripping away their hydrogen-rich envelopes. This affects the explodability of the progenitor and, therefore, the formation of the BH, which in turn may have implications for the resulting mass distribution of BHs \citep[\eg][]{timmes1996a, brown2001a, podsiadlowski2004a, woosley2019a, ertl2020a, schneider2021a}. \citet{schneider2021a, schneider2023a} show that such binary evolution may lead to a bimodal mass distribution of stellar-mass BHs, potentially explaining the observed features in the chirp mass distribution of GW-detected BHs \citep{willcox2025a}. Based on the fifth GW Transient Catalogue, \citet{guttman2026a} find evidence for at least four sub-populations of binary BH mergers and a gap in the individual BH mass distributions at about the location predicted by theory.

In this study, we compile an updated sample of BHs identified via EM observations to investigate their mass distribution. We revisit\footnote{During the review process of this work, another paper was published that also re-analysed the BH mass spectrum from EM observations and make similar conclusions \citep{rocha2026a}.} the work of \citet{farr2011a} and \citet{kreidberg2012a} with a larger and more diverse set of BHs, incorporating new systems discovered in X-ray, optical, and lensing surveys, and updating previous mass measurements with improved observations \citep[\eg][]{el-badry2023a, sahu2025a, casares2027a}. By applying the same hierarchical Bayesian framework, we aim to assess whether the BH mass distribution inferred from EM observations exhibits evidence for an NS-BH mass gap, for bimodality as suggested by \citet{schneider2023a}, or for other features seen in GW data \citep[\cf][]{ray2025a}.

This paper is structured as follows. Section~\ref{sec:obs-data} presents the updated sample of EM-detected BHs, and Sect.~\ref{sec:methods} outlines the Bayesian framework and the models we fit to the BH masses. We present our main results in Sect.~\ref{sec:bh-mass-distr}, before discussing the implications of our findings in Sect.~\ref{sec:discussion}, and summarizing the main conclusions in Sect.~\ref{sec:conclusions}. We provide a full list of the BH systems and their assumed observational properties in Appendix~\ref{appendix:masses}.

\section{Methods}
\subsection{\label{sec:obs-data}Black-hole sample}
Our statistical analysis relies on 40 observed BHs in EM binary systems in the Milky Way and nearby galaxies. The sample includes 25 low-mass X-ray binaries (LMXBs, \citealt{casares2027a}), six high-mass X-ray binaries (HMXBs, \citealt{silverman2008a,crowther2010b, farr2011a,  miller-jones2021a,binder2021a, casares2027a, ramachandran2025a, an2025a}), eight inert BHs\footnote{These are X-ray quiet, and sometimes are referred to as ``dormant'' BHs, although it should be noted that this term is traditionally used for transient X-ray binaries in quiescence.} \citep{giesers2018a, giesers2019a,shenar2022a,mahy2022a, el-badry2023b, el-badry2023a, panuzzo2024a, wang2024b} and one BH identified through gravitational microlensing \citep{sahu2022a, sahu2025a}. We use the updated mass measurements of BHs in X-ray binaries from \citet{casares2027a}, who also present a Bayesian analysis of the BH mass distribution, but focused on BHs in X-ray transients.

\paragraph{Spectroscopic masses.}
For the X-ray binaries and several inert systems, BH mass estimates rely on spectroscopic observations of the companion star \citep[\eg][]{miller-jones2021a, wang2024b}. The orbital period $P$ and the radial velocity semi-amplitude $K$ yield the spectroscopic mass function,
\begin{equation}
    f_\mathrm{spec}(M_\mathrm{BH})
    =\frac{P K^3}{2 \pi G}\left(1-e^2\right)^{3/2}
    =\frac{M_\mathrm{BH} \sin ^3 i}{\left(1+q\right)^2} 
    =\frac{M_\mathrm{BH}^3 \sin ^3 i}{\left(M_\mathrm{BH}+M_\mathrm{*}\right)^2},
    \label{eq:mass-spectroscopy}
\end{equation}
where $e$ is the orbital eccentricity (assumed negligible for most X-ray binaries), $i$ the orbital inclination and $q=M_\mathrm{*}/M_\mathrm{BH}$ the mass ratio between the companion star and the BH. Given estimates of $i$ and $q$ or $M_*$, we numerically solve Eq.~(\ref{eq:mass-spectroscopy}) for $M_\mathrm{BH}$.

\paragraph{Astrometric masses.}
For some inert systems, the orbit is constrained by astrometric measurements (\eg Gaia BH1 \citealt{el-badry2023b}, Gaia BH3 \citealt{panuzzo2024a}). These allow an estimate of the BH mass via the astrometric mass function,
\begin{equation}
    f_\mathrm{ast}(M_\mathrm{BH})
    =\left(\frac{a_1}{1\mathrm{\,AU}}\right)^3\left(\frac{P}{\mathrm{yr}}\right)^{-2} 
    =\frac{M_\mathrm{BH}^3}{\left(M_\mathrm{BH}+M_\mathrm{*}\right)^2},
    \label{eq:mass-astrometry}
\end{equation}
where $a_1$ is the semi-major axis of the photocenter orbit. 
    
\paragraph{Lensing masses.}
We derive the mass of lensing BHs from the Einstein radius $\theta_\mathrm{E}$ and the relative parallax of the lens $\pi_\mathrm{E}$ \citep[\eg][]{sahu2022a}. This yields
\begin{equation}
    M_\mathrm{BH} = \frac{\theta_\mathrm{E}}{\kappa \pi_\mathrm{E}},
    \label{eq:mass-grav-lensing}
\end{equation}
with $\kappa = 4G/[c^2(1\mathrm{\,AU})] \approx 8.144 \mathrm{\,mas\,\msun^{-1}} $. 

\begin{figure*}
    \centering
    \includegraphics[width=\linewidth]{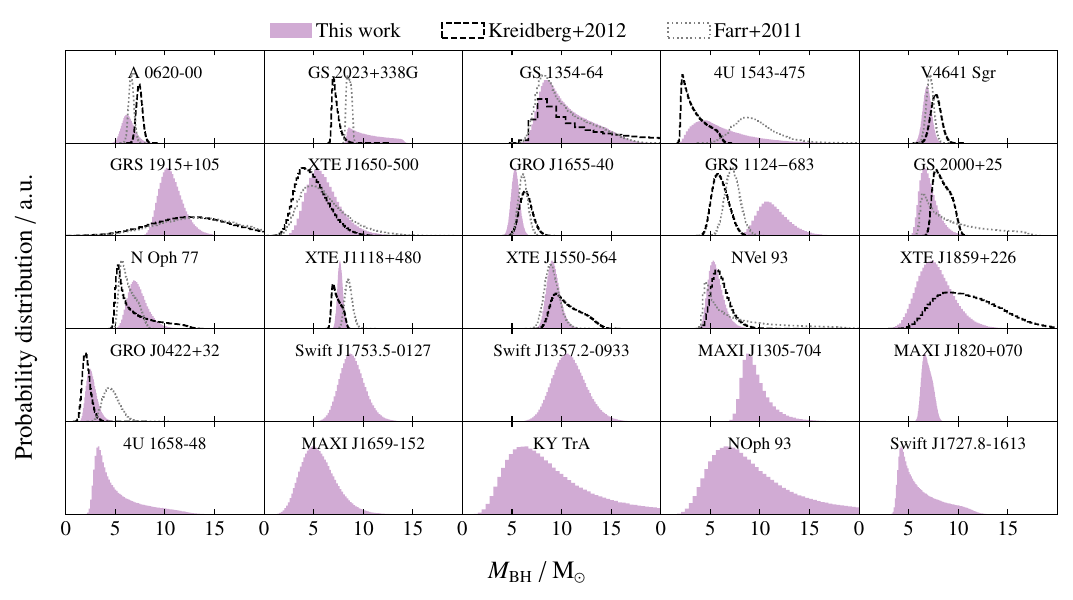}
    \caption{Individual mass distributions for the black holes in LMXBs inferred with Eq.~(\ref{eq:mass-spectroscopy}) from the parameters in Table \ref{tab:XRB} (filled histograms). The gray dashed and the black dotted lines indicate the distributions used in \cite{farr2011a} and \cite{kreidberg2012a}, respectively. Each panel is normalized to unit area.}
    \label{fig:LMXB}%
\end{figure*}

\begin{figure}
    \centering
    \includegraphics[width=\columnwidth]{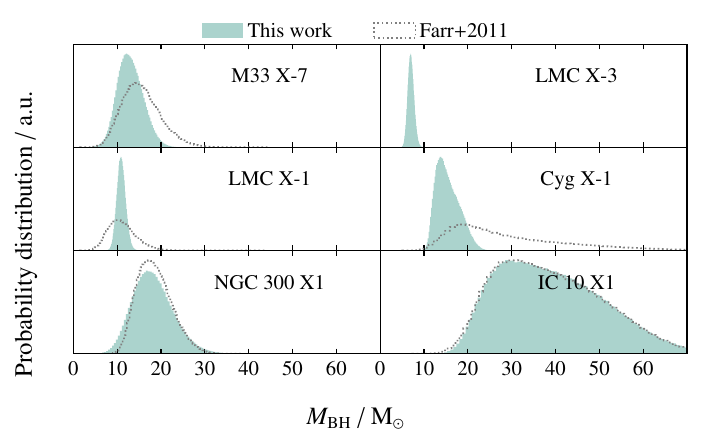}
    \caption{Individual mass distributions for the black holes in HMXBs inferred with Eq.~(\ref{eq:mass-spectroscopy}) from the parameters in Table \ref{tab:XRB}. The gray dashed lines represent the distributions used in \cite{farr2011a}. Each panel is normalized to unit area.}
    \label{fig:HMXB}%
\end{figure}

\begin{figure}
    \centering
    \includegraphics[width=\columnwidth]{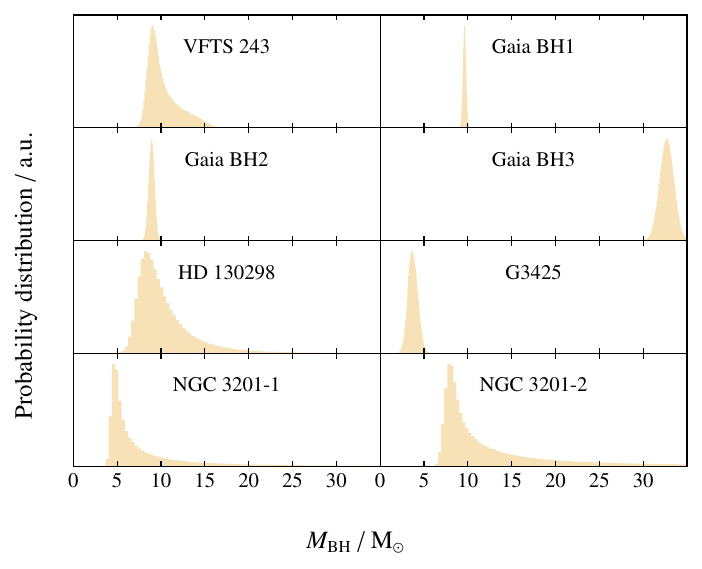}
    \caption{Individual mass distributions for the inert BHs inferred with Eqs.~(\ref{eq:mass-spectroscopy}) or (\ref{eq:mass-astrometry}) from the parameters in Table \ref{tab:Dormant} (filled histograms). Each panel is normalized to unit area.}
    \label{fig:dormant}%
\end{figure}

\begin{figure}
    \centering
    \includegraphics[width=\hsize]{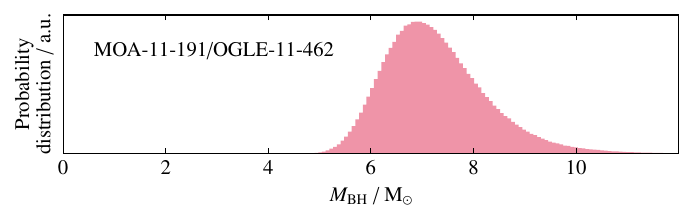}
    \caption{Individual mass distribution of the gravitational lensing black hole inferred with Eq.~(\ref{eq:mass-grav-lensing}) from the parameters in Table \ref{tab:gravL}. The probability distribution is normalized to 1.}
    \label{fig:gravL}%
\end{figure}

\paragraph{Probability density function of black hole masses.}
To account for measurement uncertainties when deriving the masses of the BHs, we apply the following steps.
\begin{itemize}
    \item For well-constrained quantities with symmetric errors, we adopt a Gaussian distribution centered around the quoted value, with a standard deviation equal to the reported uncertainty.
    \item For asymmetric uncertainties, we use split-normal distributions, with a mode at the measured value and separate widths for the lower and upper errors.
    \item For poorly constrained parameters, we assume uniform distributions (or isotropic distributions for inclination angles).
\end{itemize}
We provide the full set of assumed distributions for the observables in Appendix~\ref{appendix:masses}. From these, we derive the BH mass distributions by performing a Monte Carlo error propagation. For each system, we sample parameter values from the adopted distributions to propagate the uncertainties to the BH mass distributions via the corresponding mass equations (Eqs.~\ref{eq:mass-spectroscopy}, \ref{eq:mass-astrometry}, and~\ref{eq:mass-grav-lensing}). These distributions are shown for the different sub-groups in Figs.~\ref{fig:LMXB}--\ref{fig:gravL}.

\subsection{\label{sec:methods}Statistical analysis}
This section outlines the Bayesian analysis used to model the observed BH mass distribution. We describe the Bayesian framework by \citet{farr2011a} in Sect.~\ref{sec:bayesian-framework}, and the four parametric models for the BH mass distributions in Sect.~\ref{sec:BH-mass-dist-models}. 

It is important to distinguish between the observed, the intrinsic and the birth mass distributions of the BHs in our EM sample. In this work, we constrain the observed BH mass distribution. In future studies, incorporating the selection function of the various BH samples used here can deliver the intrinsic mass distribution of the BHs in the Milky Way and nearby galaxies, and further accounting for the evolutionary history of the binary systems (\eg correcting for mass accretion onto some of the BHs) may give access to the birth mass distribution of BHs.

\subsubsection{Bayesian inference}\label{sec:bayesian-framework}
Let a set of parameters $\vec{\theta}$ govern the observed black hole mass distribution, and $D$ be the observational data. Bayes' theorem states that 
\begin{equation}\label{eq:Bayes}
    p(\vec{\theta} \vert D)=\frac{p(D \vert \vec{\theta}) p(\vec{\theta})}{p(D)},
\end{equation}
where the posterior $p(\vec{\theta} \vert D)$ expresses the probability distribution of the parameters $\vec{\theta}$ given the data $D$ \citep[\eg][]{sharma2017a}. The likelihood $p(D \vert \vec{\theta})$ is the probability of observing $D$ if $\vec{\theta}$ characterizes the population BH mass distribution. The prior distribution $p(\vec{\theta})$ encodes our initial knowledge about the model in the absence of data, and the evidence $p(D)$ acts as a normalizing constant.
Our data $D = \{p_j(M): j=1, ..., 40\}$ comprises the individual black hole mass distributions shown in Figs.~\ref{fig:LMXB}--\ref{fig:gravL}. For each black hole $j$, the likelihood is defined as the product of its mass distribution $p_j(M)$ and the probability for a black hole of that mass $M$ to exist, given the population BH model distribution $p(M\vert \vec{\theta})$, integrated over all masses. The product over the independent measurements $j$ gives the total likelihood,
\begin{equation}\label{eq:likelihood}
    p(D \vert\vec{\theta})=\prod_j p(M_j\vert \vec{\theta})=\prod_j \int dM\, p_j(M)\, p(M \vert \vec{\theta}). 
\end{equation}
To reduce computational cost, we approximate the integrals as averages over Monte Carlo mass samples from Tables \ref{tab:XRB}--\ref{tab:gravL},
\begin{equation}\label{eq:approx}
    p(D \vert \vec{\theta}) \approx \prod_j \frac{1}{N_j} \sum_{k=1}^{N_j} p\left(M_{jk} \vert \vec{\theta}\right),
\end{equation}
following \citet{farr2011a}. $M_{jk}$ is the $k$-th of $N_j$ samples from the $j$-th black hole mass distribution. We found $N_j =100\,000$ guarantees convergence at reasonable computational costs.

Evaluating the full posterior (Eq.~\ref{eq:Bayes}) for all possible $\vec{\theta}$ values becomes computationally expensive in multidimensional parameter spaces. We therefore employ the Markov Chain Monte Carlo (MCMC) method, using the parallel stretch-move algorithm implemented in \texttt{emcee} \citep{foreman-mackey2013a}. This generates a Markov chain of samples ${\vec{\theta}}_{l=1,\dots}$ via a random walk through parameter space, guided by the posterior probability density (Eq.~\ref{eq:Bayes}). As a result, parameter space points with higher posterior probability appear more frequently in the chain. Each step only depends on the previous one, \ie, $p(\theta_{l}\vert \theta_1,\dots, \theta_{l-1}) = p(\theta_{l}\vert \theta_{l-1})$. With 64 walkers and 100,000 steps, the chain samples the posterior distribution, balancing quality and computational cost.

\subsubsection{Models and their parameters for the black hole mass distribution}\label{sec:BH-mass-dist-models}
Using this Bayesian tool, we fit four parametric models to best describe the population mass distribution of our observed EM BHs from Sect.~\ref{sec:obs-data}. The priors for the parameters are all flat within a broad range, remaining agnostic about the mass distribution of BH binary systems. We do not impose a lower limit on BH masses based on a maximum NS mass.

\paragraph{Gaussian.}
Most of the individual BH mass distributions (\cf Fig.~\ref{fig:all-masses} below) concentrate in a narrow range around $\sim 9\,\msun$. A simple approach is to describe them with a Gaussian distribution,
    \begin{align}\label{eq:gauss}
        p(M_\mathrm{BH} \vert \vec{\theta}) & =p\left(M_\mathrm{BH} \vert\left\{\mu, \sigma\right\}\right) \\
        & = \frac{1}{\sqrt{2\pi\sigma^2}}\exp{\left[-\frac{(M_\mathrm{BH}-\mu)^2}{2\sigma^2}\right]},
    \end{align}
centered around $\mu$ and with a standard-deviation $\sigma$, both measured in solar masses. The prior function
    \begin{equation}\label{eq:prior_gaussl}
        p(\left\{\mu,\sigma\right\})= \begin{cases}\frac{1}{50} \frac{1}{30}&  0\leqslant \mu \leqslant 50, \quad 0\leqslant\sigma\leqslant 30 \\ 0 & \text { otherwise, }\end{cases}
    \end{equation}
ensures a peak below $50\,\msun$, where most samples are located. 

\paragraph{Split normal.}
As the individual mass distributions appear somewhat antisymmetric (\cf Fig.~\ref{fig:all-masses} below), with more narrowly constrained masses at lower masses than at high masses, we can also fit a split-normal distribution,
\begin{align}\label{eq:split-normal}
    p(M_\mathrm{BH}\vert\vec{\theta})&=p\left(M_\mathrm{BH} \vert\left\{\mu, \sigma_1, \sigma_2\right\}\right)\\&= \frac{\sqrt{2}}{\sqrt{\pi}\left(\sigma_1 +\sigma_2\right)}\exp\left[-\frac{(M_\mathrm{BH}-\mu)^2}{2\sigma_i^2}\right], \\\quad i &= \begin{cases}1& \text{if } 0\leqslant M_\mathrm{BH}< \mu, \\ 2  & \text{if }\mu\leqslant M_\mathrm{BH},\end{cases}
\end{align}
with a peak $\mu$ and different standard-deviations on the left $\sigma_1$ and right side $\sigma_2$ of the peak. With the prior function
    \begin{equation}\label{eq:prior_split-normal}
        p(\left\{\mu,\sigma_1, \sigma_2\right\})= \begin{cases}\frac{1}{50} \frac{1}{15}\frac{1}{30}&  0\leqslant \mu \leqslant 50, \quad 0\leqslant\sigma_1\leqslant 15,\\&  0\leqslant\sigma_2\leqslant 30 \\ 0 & \text { otherwise, }\end{cases}
    \end{equation}
we ensure again that the peak is below $50\,\msun$ and accept larger tails on the right side of the peak.
        
\paragraph{Power law.}\label{sec:PL}
Since a power-law distribution describes the stellar initial mass function well, it is natural to assume that the BH masses could also follow a power-law function, as is assumed in \citet{farr2011a} and \citet{kreidberg2012a},
\begin{align}\label{eq:powlaw}
    p(M_\mathrm{BH} \vert \vec{\theta}) & =p\left(M_\mathrm{BH} \vert\left\{M_{\min }, M_{\max }, \alpha\right\}\right) \\
    & = \begin{cases}A M_\mathrm{BH}^\alpha & M_{\min } \leqslant M_\mathrm{BH}\leqslant M_{\max } \\
    0 & \text { otherwise }\end{cases},
\end{align}
with a sharp cutoff below a threshold $M_\mathrm{min}$ and above a maximal mass $M_\mathrm{max}$, both given in solar masses. The exponent $\alpha$ is dimensionless, and $A$ is a normalizing constant,
\begin{equation}\label{eq:powlaw_norm}
    A=\frac{1+\alpha}{M_{\max }^{1+\alpha}-M_{\min }^{1+\alpha}}.
\end{equation}
We use the prior function
\begin{equation}\label{eq:prior_pl}
    p(\left\{M_{\min }, M_{\max }, \alpha\right\})= \begin{cases}\frac{2}{100^2} \frac{1}{28}&  0\leqslant M_{\min } \leqslant M_{\max }\leqslant 100,\\&  -15\leqslant\alpha\leqslant 13 \\ 0 & \text { otherwise, }\end{cases}
\end{equation}
accepting larger masses than in \cite{farr2011a} to account for all individual BH masses in our updated sample. 
            
\paragraph{Bimodal distribution: Gaussian + power law.} \label{sec:Bim}
According to \cite{schneider2021a, schneider2023a}, the stellar-mass BH mass distribution may exhibit a bimodal structure: a first peak near $9\,\msun$, a gap between $10$ and $15\,\msun$, and another peak near $16\,\msun$ that gradually decays at higher masses. This motivates a mixed model combining a Gaussian with a power-law function, \ie
\begin{align}\label{eq:bimodal}
    p(M_\mathrm{BH} \vert \vec{\theta}) & =p\left(M_\mathrm{BH} \vert\left\{\mu, \sigma, M_{\min }, M_{\max }, \alpha, r\right\}\right) \\
    & = \begin{cases} 
        r B  \exp \left[-\frac{\left(M-\mu\right)^2}{2 \sigma^2}\right] & 0\leqslant M_\mathrm{BH} \leqslant M_{\min} \\
        (1-r)A M^\alpha & M_{\min } \leqslant M_\mathrm{BH}\leqslant M_{\max } \\
        0 & \text { otherwise. } \end{cases}
\end{align}
$B$ ensures the normalization of the Gaussian with a hard cut at 0 and $M_{\min}$,
\begin{equation}
    B = \frac{\sqrt{2}}{\sqrt{\pi}\sigma}\left[ \mathrm{erf}\left(\frac{\mu}{\sqrt{2}\sigma}\right)-  \mathrm{erf}\left(\frac{\mu-M_{\min}}{\sqrt{2}\sigma}\right)\right]^{-1}.
\end{equation}
$A$ normalizes the power law as in Eq.~(\ref{eq:powlaw_norm}). The additional parameter $r$ indicates the mixing ratio between the two models. We use a uniform prior distribution on all parameters $\vec{\theta}_\mathrm{bimodal} = \left\{\mu, \sigma, M_{\min }, M_{\max }, \alpha, r\right\}$ within generous ranges,
\begin{equation}\label{eq:prior_bimodal}
    p(\vec{\theta}_\mathrm{bimodal})= \begin{cases}\frac{6}{100^3} \frac{1}{15}\frac{1}{20} & 0\leqslant \mu\leqslant M_{\min } \leqslant M_{\max }\leqslant100,\\
        &  -10\leqslant\alpha\leqslant 10,\\ & 0<r<1, \\&0<\sigma<15
        \\ 0 & \text { otherwise. }\end{cases}
\end{equation}
This prior ensures the Gaussian is applied to lower masses than the power law, over the range $0$ to $100\mathrm{\,M_\odot}$. As we have no prior information on the mixing ratio $r$, it is uniform over [0,1].

\section{\label{sec:bh-mass-distr}Mass distribution of black holes}
In this section, we analyze the trends in the observed black hole masses from Sect.~\ref{sec:obs-data}, then discuss the results of our hierarchical Bayesian analysis and present the posterior distributions of the parameters for the models from Sect.~\ref{sec:methods}.

\subsection{Observed BH mass distributions}

\begin{figure*}
    \centering
    \includegraphics[width=0.85\linewidth]{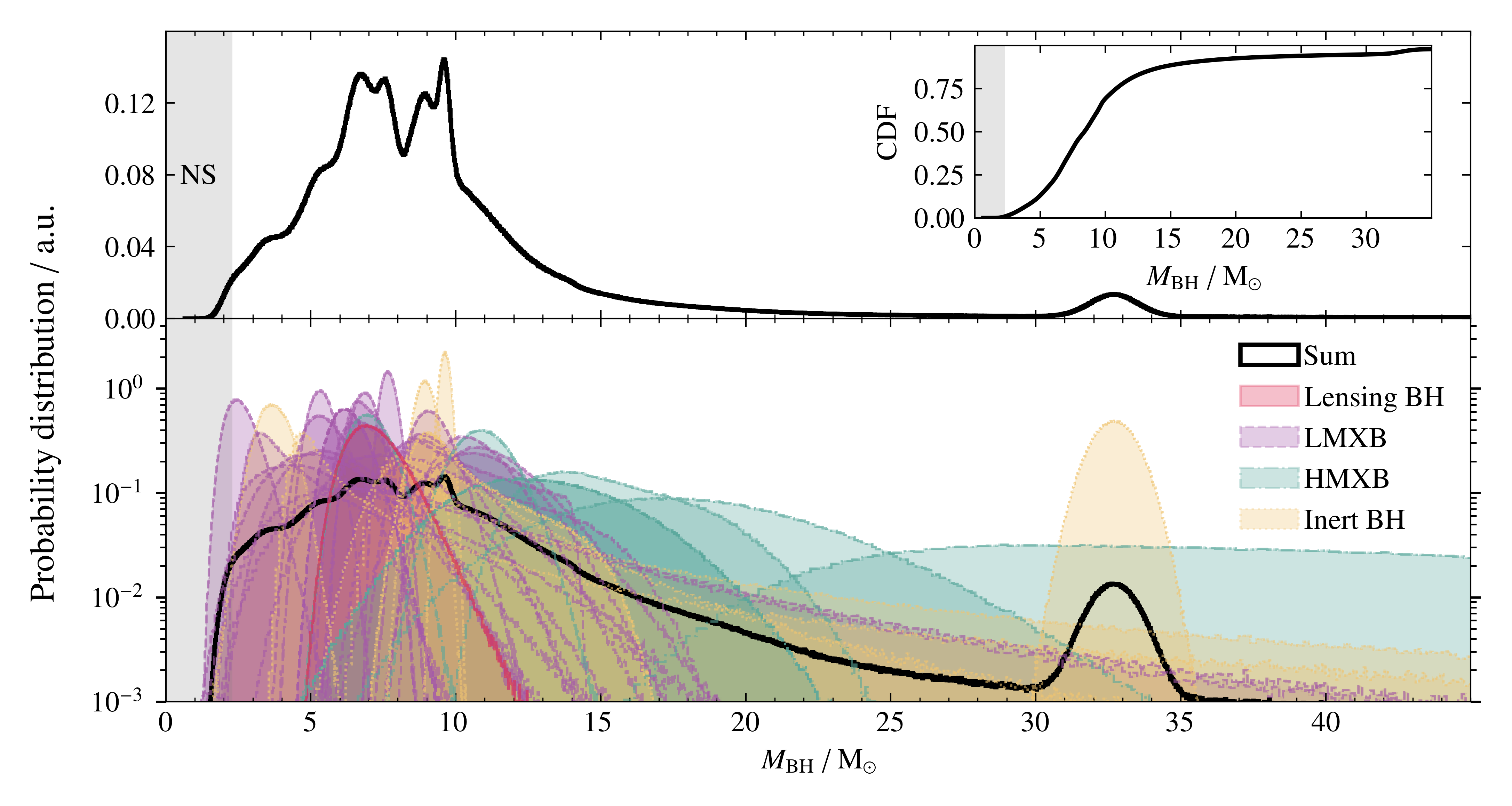}
    \caption{Mass distributions of all BHs included in our sample. Each probability density function (PDF) is normalized to unit area. The black curve shows the sum of these PDFs (also normalized to 1). The top panel displays the summed distribution on a linear y-axis, along with the cumulative distribution functions (CDF). The bottom panel shows the summed PDF on a logarithmic scale, overlaid with the individual distributions. The gray-filled area indicates the mass range of neutron stars ($\lesssim 2.3\,\msun$ \citealp{bailyn1998a}).}
    \label{fig:all-masses}%
\end{figure*}

We show the posterior mass distributions of individual BHs in Figs.~\ref{fig:LMXB}, \ref{fig:HMXB}, \ref{fig:dormant}, and~\ref{fig:gravL} for the LMXBs, HMXBs, inert/X-ray quiet binaries, and gravitational lensing black holes, respectively.

Figure~\ref{fig:LMXB} compares the LMXB mass measurements used in our analysis (filled histograms) with those used in the analyses of \citet{farr2011a} (gray dotted lines) and \citet{kreidberg2012a} (black dashed lines). We expand the previous samples with nine additional systems and update existing measurements using recent constraints from \citet{casares2027a}. Of particular interest is GRO J0422+32, which previously hovered near the $5\,\msun$ threshold \citep{farr2011a}, while more recent results place it below it \citep{kreidberg2012a, casares2022a}, challenging the existence of a sharp NS-BH mass gap. Additional systems such as 4U 1658-48 peak below $5\,\msun$, or exhibit extended tails into to the $2\text{--}5\,\msun$ regime (\eg~MAXI J1659-152, KY TrA, N Oph 93, and Swift J1727.8-1613) .

Similarly, Fig.~\ref{fig:HMXB} presents the mass posteriors of HMXBs, updating earlier work by \citet{farr2011a} with revised or newly included measurements. These include LMC X-1, LMC X-3 and M33 X-7 (from \citealt{casares2027a}), as well as Cyg X-1 \citep{miller-jones2021a, ramachandran2025a} and NGC 300 X1 \citep{crowther2010b, binder2021a}. The updated values notably narrow the distributions in most cases, except for IC 10 X1, for which no new constraints exist. 

The newly discovered inert BHs (Fig.~\ref{fig:dormant}) span a broad mass range from systems like G3425 \citep{wang2024b} below $5\,\msun$, up to the massive Gaia BH3 at ${\sim}\, 33\,\msun$  \citep{panuzzo2024a}, although most BHs lie between $5\text{--}10\,\msun$. Lastly, Fig.~\ref{fig:gravL} shows the BH confidently detected via gravitational lensing, MOA-11-191/OGLE-11-462 \citep{sahu2022a, sahu2025a}, with an inferred mass also in the $5\text{--}10\,\msun$ range.
    
Figure~\ref{fig:all-masses} summarizes the individual mass distributions in our sample, along with their cumulative distribution functions (CDFs). We observe a pronounced peak in the $5\text{--}10\,\msun$ range, where most of the sources are located. This range contains approximately $56\%$ of the total integrated probability (\cf the CDF). At higher masses, the distribution steadily decreases, except for a well-defined peak near ${\sim}\,33\,\msun$, caused by a single source: Gaia BH3. The high-mass regime (${\gtrsim}\,10\,\msun$) comprises roughly $30\%$ of the sample.

At the low-mass end, BHs become even scarcer, and the summed PDF shows a smooth tail extending below $5\,\msun$ with about $13\%$ of the data in this regime, contrary to earlier claims of a sharp lower-mass cutoff \citep{bailyn1998a, ozel2010a, farr2011a, kreidberg2012a}. Several BH masses, such as the LMXB GRO J0422+32 and the inert BH G3425, fall entirely below this threshold, while others (\eg 4U 1658-48, 4U 1543-475, MAX J1659-152, KY TrA, N Oph 93, and Swift J1727.8-1613) have distributions extending into this region. 

Interestingly, BHs in X-ray binaries, in inert BH systems and from gravitational lensing have similar masses. Also, their mass distributions seem to share the same qualitative features, possibly implying similar formation channels (see Sect.~\ref{sec:bh-subsamples}).

\subsection{Inferred parametric mass distribution of BHs}

\begin{table}
\caption{Best-fit parameters for the BH mass distribution models.}
\label{tab:Res_all}
\begingroup
\renewcommand{\arraystretch}{1.5} 
\small
	\begin{tabular}{l l l l }
		\hline \hline 
		Model & Parameter & Best fit & \\ \hline 
		Gaussian (Eq.\ref{eq:gauss}) &$\mu/\msun$ & $8.2_{-0.5}^{+0.5}$\\
		& $\sigma/\msun$&$2.6_{-0.4}^{+0.5}$&\\\hline

       Split Normal (Eq. \ref{eq:split-normal}) &$\mu/\msun$ &$7.0_{-1.4}^{+1.4}$\\
		& $\sigma_1/\msun$ & $1.9_{-0.8}^{+0.9}$&\\
		& $\sigma_2/\msun$ &$3.6_{-1.1}^{+1.3}$&\\\hline
		
		Power law$^{a}$ (Eq. \ref{eq:powlaw}) &$M_{\min}/\msun$ &$5.0_{-1.0}^{+0.8}$&$3.6_{-2.1}^{+1.1}$\\
		& $M_{\max}/\msun$ & $19.2_{-5.7}^{+68.7}$&$12.8_{-0.7}^{+3.1}$\\
		& $\alpha$ &$-2.9_{-1.0}^{+1.4}$&$0.2_{-1.6}^{+0.9}$\\\hline
		Bimodal$^{b}$ (Eq. \ref{eq:bimodal}) &$\mu/\msun$ & $7.9_{-0.7}^{+0.8}$& $8.0_{-1.1}^{+2.2}$\\
		& $\sigma/\msun$& $2.2_{-0.5}^{+0.7}$& $2.4_{-0.8}^{+1.5}$ \\
		&$M_{\min}/\msun$ & $22.5_{-6.6}^{+26.2}$& $10.3_{-2.0}^{+4.5}$\\
		& $M_{\max}/\msun$& $52_{-21}^{+44}$& $37_{-17}^{+54}$ \\
		& $\alpha$  &$-9.4_{-0.6}^{+16.3}$ & $-4.2_{-5.5}^{+10.3}$\\
		& $r$  & $0.97_{-0.07}^{+0.03}$&  $0.96_{-0.34}^{+0.04}$\\\hline
	\end{tabular}
    \endgroup
    \tablefoot{We indicate the mode and the 68\% interval of confidence of the marginalized posterior distributions.\\
    \tablefoottext{a}{As the power law model finds two different solutions, we report the parameter values around each.}
    \tablefoottext{b}{We find two solutions for $M_\mathrm{min}$, and we indicate the parameter values from the marginalized distributions around the two different solutions.}}
\end{table}
    
\begin{figure}
    \centering
    \includegraphics[width=\linewidth]{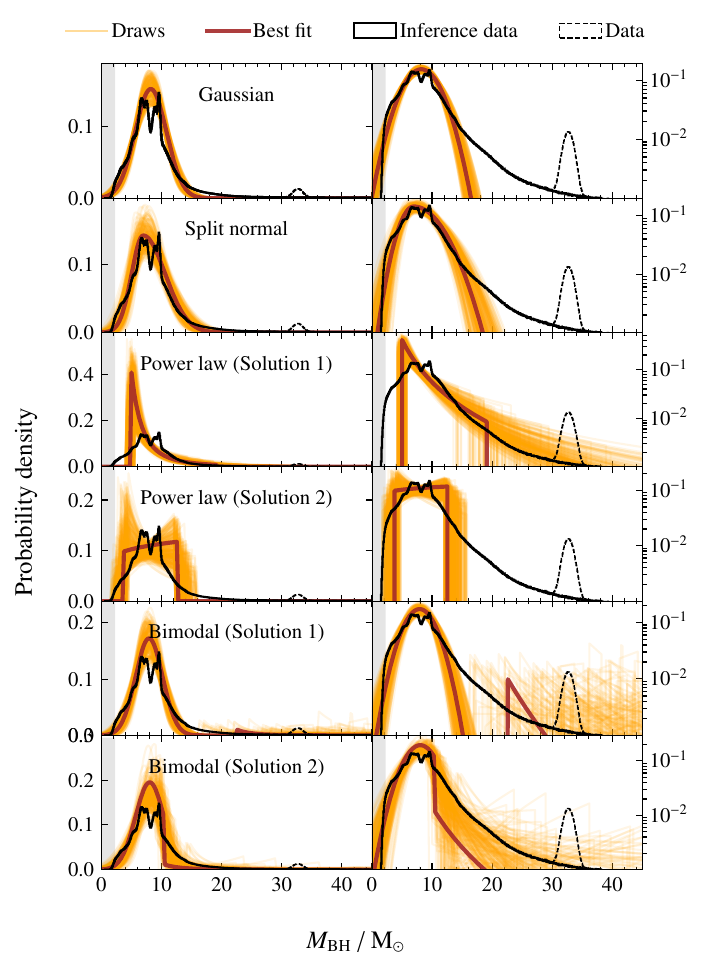}
    \caption{BH mass distribution $p(M\vert\vec{\theta})$ implied by the posterior $p(\vec{\theta}\vert M)$ and the observational data from Sect.~\ref{sec:obs-data} for the models discussed in Sect.~\ref{sec:methods}. The brown line gives the distributions implied by the best-fit parameters, and the orange lines use parameters sampled from the 68\% interval of confidence listed in Table \ref{tab:Res_all}. The black histogram indicates the summed probability density function shown in Fig.~\ref{fig:all-masses}. The shaded region indicates the mass range of neutron stars. }
    \label{fig:res}
\end{figure}

As our sample only comprises 40 BH mass measurements with varying uncertainties, it is essential to characterize the population mass distribution using a statistical model. To this end, we apply the Bayesian inference framework detailed in Sect.~\ref{sec:bayesian-framework} and investigate the posterior distributions of model parameters via MCMC sampling. 
Table~\ref{tab:Res_all} summarizes the modes and $68\%$ highest density intervals (HDIs) of the marginalized posterior distributions of the model parameters implied by the observational data. Figure~\ref{fig:res} pictures the resulting mass distributions: we plot the given models with parameter values sampled within the $68\%$ HDI over the sum of the observed PDFs for comparison. 

For this analysis, we exclude Gaia BH3 because of its exceptionally low metallicity \citep{panuzzo2024a, marinpina2024a}. Including this source would introduce a strong bias, as the other BHs are most likely from stars of higher metallicity. For completeness, we provide a full inference including Gaia BH3 in Appendix~\ref{appendix:GaiaBH3} that also shows how this single BH negatively influences the fit quality.

\subsubsection{Gaussian}
Figure~\ref{fig:Gauss-corner} displays the one-dimensional marginalized posteriors for the Gaussian model parameters and their correlations. The analysis yields a peak mass of $\mu = 8.2_{-0.5}^{+0.5}\,\msun$ and a dispersion of $\sigma = 2.6_{-0.4}^{+0.5}\,\msun$ (68\% HDI). The Gaussian thus provides a reasonable description of the main component in the BH mass distribution, but it fails to reproduce the observed BH mass distribution at ${\gtrsim}\, 15\,\msun$ (Fig.~\ref{fig:res}).

\subsubsection{Split-normal model}
Figure~\ref{fig:SN-corner} shows the marginalized posterior distributions of the parameters $\mu$, $\sigma_1$, and $\sigma_2$ of the split-normal model, as well as their mutual correlations. In contrast to the symmetric Gaussian, the split-normal distribution introduces skewness, allowing a steeper decline at low masses and a more gradual one toward higher masses. We infer a peak at $7.0_{-1.4}^{+1.4}\,\msun$, with a sharp low-mass tail ($\sigma_1 =1.9_{-0.8}^{+0.9} \,\msun$) and a broader high-mass tail ($\sigma_2 = 3.6_{-1.1}^{+1.3}\,\msun$), all quoted as 68\% HDIs. 

The split-normal therefore provides a good description of the main component at $5\text{--}10\,\msun$ and reproduces the data's behavior in the low- and intermediate-mass ranges within the stated uncertainties. Its main limitation lies in the highest mass regime, where the model predicts fewer systems than observed (Fig.~\ref{fig:res}); however, this region is only sparsely populated with BH measurements with large uncertainties.

\subsubsection{Power-law model}
In Fig.~\ref{fig:PL-corner}, we plot the posterior distributions of the power-law model parameters: the minimum and maximum BH masses, $M_\mathrm{min}$ and $M_\mathrm{max}$, and the slope $\alpha$. We identify two distinct power-law solutions. Solution 1 has a negative slope and accounts for about $60\%$ of the posterior samples, while Solution 2 has a positive slope and comprises the remaining $40\%$ (\cf Fig.~\ref{fig:res}).

Solution 1 truncates the distribution at the low-mass end, with a minimum mass at $M_\mathrm{min,1}= 5.0_{-1.0}^{+0.8}\,\msun$. We find a well-constrained, negative slope $\alpha_1 =-2.9_{-1.0}^{+1.4}$. The maximum mass, although showing a distinct peak at $M_\mathrm{max,1}= 19_{-6}^{+69}\,\msun$ has no clear upper constraint, resulting in a broad range of possible masses ($16\text{--}88\,\msun$), extending towards even higher masses if allowed for by the prior distribution.

Solution 2 instead places the minimum mass at $M_\mathrm{min,2}= 3.6_{-2.1}^{+1.1}\,\msun$ and yields a slope $\alpha_2 =0.2_{-1.6}^{+0.9}$, capturing a nearly constant BH population between the NS mass range and the BH peak. However, this solution imposes an unphysical cutoff at $M_\mathrm{max,2}= 12.8_{-0.7}^{+3.1}\,\msun$, excluding all BHs with higher masses.

Compared to the two previous models, the power-law Solutions 1 and 2 more accurately capture the low-mass increase and the high-mass tail of the observed distribution, respectively. However, each performs poorly at the opposite end of the mass spectrum, due to the unphysical sharp cutoff beyond the peak, which fails to account for $13\text{--}24\%$ of the data. 

Notably, the negative-exponent solution is also highly sensitive to individual systems: excluding single BHs such as IC 10 X-1 --- whose mass estimate carries large uncertainties and remains debated \citep[][]{laycock2015a} --- eliminates the negative solution, leaving only a positive-exponent power-law.

\subsubsection{Bimodal distribution: Gaussian + power law}\label{res:bimodal}

Figure~\ref{fig:bi-corner} shows the marginalized posterior distributions and correlations for the parameters of the bimodal model, which combines a Gaussian component (with parameters $\mu$ and $\sigma$), followed by a power law (defined by $M_\mathrm{min}$, $M_\mathrm{max}$ and $\alpha$). These two components are mixed according to a weight $r$, governing the relative contributions of the Gaussian ($r$) versus the power law ($1-r$) component. The MCMC samplings find three distinct solutions for the minimum mass of the power-law component. The two dominant modes comprise approximately 65\% and 32\% of the posterior samples, placing the minimum masses of the power-law component above $15.9\,\msun$ (Solution 1) and between $8.3\text{--}14.8\,\msun$ (Solution 2), respectively (Fig.~\ref{fig:res}). The third mode, with $M_\mathrm{min} < 7.8\,\msun$, accounts for only about $4\%$ of the posterior and essentially overlaps with the simple power law model. We therefore only report on the two primary solutions.

Both solutions heavily favor the Gaussian component, with mixing ratios of $r_1=0.97_{-0.07}^{+0.03}$ and $r_2=0.96_{-0.34}^{+0.04}$. The Gaussian peaks are consistent across both modes with $\mu_1=7.9_{-0.7}^{+0.8}\,\msun$ and $\mu_2=8.0_{-1.1}^{+2.2}\,\msun$. The standard-deviation of Solution 1 is $\sigma_1=2.2_{-0.5}^{+0.7}\,\msun$, while Solution 2 has a larger one, $\sigma_2=2.4_{-0.8}^{+1.5}\,\msun$. The two solutions differ mostly in the onset of the power-law tail. In Solution 1, the power law begins at $M_\mathrm{min,1}=22.5_{-6.6}^{+26.2}\,\msun$, roughly $5\sigma$ above the Gaussian peak, leaving a gap between $14\text{--}20\,\msun$; a similar feature may exist in the GW data (\cf Sect.~\ref{sec:discussion-gw-data}). In contrast, Solution 2 starts the power law only $1\sigma$ above the mean namely at $M_\mathrm{min,2}=10.3_{-2.0}^{+4.5}\,\msun$. In both cases, the maximum is loosely constrained at $M_\mathrm{max,1} = 52_{-21}^{+44}\,\msun$ and $M_\mathrm{max,2} = 37_{-17}^{+54}\,\msun$, similar to the single-component power law model with negative slope. The resulting power law slopes span wide ranges with $\alpha_1 = -9.4_{-0.6}^{+16.3}$ and $\alpha_2 = -4.2_{-5.5}^{+10.3}$, allowing for both positive and negative slope solutions within the 68\% HDI. These intervals remain broad due to the limited number of available high-mass black hole measurements.

In summary, the higher number of degrees of freedom of the bimodal distribution allows for a more flexible and qualitatively better fit to the observed mass distribution than single-component models. However, the existence of multiple viable solutions due to wide parameter uncertainties, especially on the second component, reflects the current limitations imposed by the sample size. Stronger constraints thus require additional data.

\subsection{Neutron star--black hole mass gap}

\begin{table}
    \centering
    \caption{Fraction of BHs in the given mass intervals for the inferred models and the current observational data. $M_{1\%}$ indicates the first percentile of BH masses in each case.}
    \label{tab:probs}
    \begingroup\renewcommand{\arraystretch}{1.5}\small
    \begin{tabular}{c|c c c|c}
    \hline\hline
         & $< 5\,\msun$ & $=5\text{--}10\,\msun$& $> 13\,\msun$& $M_{1\%}$ \\\hline
         Data & $0.13$&$0.56$& $0.15$&$2.3\,\msun$\\
         Gauss  & $0.09_{-0.03}^{+0.06}$ & $0.62_{-0.08}^{+0.07}$ & $0.02_{-0.02}^{+0.04}$&$2.2_{-1.1}^{+0.9}\,\msun$\\
         SN & $0.11_{-0.05}^{+0.06}$ & $0.59_{-0.08}^{+0.08}$&$0.05_{-0.04}^{+0.05}$ &$2.7_{-1.2}^{+0.7}\,\msun$ \\
         PL (Sol. 1) & $0.003_{-0.003}^{+0.105}$ & $0.78_{-0.18}^{+0.06}$& $0.12_{-0.05}^{+0.05}$ & $5.0_{-0.6}^{+0.4}\,\msun$\\
         PL (Sol. 2)& $0.14_{-0.05}^{+0.06}$ &$0.50_{-0.07}^{+0.05}$ & $0.0_{-0.0}^{+0.08}$ & $3.7_{-1.0}^{+0.6}\,\msun$\\ 
         Bi (Sol. 1) & $0.08_{-0.04}^{+0.05}$ & $0.68_{-0.09}^{+0.06}$& $0.05_{-0.03}^{+0.04}$& $2.8_{-0.9}^{+0.8}\,\msun$ \\     
         Bi (Sol. 2) & $0.09_{-0.04}^{+0.05}$ & $0.68_{-0.1}^{+0.08}$& $0.03_{-0.02}^{+0.05}$ & $2.5_{-1.1}^{+0.9}\,\msun$\\
         \hline
    \end{tabular}
    \endgroup
    \tablefoot{``Gauss'', ``SN'', ``PL'' and ``Bi'' refer to the Gaussian, split-normal, power-law and bimodal BH mass model. The ``data'' row has been calculated from the stacked posterior distributions of the individual BHs. Measurement uncertainties of the BHs thus influence these values. For example, the low-mass (${<}\,5\msun$) and high-mass (${>}\,13\msun$) BHs may thus partly reflect measurement scatter.}
\end{table}

Previous studies based on X-ray binaries \citep[\eg][]{ozel2010a, farr2011a, kreidberg2012a} have inferred a lower limit to stellar BH masses, suggesting a mass gap between the maximum NS mass and the minimum BH mass. However, more recent statistical analyses on BHs detected via GWs, \citep[\eg][]{ray2025a}, simulations of LMXB populations \citep{siegel2023a}, as well as new BH detections \citep{abac2024a} have challenged this picture, showing BHs within the previously presumed ${\sim}\,3\text{--}5\,\msun$ gap. We present a quantitative analysis of our BH sample (doubling the size of that used in \citealt{farr2011a}) and of the statistical models fitted to it, both of which are inconsistent with a completely empty $3\text{--}5\,\msun$ region.

Table~\ref{tab:probs} quantifies the probabilities of finding a BH within specific mass intervals under each model and reports the 1\% quantile of the CDF, $M_{1\%}$, defined as the mass below which ${\leq}\,1\%$ of BHs are expected. In both the raw data and most models, the probability of finding a BH with $M\,{\leq}\, 5\,\msun$ is non-negligible, ranging from ${\sim}\,9\%$ for the Gaussian and bimodal solutions to ${\sim}\,13\%$ in the data. Almost all models predict a minimum BH mass between $M_{1\%,\mathrm{Gauss}}=2.2_{-1.1}^{+0.9}\,\msun$ and $M_{1\%,\mathrm{PL2}}=3.7_{-1.0}^{+0.6}$, near or below the maximum NS mass, implying no mass gap.

Only the power-law solution with a negative slope suppresses low-mass BHs, assigning a probability of ${\lesssim}\,1\%$ below $M_{1\%, \mathrm{PL1}} = 5.0_{-0.6}^{+0.4}\,\msun$, thus treating masses below that threshold as statistical outliers. While this can indicate that there is a mass gap after all, it may also mean that the historically favored power law does not describe the BH mass distribution sufficiently well. Indeed, this behavior strongly contrasts with both the data and the other models, particularly the bimodal mass distribution, which, even with six degrees of freedom, consistently finds minimum masses well below $5\,\msun$. 

A few systems stand out against the existence of a mass gap. The LMXB GRO J0422+32 and the inert BH system G3425, for example, have masses below $5\,\msun$ within $95\%$ confidence. While past low-mass BH candidates have sometimes been reclassified (\eg, V723 Mon, first assumed to be a BH in \citet{jayasinghe2021a}, was later found to be a hot star with a stripped red giant companion \citep[][]{el-badry2022a, kochanek2025a}, more recent GW detections such as GW230529 \citep{abac2024a} confirm the existence of BHs in the $3\text{--}5\,\msun$ region. Meanwhile, LMXB population synthesis studies \citep[\eg][]{siegel2023a} also find a fair number of low-mass BHs and suggest that the original gap may have stemmed from selection biases. These observational biases imply that some systems in this mass range evaded detection (\cf Sect.~\ref{sec:discussion-selection-effects}).

Although the $3\text{--}5\,\msun$ mass range is populated, BHs do become scarcer at these low masses (see Table~\ref{tab:probs}). Over half of our sample lies in the $5\text{--}10\,\msun$ range, with about $33\%$ falling within the $5\text{--}8\,\msun$. In contrast, only $13\%$ lie below $5\,\msun$ (and presumably above NS upper mass limit $\sim2\text{--}3\,\msun$). The statistical models mirror this trend. For instance, both bimodal solutions place over $30\%$ of their probabilities between $5$ and $8\,\msun$, but only $\sim 8\%$ below $5\,\msun$. The Gaussian and the split-normal models similarly predict declining frequencies at lower masses.

Mass measurements of BHs are notoriously challenging. \citet{casares2027a} introduce a ``golden'' sample of BHs in LMXBs with the best mass measurements. This subsample is more sharply truncated at $5\,\msun$ and none of the systems in this sample lies in the $3\text{--}5\,\msun$ mass range\footnote{GRO J0422+32 is not part of the golden LMXB sample because its inclination angle is not robustly constrained due to the large flickering contamination that affects its ellipsoidal light curve modeling; instead, the inclination angle is based on the depth of the double-peaked H$\alpha$ trough, which is considered less robust \citep{casares2027a}.}. Limiting the XRB BH sample to these more robust BH masses from LMXBs, there would only be one source, the inert BH G3425, in the $3\text{--}5\,\msun$ mass range, showcasing how sensitive any conclusions are given the still low number of BHs from EM observations.

In conclusion, the current EM data and fitted models appear inconsistent with a completely empty $3\text{--}5\,\msun$ region, \ie a sharp NS-BH mass gap. Instead, they support a steep but continuous decrease in BH numbers below $5\,\msun$, consistent with recent GW-based findings \citep[\eg][]{ray2025a} and bias-corrected simulations of LMXB formation \citep{siegel2023a}.

\subsection{High mass BHs}
At higher masses, one-component models fail to capture the structure of the observed BH mass distribution. The EM sample shows a peak at $5\text{--}10\,\msun$, with a steep decline below and a more gradual tail above, including a handful of BHs with $M\,{\gtrsim}\,20\,\msun$. 

A single Gaussian captures the central peak, but, being symmetric, fails to reproduce the high-mass tail (Fig.~\ref{fig:res}) and underpredicts the fraction of BHs above $13\,\msun$ (Table~\ref{tab:probs}). The split-normal introduces asymmetry and captures both the main peak and the low-mass decline. However, this model still underestimates the high-mass end. A power law with a positive slope similarly provides a good description of the low-mass end but excludes all measurements beyond the peak. Conversely, a power law with a negative slope better accommodates the high-mass tail but truncates the distribution at ${\sim}\,5\,\msun$. As a result, it classifies more than ${\sim}\,14\%$ of the data at ${\lesssim}\,5\,\msun$ as statistical outliers and overpredicts the $5\text{--}6\,\msun$ mass range by a factor of two. These limitations across the applied models indicate that the data likely demand a more flexible, possibly multi-component description. Indeed, the presence of a main peak and a sparsely populated high-mass end may suggest an inherently multimodal/multi-component BH mass distribution \citep[\eg the bimodal model of][]{schneider2023a} and/or possibly reflecting distinct BH formation channels and environments (\eg, LMXBs vs.\ HMXBs, field binaries vs.\ dynamically formed systems, or different metallicities).

A bimodal model---such as the Gaussian peak followed by a power law tail---provides a viable phenomenological description of the observed structure. We identify two possible bimodal solutions, both emphasizing the need for a multi-component structure. However, the high-mass tail remains poorly constrained and even the bimodal models underpredict the fraction of ${\geq}\,13\,\msun$ BHs (Table~\ref{tab:probs}). The slope of the power law is uncertain, as even the sign is ambiguous---\ie, the data are consistent with both increasing and decreasing BH numbers with mass. Similarly, the high-mass cutoff is unconstrained, mostly due to the limited number of BHs in this mass region. Robustly constraining the high-mass component will require significantly more BH detections. The upcoming Gaia DR4, scheduled for the end of 2026, holds promise here, as new astrometric detections of inert BHs \citep[\eg][]{el-badry2023a, el-badry2023b, panuzzo2024a} may significantly expand the sample.

In summary, single-component models struggle to explain the observed distribution. In particular, the shape of the high-mass part of the distribution remains uncertain given the low number of high-mass BHs detected with EM observations.

\subsection{Quantitative model comparison}
We confirm our qualitative model comparison quantitalively by calculating the Aikaike and Bayesian Information Criteria (AIC, BIC) of the different models, with
\begin{equation}\label{eq:AIC}
    \mathrm{AICc} = 2k-2\ln L+\frac{2k^2+2k}{n-k-1},
\end{equation}
and 
\begin{equation}\label{eq:BIC}
    \mathrm{BIC} = k\ln n -2\ln L,
\end{equation}
where $k$ is the number of free parameters of a model, $n$ the sample size and $L$ the maximum likelihood. The last term in the AICc corrects for small sample sizes. We report the results in Table~\ref{tab:AIC-BIC}.

\begin{table}
    \centering
    \caption{Akaike- and Bayesian Information Criteria (AICc and BIC) of the models with respect to the best model (bold).}
    \label{tab:AIC-BIC}\begingroup \small
     \begin{tabular}{c| c| c c }
    \hline\hline
        & $k$ &   $\Delta\mathrm{AICc}$& $\Delta\mathrm{BIC}$ \\\hline
         Gaussian & 2 & \textbf{0.0} & \textbf{0.0}\\
         Split-normal & 3 &  1.3 & 2.6\\ 
         Power-law & 3 & 7.7 & 9.1\\ 
         Bimodal & 6 & 1.9  & 6.2\\ \hline
    \end{tabular}
    \endgroup
\end{table}

The power law is disfavored by both AICc and BIC relative to the other three models ($\Delta\mathrm{AIC}=7.7$, $\Delta\mathrm{BIC}=9.1$). Among the remaining models, AICc shows marginal preference for the Gaussian over the bimodal model ($\Delta\mathrm{AIC}=1.9$), while BIC---which penalizes model complexity more strongly and is more conservative at our sample size ($n=39$)---clearly favors the Gaussian relative to the bimodal model ($\Delta\mathrm{BIC}=6.2$). This shows that although the bimodal model improves the fit to the data, the improvement is modest relative to the four additional parameters it introduces. This is expected, as these two models differ only in how they treat the high-mass component, and both underpredict the number of high-mass BHs (Table~\ref{tab:probs}). 

In summary, the model comparisons show that, of the tested models, a Gaussian best describes the data as it reproduces well the main BH component while requiring the least free parameters. Nevertheless, it underpredicts the number of BHs at ${>}\,13\,\msun$ and thus fails to fully describe the observations.

\subsection{BH subsamples}\label{sec:bh-subsamples}
Mass measurements of BHs are inherently uncertain and often tied to the mass of a companion star. To test the robustness of our main findings and probe possible differences in the observed BH mass distributions across the different BH groups, we repeat our inference on the following subsamples: the ``golden'' subsample of LMXBs that have the most reliable BH mass measurements following \citet{casares2027a}, all LMXBs, HMXBs, inert BH binaries (without Gaia BH3), only solar-like systems\footnote{Defined generously to be within a factor of about 3 within solar metallicity. This sample thus includes systems from the LMC and similar environments but excludes Gaia BH3 and the two BHs in NGC\,3201. However, many BH systems have unknown metallicities, so it remains unclear how well this sample is defined.}, without IC10X1, and only with BHs with mass uncertainties ${<}\,5\,\msun$; the one lensing BH is not considered (see Appendix~\ref{app:subsamples}). 

Qualitatively, all inferences with the parametric models reveal a similar picture for all subsamples (see Appendix~\ref{app:subsamples} and Fig.~\ref{fig:subsamples} and Table~\ref{tab:subsamples}). There is a main component at $5\text{--}10\,\msun$ that a Gaussian function can best describe; the Gaussian model is preferred given the AICc and BIC. The Golden LMXBs sample is more sharply truncated at ${<}\,5\,\msun$ than the full LMXB sample, and the main component of the inert BH binaries leans towards $8\text{--}10\,\msun$ and lacks BHs at ${\sim}\,6\text{--}7\,\msun$. Intriguingly, both the Golden and full LMXB samples show evidence for $6\text{--}7\,\msun$ and $8\text{--}10\,\msun$ BHs (\cf apparent bimodality at these masses in Fig.~\ref{fig:all-masses}). The HMXBs and inert BH binaries have the largest high-mass components. Overall, the inference on these subsamples agrees with that on the entire EM BH sample.

In all subsamples, simple functions such as the Gaussian or power-law models cannot simultaneously reproduce the main BH component at $5\text{--}10\,\msun$ and the BHs at lower and higher masses. BHs with masses above this main component are found mostly in HMXBs and inert BH systems (Gaia BH3). Interestingly, all these massive BHs except for the one in Cyg X-1 formed from stars with sub-solar metallicities (see Tables~\ref{tab:XRB}--\ref{tab:gravL}). This could indicate that massive BHs form more frequently at lower metallicities, as expected from metallicity-dependent stellar winds. Given the massive BH in Cyg X-1, such BHs seem to also form at solar metallicity. Theoretically, one expects a metallicity threshold beyond which (massive) BHs no longer form because of the large wind mass loss \citep{heger2023a}.

\section{\label{sec:discussion}Discussion}

During the refereeing process, \citet{rocha2026a} published a similar study to this one and came to qualitatively similar conclusions. They prefer a power-law model to describe the observed BH mass distribution and also find no evidence for an empty $3\text{--}5\,\msun$ mass range. The difference in their preferred power-law model is likely related to the inclusion of Gaia BH3 in the inference, as well as the assumed mass measurements of BHs. For example, they include the BH imposter 2MASS J0521+4359, a stripped red giant with a subgiant companion \citep{bianchi2024a}, and use the higher-mass estimate for the BH in Cyg X-1 from \citet{miller-jones2021a} rather than the improved BH mass estimate from \citet{ramachandran2025a}. Intriguingly, \citet{rocha2026a} also employ a bimodal model of two Gaussians that can represent the data but is disfavoured in comparison to models with fewer free parameters.

\subsection{\label{sec:discussion-selection-effects}Observational biases and selection effects}
When determining mass distributions of any astrophysical objects, observational biases and selection effects may influence the interpretation of the data \citep[see \eg][for biases affecting the BH mass distribution of X-ray binaries]{ozel2010a, kreidberg2012a, jonker2021a, siegel2023a}. The aggregated observational data of the individual BHs in our sample, shown in Fig.~\ref{fig:all-masses}, and the inferred BH mass distributions in Sect.~\ref{sec:bh-mass-distr} are subject to such observational biases that we discuss in this section and that have not been corrected for.

Many X-ray binaries studied here, especially the LMXBs that dominate this sample, are discovered during their transient phase. At first sight, these discoveries appear almost bias-free as such transient outbursts are likely observable throughout the Galaxy. However, \citet{siegel2023a} show that the duty cycle and transient duration above a threshold detection flux introduce a bias against finding the lowest-mass BHs that may exist in the NS-BH mass gap. Hence, the true number of BHs in X-ray binaries in this mass gap is likely higher than known today, but this bias is unlikely to create a gap or other sharp features. 

Once transient X-ray binaries are detected, follow-up monitoring of the donor star (\ie the BH companion) is required to allow for the inference of the BH mass. This itself introduces another bias \citep[\eg][]{ozel2010a}: more luminous (and hence more massive) donor stars or longer period binaries are observable at greater distances. Furthermore, higher BH masses lead to larger radial-velocity variations, making more massive BHs, in principle, easier to detect. However, \citet{jonker2021a} show that the LMXB population is actually biased against massive BHs because more massive LMXBs likely receive smaller momentum kicks and thus are more difficult to observe because they are found closer to their birthplaces in the Galactic plane, where interstellar absorption disfavors BH mass measurements.

The inert BH binaries and the one lensing BH are subject to different selection functions. Inert BH binaries are either found in spectroscopic surveys or via astrometry with, \eg, Gaia. To zeroth order, spectroscopic surveys are biased towards large companion and BH masses (as detailed above), and astrometric surveys are biased towards more nearby sources and high BH masses, which both create larger astrometric signals. However, our sample of inert BHs from spectroscopic surveys is very heterogeneous, and each survey has its own observational bias. The Gaia selection function for compact object binaries is also complex \citep[\eg][]{el-badry2024a, lam2025a}. The same holds for lensing BHs: such surveys are likely biased against massive BHs as these have longer transient lightcurves that can easily exceed survey durations. In all cases, even if the selection function as a function of BH mass was known, one could not readily apply it because of the still rather low number of observed BHs. 

We conclude that biases are clearly present in the inferred BH mass distribution. Most likely, the current BH sample misses some low-mass BHs (possibly also at ${<}\,5\,\msun$, \ie within the purported NS-BH mass gap) while the net bias for massive (${>}\,10\,\msun$) BHs remains uncertain. As LMXBs dominate our BH sample, we are perhaps biased against massive BHs. Clearly, future work should aim to quantify the selection function.

\subsection{Present-day BH masses and evolutionary biases}
As explained in Sect.~\ref{sec:methods}, the inferred BH mass distribution represents the present-day masses of the observed systems, not their birth masses. This distinction is important, as BHs can accrete mass from their companion or surroundings \citep[\eg][]{tauris2023a}. Although accretion in X-ray binaries is typically sub-Eddington, it may persist over long periods or reach super-Eddington values, and thus become substantial in some systems \citep[\eg][]{tauris2006a, bahramian2023a, fornasini2023a}. Consequently, the observed masses should be interpreted as upper limits (\citealt{fragos2015a} suggest an average mass shift of ${\approx}\,1.5\,\msun$ based on of fully conservative mass transfer models), and one may expect a corresponding systematic shift to lower values for the BH birth mass distribution. Moreover, accreting NSs (even if limited by their Eddington rate) could reach masses of $2\text{--}4\,\msun$ and collapse into low-mass BHs \citep[\eg][]{pfahl2003a}, populating the $3\text{--}5\,\msun$ mass range between NSs and the main BH component. 

Moreover, our sample comprises BHs from different evolutionary channels. LMXBs and HMXBs, for instance, differ both in stellar evolution and accretion mechanisms: in LMXBs, a low-mass (${\lesssim}\, 1\,\msun$) companion feeds the BH via Roche-lobe overflow, whereas HMXBs accrete via stellar winds from a massive ($\gtrsim 10\,\msun$) companion \citep[\eg][]{mandel2021a, tauris2023a, bahramian2023a, fornasini2023a}. Additionally, in stellar dense environments, such as globular clusters, binaries can form dynamically, where BHs may tidally capture another field star or disrupt another binary system, ejecting the least massive star \citep[][]{bahramian2023a}. Such systems do likely not share the same evolution as co-evolved binaries. This implies that even separating binaries by class (\eg LMXB, HMXB, inert binaries) may still combine different populations with different evolution histories and BH birth mass distributions.

BH formation also depends on metallicity, which, \eg, influences stellar winds and, in turn, the star's remnant mass\footnote{The metallicity also affects the radius evolution of stars and hence the outcome of binary mass transfer.}. Typically, LMXBs reside in the metal-poor galactic halo or bulge, whereas HMXBs are found in the metal-rich spiral arms \citep[\eg,][]{vanparadijs1998a, coleiro2013a, casares2017a}. However, the nearby sample of LMXBs with dynamical mass measurements used here has companions with solar-like metallicities consistent with stars in the local neighborhood \citep{casares2017a}. The link between metallicity and BH mass is most obvious in our inert BH sample: while most BHs have masses of ${\sim}\,8\text{--}10 \,\msun$ or below (see Table~\ref{tab:Dormant}) and solar-like metallicities, Gaia BH3, for example, formed at low metallicity\footnote{Also the BHs in the LMC and M33 formed at a different metallicity.} ($[\text{Fe/H}]=-2.56\pm0.11\,\mathrm{dex}$, \citealt{panuzzo2024a}) and far exceeds that range with ${\sim}\,33\,\msun$. This suggests the maximum BH mass is not a fixed parameter but is sensitive to metallicity and star formation environment.

In summary, our statistical analysis provides a snapshot of the currently observed BH population rather than a reconstruction of the birth mass function. As the sample grows (\eg, with Gaia DR4 in 2026), future work should aim to isolate the effects of metallicity and stellar evolution, \eg using hierarchical models allowing for population-dependent substructures.

\subsection{\label{sec:discussion-gw-data}Qualitative comparison with GW data}
The LIGO-VIRGO-KAGRA (LVK) collaboration has published a 5th catalog of GW observations that firmly reveal structure in the mass distribution of NSs and BHs \citep{theligoscientificcollaboration2026a,theligoscientificcollaboration2026b}. At low masses, the emerging picture is that of peaks at ${<}\,2\,\msun$ and about $9\,\msun$ in the NS-BH mass distribution and a lower number of compact objects in between these peaks. In particular, there is no evidence for a gap between NSs and BHs at $2\text{--}5\,\msun$. At ${>}\,9\,\msun$, the BH mass distribution declines. This qualitative picture is consistent with our findings for the EM BH mass distribution.

The chirp mass distribution of binary BH mergers shows a clear peak at about $8\,\msun$, a prominent gap at about $10\,\msun$, and an increase thereafter up to another peak at about $27\,\msun$. \citet{willcox2025a} shows that such a structure can be explained by a bimodal BH mass distribution as proposed by \citet{schneider2021a, schneider2023a} for BHs formed from binary stripped stars. Such a bimodality is also consistent with the EM BH sample studied in the present work.

These similarities across the inferred BH mass distributions suggest the intriguing possibility that both BH populations share a common origin. Given that the EM samples studied here are sensitive to different environments, metallicities, and redshifts, and that the two populations are affected by very different selection effects, this qualitative similarity does not constitute evidence that both populations sample the same underlying distribution. A more complete analysis taking into account selection effects and the above-mentioned evolutionary biases is needed to substantiate this qualitative impression.

\subsection{Implications of the inferred BH mass distributions}
The main component of the BH mass distribution is at $5\text{--}10\,\msun$, qualitatively similar to what is seen in GW data, and there is a dearth towards lower masses and the NS mass regime. If confirmed, such a feature would have several implications and may allow us to, for example, constrain binary star evolution, nuclear physics and the supernova explosion mechanism. Stars reach core collapse with a continuous mass distribution. A rapid explosion with central NS formation and no additional mass accretion onto the NS would create an NS-BH mass gap. In this case, the exploding stars produce an NS of mass $M_\mathrm{NS}\,{\approx}\,1\text{--}2\,\msun$ and eject a mass of $\Delta M_\mathrm{ej}$. In contrast, BH-forming stars would leave behind a compact object remnant with a mass of $M_\mathrm{NS}$ plus a fraction of $\Delta M_\mathrm{ej}$, \ie a mass greater than $M_\mathrm{NS}$, and thus produce an NS-BH mass gap. Continuous and prolonged accretion onto the central proto-NS remnant and/or the fallback of a fraction of the SN ejecta mass are required to prevent the formation of a sharp NS-BH mass gap \citep[see \eg][]{heger2023a}. Alternatively, it is possible to populate an intrinsic NS-BH mass gap with mergers of second-generation NSs and low-mass BHs, and thereby explain the observations. 

The explodability of stars remains unsolved \citep[\eg][]{oconnor2018b, janka2025a}, and various proxies of explodability have been suggested in the literature \citep[\eg][]{oconnor2011a, ertl2016a, muller2016a, schneider2021a, temaj2024a, laplace2025a}. The so-called compactness parameter or the central specific entropy of stars at core collapse are commonly used proxies summarizing the pre-SN structure of stars. Stellar evolution models have shown that these proxies follow a pattern of multiple peaks as a function of initial or core mass of stars, each possibly signalling a higher likelihood of BH formation \citep[][]{oconnor2011a, sukhbold2014a, limongi2018a, schneider2021a, aguilera-dena2023a, takahashi2023a, temaj2024a, laplace2025a}. \citet{schneider2023a} found that binary-stripped stars, which are the progenitors of most BHs in binary systems, predominantly produce BHs of ${\approx}\,9\,\msun$ across a large range of metallicities because of one of these compactness peaks. This structural feature found by many groups across different stellar evolution codes is sensitive to a variety of physical processes such as the important $^{12}\mathrm{C}(\alpha, \gamma)^{16}\mathrm{O}$ and other nuclear reaction rates, neutrino physics, convection, etc.\ \citep{schneider2023a, laplace2025a}. If the $5\text{--}10\,\msun$ feature in the EM and GW data of BHs is indeed linked to this compactness peak, one could constrain these physical processes.

The models of \citet{schneider2021a, schneider2023a} further predict BHs of ${\gtrsim}\,15\,\msun$, and together with the ${\approx}\,9\,\msun$ BHs, this creates a bimodal BH mass distribution. Evidence for the existence of such a bimodal or otherwise multimodal BH mass distribution has been presented by \citet{willcox2025a} for GW binary BH mergers. The evidence for the existence of a high-mass BH component presented in this work could support this hypothesis. 

\section{\label{sec:conclusions}Conclusions}
We presented a statistical analysis of the mass distribution of stellar-mass black holes. We considered a sample of 40 electromagnetic detections comprising 25 LMXBs, six HMXBs, eight inert BHs and one microlensing event. We used MCMC methods to sample the posterior distribution of parameters implied by the data for several parametric models of the observed BH mass distributions. As EM mass measurements of BHs are challenging and the sample size is still rather small, conclusions from this sample remain uncertain. From the available data and the Bayesian inference, we find the following.
\begin{itemize}
    \item The main population of the observed BHs is concentrated in the mass range $5\text{--}10\,\msun$ (about half the sample).
    \item We find no evidence for a sharp NS-BH mass gap at ${\sim}\,3\text{--}5\,\msun$. Instead, there are several sources in our updated and extended BH sample that have masses in this purported gap. Moreover, our BH sample at this mass range is likely biased against low-mass BHs, further arguing against the existence of a sharp NS-BH mass gap. We do, however, observe a dearth of BHs at these masses in our sample, qualitatively consistent with GW observations. Restricting the sample to the ten LMXBs with the best BH mass measurements, there is an NS-BH mass gap.
    \item At higher masses (${\gtrsim}\,10\,\msun$), the number of BHs steadily decreases, and about 15\% of BHs in our sample have a mass of ${>}\,13\,\msun$. 
    \item A Gaussian or a split-normal distribution designed to fit the $5\text{--}10\,\msun$ main component captures it well. However, they fail to describe the high-mass BH component. 
    \item The historically favored power-law model finds two solutions, one with a positive and one with a negative slope. Both fits would designate parts of the main component as statistical outliers. The negative power-law slope appears to fit well the high-mass end of the BH mass distribution at first glance, but the fit sensitively depends on a few individual BHs (\eg IC 10 X-1) whose masses remain highly debated \citep{laycock2015a}.
    \item Overall, simple functional forms, such as power-law and Gaussian functions, struggle to simultaneously describe the low-mass and high-mass data well. A bimodal model, such as a Gaussian followed by a power-law function, provides a viable phenomenological description of the observed BH masses and has been suggested on theoretical grounds from binary star evolution. However, the current data do not statistically require such a second component. Given the low numbers of BHS, the high-mass component remains unconstrained. Significantly more BHs are required for a robust inference of the high-mass part of the BH mass distribution. 
    \item While the number of X-ray quiet BHs (\ie those in inert BH binaries and the one lensing BH) is still low, their masses are similar to those in X-ray binaries. Also, the GW BHs of comparable masses have a qualitatively similar mass distribution as found here. If corrobated by future studies, this could indicate all these BHs follow similar formation channels.
\end{itemize}
With upcoming spectroscopic surveys and future Gaia data releases expected to identify significantly more BHs, future work needs to address selection biases \citep[\eg][]{jonker2021a, siegel2023a} and disentangle metallicity dependencies and other evolutionary effects, for instance via hierarchical models allowing for population-dependent substructures and non-parametric models. In this way, one can then constrain the birth mass distribution of BHs in the Milky Way and nearby galaxies.

\begin{acknowledgements}
We thank the reviewer for constructive comments and suggestions that helped improve this work. ASN, FRNS and EL acknowledge support by the Klaus Tschira Foundation. This work has received funding from the European Research Council (ERC) under the European Union’s Horizon 2020 research and innovation programme (Grant agreement No.\ 945806) and is supported by the Deutsche Forschungsgemeinschaft (DFG, German Research Foundation) under Germany’s Excellence Strategy EXC 2181/1-390900948 (the Heidelberg STRUCTURES Excellence Cluster). EL acknowledges support through a start-up grant from the Internal Funds KU Leuven (STG/24/073), through a Veni grant (VI.Veni.232.205) from the Netherlands Organization for Scientific Research (NWO), and through the Research Foundation – Flanders (FWO) under the Odysseus Program, Type II (G0AT525N). JC acknowledges support by the Spanish Ministry of Science via the Plan de Generaci\'on de Conocimiento through grants PID2022-143331NB-100 and PID2021-124879NB-I00, respectively. T.M.-D. and MAP acknowledge support by the Spanish \textit{Agencia estatal de investigaci\'on} via PID2021-124879NB-I00 and PID2024-161863NB-I00. GJE acknowledges support from the Spanish Ministry of Science via the Plan de Generación de Conocimiento through grant PID2022-143331NB-100.
\end{acknowledgements}

\bibliographystyle{aa}
\bibliography{references}

@misc{theligoscientificcollaboration2025a,
  author =        {{The LIGO Scientific Collaboration} and
                   {the Virgo Collaboration} and
                   {the KAGRA Collaboration} and Abac, A. G. and
                   Abouelfettouh, I. and Acernese, F. and Ackley, K. and
                   Adamcewicz, C. and Adhicary, S. and Adhikari, D. and
                   Adhikari, N. and Adhikari, R. X. and Adkins, V. K. and
                   Afroz, S. and Agapito, A. and Agarwal, D. and
                   Agathos, M. and Aggarwal, N. and Aggarwal, S. and
                   Aguiar, O. D. and Ahrend, I. -L. and Aiello, L. and
                   Ain, A. and Ajith, P. and Akutsu, T. and Albanesi, S. and
                   Ali, W. and {Al-Kershi}, S. and All{\'e}n{\'e}, C. and
                   Allocca, A. and {Al-Shammari}, S. and Altin, P. A. and
                   {Alvarez-Lopez}, S. and Amar, W. and Amarasinghe, O. and
                   Amato, A. and Amicucci, F. and Amra, C. and
                   Ananyeva, A. and Anderson, S. B. and Anderson, W. G. and
                   Andia, M. and Ando, M. and {Andr{\'e}s-Carcasona}, M. and
                   Andri{\'c}, T. and Anglin, J. and Ansoldi, S. and
                   Antelis, J. M. and Antier, S. and Aoumi, M. and
                   Appavuravther, E. Z. and Appert, S. and Apple, S. K. and
                   Arai, K. and Araya, A. and Araya, M. C. and
                   Arca Sedda, M. and Areeda, J. S. and Aritomi, N. and
                   Armato, F. and Armstrong, S. and Arnaud, N. and
                   Arogeti, M. and Aronson, S. M. and Arun, K. G. and
                   Ashton, G. and Aso, Y. and Asprea, L. and Assiduo, M. and
                   {Assis de Souza Melo}, S. and Aston, S. M. and
                   Astone, P. and Attadio, F. and Aubin, F. and
                   AultONeal, K. and Avallone, G. and Avila, E. A. and
                   Babak, S. and Badger, C. and Bae, S. and Bagnasco, S. and
                   Baiotti, L. and Bajpai, R. and Baka, T. and
                   Baker, A. M. and Baker, K. A. and Baker, T. and
                   Baldi, G. and Baldicchi, N. and Ball, M. and
                   Ballardin, G. and Ballmer, S. W. and Banagiri, S. and
                   Banerjee, B. and Bankar, D. and Baptiste, T. M. and
                   Baral, P. and Baratti, M. and Barayoga, J. C. and
                   Barish, B. C. and Barker, D. and Barman, N. and
                   Barneo, P. and Barone, F. and Barr, B. and
                   Barsotti, L. and Barsuglia, M. and Barta, D. and
                   Bartoletti, A. M. and Barton, M. A. and Bartos, I. and
                   Basalaev, A. and Bassiri, R. and Basti, A. and
                   Bawaj, M. and Baxi, P. and Bayley, J. C. and
                   Baylor, A. C. and Baynard, II, P. A. and Bazzan, M. and
                   Bedakihale, V. M. and Beirnaert, F. and Bejger, M. and
                   Belardinelli, D. and Bell, A. S. and Bellie, D. S. and
                   Bellizzi, L. and Benoit, W. and Bentara, I. and
                   Bentley, J. D. and Ben Yaala, M. and Bera, S. and
                   Bergamin, F. and Berger, B. K. and Bernuzzi, S. and
                   Beroiz, M. and Berry, C. P. L. and Bersanetti, D. and
                   Bertheas, T. and Bertolini, A. and Betzwieser, J. and
                   Beveridge, D. and Bevilacqua, G. and Bevins, N. and
                   Bhandare, R. and Bhatt, R. and Bhattacharjee, D. and
                   Bhattacharyya, S. and Bhaumik, S. and Biancalana, V. and
                   Bianchi, A. and Bilenko, I. A. and Billingsley, G. and
                   Binetti, A. and Bini, S. and Binu, C. and Biot, S. and
                   Birnholtz, O. and Biscoveanu, S. and Bisht, A. and
                   Bitossi, M. and Bizouard, M. -A. and Blaber, S. and
                   Blackburn, J. K. and Blagg, L. A. and Blair, C. D. and
                   Blair, D. G. and Bode, N. and Boettner, N. and
                   Boileau, G. and Boldrini, M. and Bolingbroke, G. N. and
                   Bolliand, A. and Bonavena, L. D. and Bondarescu, R. and
                   Bondu, F. and Bonilla, E. and Bonilla, M. S. and
                   Bonino, A. and Bonnand, R. and Borchers, A. and
                   Borhanian, S. and Boschi, V. and Bose, S. and
                   Bossilkov, V. and Bothra, Y. and Boudon, A. and
                   Bourg, L. and Boyle, M. and Bozzi, A. and
                   Bradaschia, C. and Brady, P. R. and Branch, A. and
                   Branchesi, M. and Braun, I. and Briant, T. and
                   Brillet, A. and Brinkmann, M. and Brockill, P. and
                   Brockmueller, E.},
  month =         aug,
  publisher =     {arXiv},
  title =         {{{GWTC-4}}.0: {{Updating}} the {{Gravitational-Wave
                   Transient Catalog}} with {{Observations}} from the
                   {{First Part}} of the {{Fourth LIGO-Virgo-KAGRA
                   Observing Run}}},
  year =          {2025},
  doi =           {10.48550/arXiv.2508.18082},
}

@misc{theligoscientificcollaboration2026b,
  author =        {{The LIGO Scientific Collaboration} and
                   {the Virgo Collaboration} and
                   {the KAGRA Collaboration}},
  month =         may,
  publisher =     {arXiv},
  title =         {{{GWTC-5}}.0: {{Population Properties}} of {{Merging
                   Compact Binaries}}},
  year =          {2026},
  doi =           {10.48550/arXiv.2605.27226},
}

@article{tiwari2021a,
  author =        {Tiwari, Vaibhav and Fairhurst, Stephen},
  journal =       {ApJ},
  month =         jun,
  number =        {L19},
  pages =         {L19},
  title =         {The Emergence of Structure in the Binary Black Hole
                   Mass Distribution},
  volume =        {913},
  year =          {2021},
  doi =           {10.3847/2041-8213/abfbe7},
}

@article{tiwari2022a,
  author =        {Tiwari, Vaibhav},
  journal =       {ApJ},
  month =         apr,
  number =        {155},
  pages =         {155},
  title =         {Exploring Features in the Binary Black Hole
                   Population},
  volume =        {928},
  year =          {2022},
  doi =           {10.3847/1538-4357/ac589a},
}

@article{tiwari2025a,
  author =        {Tiwari, Vaibhav},
  journal =       {ApJ},
  month =         dec,
  pages =         {177},
  publisher =     {IOP},
  title =         {Examining the {{Gap}} in the {{Chirp Mass
                   Distribution}} of {{Binary Black Holes}}},
  volume =        {995},
  year =          {2025},
  doi =           {10.3847/1538-4357/ae2260},
  issn =          {0004-637X},
}

@article{abbott2023a,
  author =        {Abbott, R. and Abbott, T. D. and Acernese, F. and
                   Ackley, K. and Adams, C. and Adhikari, N. and
                   Adhikari, R. X. and Adya, V. B. and Affeldt, C. and
                   Agarwal, D. and Agathos, M. and Agatsuma, K. and
                   Aggarwal, N. and Aguiar, O. D. and Aiello, L. and
                   Ain, A. and Ajith, P. and Akutsu, T. and
                   {de Alarc{\'o}n}, P. F. and Akcay, S. and
                   Albanesi, S. and Allocca, A. and Altin, P. A. and
                   Amato, A. and Anand, C. and Anand, S. and
                   Ananyeva, A. and Anderson, S. B. and Anderson, W. G. and
                   Ando, M. and Andrade, T. and Andres, N. and
                   Andri{\'c}, T. and Angelova, S. V. and Ansoldi, S. and
                   Antelis, J. M. and Antier, S. and Antonini, F. and
                   Appert, S. and Arai, Koji and Arai, Koya and Arai, Y. and
                   Araki, S. and Araya, A. and Araya, M. C. and
                   Areeda, J. S. and Ar{\`e}ne, M. and Aritomi, N. and
                   Arnaud, N. and Arogeti, M. and Aronson, S. M. and
                   Arun, K. G. and Asada, H. and Asali, Y. and
                   Ashton, G. and Aso, Y. and Assiduo, M. and
                   Aston, S. M. and Astone, P. and Aubin, F. and
                   Austin, C. and Babak, S. and Badaracco, F. and
                   Bader, M. K. M. and Badger, C. and Bae, S. and
                   Bae, Y. and Baer, A. M. and Bagnasco, S. and Bai, Y. and
                   Baiotti, L. and Baird, J. and Bajpai, R. and Ball, M. and
                   Ballardin, G. and Ballmer, S. W. and Balsamo, A. and
                   Baltus, G. and Banagiri, S. and Bankar, D. and
                   Barayoga, J. C. and Barbieri, C. and Barish, B. C. and
                   Barker, D. and Barneo, P. and Barone, F. and Barr, B. and
                   Barsotti, L. and Barsuglia, M. and Barta, D. and
                   Bartlett, J. and Barton, M. A. and Bartos, I. and
                   Bassiri, R. and Basti, A. and Bawaj, M. and
                   Bayley, J. C. and Baylor, A. C. and Bazzan, M. and
                   B{\'e}csy, B. and Bedakihale, V. M. and Bejger, M. and
                   Belahcene, I. and Benedetto, V. and Beniwal, D. and
                   Bennett, T. F. and Bentley, J. D. and Benyaala, M. and
                   Bergamin, F. and Berger, B. K. and Bernuzzi, S. and
                   Berry, C. P. L. and Bersanetti, D. and Bertolini, A. and
                   Betzwieser, J. and Beveridge, D. and Bhandare, R. and
                   Bhardwaj, U. and Bhattacharjee, D. and Bhaumik, S. and
                   Bilenko, I. A. and Billingsley, G. and Bini, S. and
                   Birney, R. and Birnholtz, O. and Biscans, S. and
                   Bischi, M. and Biscoveanu, S. and Bisht, A. and
                   Biswas, B. and Bitossi, M. and Bizouard, M. -A. and
                   Blackburn, J. K. and Blair, C. D. and Blair, D. G. and
                   Blair, R. M. and Bobba, F. and Bode, N. and Boer, M. and
                   Bogaert, G. and Boldrini, M. and Bonavena, L. D. and
                   Bondu, F. and Bonilla, E. and Bonnand, R. and
                   Booker, P. and Boom, B. A. and Bork, R. and
                   Boschi, V. and Bose, N. and Bose, S. and
                   Bossilkov, V. and Boudart, V. and Bouffanais, Y. and
                   Bozzi, A. and Bradaschia, C. and Brady, P. R. and
                   Bramley, A. and Branch, A. and Branchesi, M. and
                   Brandt, J. and Brau, J. E. and Breschi, M. and
                   Briant, T. and Briggs, J. H. and Brillet, A. and
                   Brinkmann, M. and Brockill, P. and Brooks, A. F. and
                   Brooks, J. and Brown, D. D. and Brunett, S. and
                   Bruno, G. and Bruntz, R. and Bryant, J. and Bulik, T. and
                   Bulten, H. J. and Buonanno, A. and Buscicchio, R. and
                   Buskulic, D. and Buy, C. and Byer, R. L. and
                   Cadonati, L. and Cagnoli, G. and Cahillane, C. and
                   Bustillo, J. Calder{\'o}n and Callaghan, J. D. and
                   Callister, T. A. and Calloni, E. and Cameron, J. and
                   Camp, J. B. and Canepa, M. and Canevarolo, S. and
                   Cannavacciuolo, M. and Cannon, K. C. and Cao, H. and
                   Cao, Z. and Capocasa, E. and Capote, E. and
                   Carapella, G. and Carbognani, F. and Carlin, J. B. and
                   Carney, M. F. and Carpinelli, M. and Carrillo, G. and
                   Carullo, G. and Carver, T. L. and Diaz, J. Casanueva and
                   Casentini, C. and Castaldi, G. and Caudill, S. and
                   Cavagli{\`a}, M. and Cavalier, F. and Cavalieri, R. and
                   Ceasar, M. and Cella, G. and
                   {Cerd{\'a}-Dur{\'a}n}, P. and Cesarini, E. and
                   Chaibi, W. and Chakravarti, K. and
                   Subrahmanya, S. Chalathadka and Champion, E. and
                   Chan, C. -H. and Chan, C. and Chan, C. L. and
                   Chan, K. and Chan, M. and Chandra, K. and Chanial, P. and
                   Chao, S. and {Chapman-Bird}, C. E. A. and
                   Charlton, P. and Chase, E. A. and
                   {Chassande-Mottin}, E. and Chatterjee, C. and
                   Chatterjee, Debarati and Chatterjee, Deep and
                   Chaturvedi, M. and Chaty, S. and Chatziioannou, K. and
                   Chen, C. and Chen, H. Y. and Chen, J. and Chen, K. and
                   Chen, X. and Chen, Y. -B. and Chen, Y. -R. and
                   Chen, Z. and Cheng, H. and Cheong, C. K. and
                   Cheung, H. Y. and Chia, H. Y. and Chiadini, F. and
                   Chiang, C. -Y. and Chiarini, G. and Chierici, R. and
                   Chincarini, A. and Chiofalo, M. L. and Chiummo, A. and
                   Cho, G. and Cho, H. S. and Choudhary, R. K. and
                   Choudhary, S. and Christensen, N. and Chu, H. and
                   Chu, Q. and Chu, Y. -K. and Chua, S. and Chung, K. W. and
                   Ciani, G. and Ciecielag, P. and Cie{\'s}lar, M. and
                   Cifaldi, M. and Ciobanu, A. A. and Ciolfi, R. and
                   Cipriano, F. and Cirone, A. and Clara, F. and
                   Clark, E. N. and Clark, J. A. and Clarke, L. and
                   Clearwater, P. and Clesse, S. and Cleva, F. and
                   Coccia, E. and Codazzo, E. and Cohadon, P. -F. and
                   Cohen, D. E. and Cohen, L. and Colleoni, M. and
                   Collette, C. G. and Colombo, A. and Colpi, M. and
                   Compton, C. M. and Constancio, M. and Conti, L. and
                   Cooper, S. J. and Corban, P. and Corbitt, T. R. and
                   {Cordero-Carri{\'o}n}, I. and Corezzi, S. and
                   Corley, K. R. and Cornish, N. and Corre, D. and
                   Corsi, A. and Cortese, S. and Costa, C. A. and
                   Cotesta, R. and Coughlin, M. W. and Coulon, J. -P. and
                   Countryman, S. T. and Cousins, B. and Couvares, P. and
                   Coward, D. M. and Cowart, M. J. and Coyne, D. C. and
                   Coyne, R. and Creighton, J. D. E. and
                   Creighton, T. D. and Criswell, A. W. and
                   Croquette, M. and Crowder, S. G. and Cudell, J. R. and
                   Cullen, T. J. and Cumming, A. and Cummings, R. and
                   Cunningham, L. and Cuoco, E. and Cury{\l}o, M. and
                   Dabadie, P. and Canton, T. Dal and Dall'Osso, S. and
                   D{\'a}lya, G. and Dana, A. and
                   Daneshgaranbajastani, L. M. and D'Angelo, B. and
                   Danila, B. and Danilishin, S. and D'Antonio, S. and
                   Danzmann, K. and {Darsow-Fromm}, C. and Dasgupta, A. and
                   Datrier, L. E. H. and Datta, S. and Dattilo, V. and
                   Dave, I. and Davier, M. and Davies, G. S. and
                   Davis, D. and Davis, M. C. and Daw, E. J. and
                   Dean, R. and Debra, D. and Deenadayalan, M. and
                   Degallaix, J. and {de Laurentis}, M. and
                   Del{\'e}glise, S. and Del Favero, V. and
                   {de Lillo}, F. and {de Lillo}, N. and Del Pozzo, W. and
                   Demarchi, L. M. and {de Matteis}, F. and D'Emilio, V. and
                   Demos, N. and Dent, T. and Depasse, A. and
                   {de Pietri}, R. and De Rosa, R. and {de Rossi}, C. and
                   Desalvo, R. and {de Simone}, R. and Dhurandhar, S. and
                   D{\'i}az, M. C. and {Diaz-Ortiz}, M. and Didio, N. A. and
                   Dietrich, T. and {di Fiore}, L. and {di Fronzo}, C. and
                   {di Giorgio}, C. and {di Giovanni}, F. and
                   {di Giovanni}, M. and {di Girolamo}, T. and
                   {di Lieto}, A. and Ding, B. and {di Pace}, S. and
                   {di Palma}, I. and {di Renzo}, F. and
                   Divakarla, A. K. and Dmitriev, A. and Doctor, Z. and
                   D'Onofrio, L. and Donovan, F. and Dooley, K. L. and
                   Doravari, S. and Dorrington, I. and Drago, M. and
                   Driggers, J. C. and Drori, Y. and Ducoin, J. -G. and
                   Dupej, P. and Durante, O. and D'Urso, D. and
                   Duverne, P. -A. and Dwyer, S. E. and Eassa, C. and
                   Easter, P. J. and Ebersold, M. and Eckhardt, T. and
                   Eddolls, G. and Edelman, B. and Edo, T. B. and
                   Edy, O. and Effler, A. and Eguchi, S. and
                   Eichholz, J. and Eikenberry, S. S. and Eisenmann, M. and
                   Eisenstein, R. A. and Ejlli, A. and Engelby, E. and
                   Enomoto, Y. and Errico, L. and Essick, R. C. and
                   Estell{\'e}s, H. and Estevez, D. and Etienne, Z. and
                   Etzel, T. and Evans, M. and Evans, T. M. and
                   Ewing, B. E. and Fafone, V. and Fair, H. and
                   Fairhurst, S. and Farah, A. M. and Farinon, S. and
                   Farr, B. and Farr, W. M. and Farrow, N. W. and
                   {Fauchon-Jones}, E. J. and Favaro, G. and Favata, M. and
                   Fays, M. and Fazio, M. and Feicht, J. and
                   Fejer, M. M. and Fenyvesi, E. and Ferguson, D. L. and
                   {Fernandez-Galiana}, A. and Ferrante, I. and
                   Ferreira, T. A. and Fidecaro, F. and Figura, P. and
                   Fiori, I. and Fishbach, M. and Fisher, R. P. and
                   Fittipaldi, R. and Fiumara, V. and Flaminio, R. and
                   Floden, E. and Fong, H. and Font, J. A. and
                   Fornal, B. and Forsyth, P. W. F. and Franke, A. and
                   Frasca, S. and Frasconi, F. and Frederick, C. and
                   Freed, J. P. and Frei, Z. and Freise, A. and Frey, R. and
                   Fritschel, P. and Frolov, V. V. and Fronz{\'e}, G. G. and
                   Fujii, Y. and Fujikawa, Y. and Fukunaga, M. and
                   Fukushima, M. and Fulda, P. and Fyffe, M. and
                   Gabbard, H. A. and Gadre, B. U. and Gair, J. R. and
                   Gais, J. and Galaudage, S. and Gamba, R. and
                   Ganapathy, D. and Ganguly, A. and Gao, D. and
                   Gaonkar, S. G. and Garaventa, B. and Garc{\'i}a, F. and
                   {Garc{\'i}a-N{\'u}{\~n}ez}, C. and
                   {Garc{\'i}a-Quir{\'o}s}, C. and Garufi, F. and
                   Gateley, B. and Gaudio, S. and Gayathri, V. and
                   Ge, G. -G. and Gemme, G. and Gennai, A. and
                   George, J. and George, R. N. and Gerberding, O. and
                   Gergely, L. and Gewecke, P. and Ghonge, S. and
                   Ghosh, Abhirup and Ghosh, Archisman and Ghosh, Shaon and
                   Ghosh, Shrobana and Giacomazzo, B. and Giacoppo, L. and
                   Giaime, J. A. and Giardina, K. D. and Gibson, D. R. and
                   Gier, C. and Giesler, M. and Giri, P. and Gissi, F. and
                   Glanzer, J. and Gleckl, A. E. and Godwin, P. and
                   Golomb, J. and Goetz, E. and Goetz, R. and Gohlke, N. and
                   Goncharov, B. and Gonz{\'a}lez, G. and Gopakumar, A. and
                   Gosselin, M. and Gouaty, R. and Gould, D. W. and
                   Grace, B. and Grado, A. and Granata, M. and
                   Granata, V. and Grant, A. and Gras, S. and
                   Grassia, P. and Gray, C. and Gray, R. and Greco, G. and
                   Green, A. C. and Green, R. and Gretarsson, A. M. and
                   Gretarsson, E. M. and Griffith, D. and Griffiths, W. and
                   Griggs, H. L. and Grignani, G. and Grimaldi, A. and
                   Grimm, S. J. and Grote, H. and Grunewald, S. and
                   Gruning, P. and Guerra, D. and Guidi, G. M. and
                   Guimaraes, A. R. and Guix{\'e}, G. and Gulati, H. K. and
                   Guo, H. -K. and Guo, Y. and Gupta, Anchal and
                   Gupta, Anuradha and Gupta, P. and Gustafson, E. K. and
                   Gustafson, R. and Guzman, F. and Ha, S. and
                   Haegel, L. and Hagiwara, A. and Haino, S. and
                   Halim, O. and Hall, E. D. and Hamilton, E. Z. and
                   Hammond, G. and Han, W. -B. and Haney, M. and
                   Hanks, J. and Hanna, C. and Hannam, M. D. and
                   Hannuksela, O. and Hansen, H. and Hansen, T. J. and
                   Hanson, J. and Harder, T. and Hardwick, T. and
                   Haris, K. and Harms, J. and Harry, G. M. and
                   Harry, I. W. and Hartwig, D. and Hasegawa, K. and
                   Haskell, B. and Hasskew, R. K. and Haster, C. -J. and
                   Hattori, K. and Haughian, K. and Hayakawa, H. and
                   Hayama, K. and Hayes, F. J. and Healy, J. and
                   Heidmann, A. and Heidt, A. and Heintze, M. C. and
                   Heinze, J. and Heinzel, J. and Heitmann, H. and
                   Hellman, F. and Hello, P. and
                   {Helmling-Cornell}, A. F. and Hemming, G. and
                   Hendry, M. and Heng, I. S. and Hennes, E. and
                   Hennig, J. and Hennig, M. H. and Hernandez, A. G. and
                   Vivanco, F. Hernandez and Heurs, M. and Hild, S. and
                   Hill, P. and Himemoto, Y. and Hines, A. S. and
                   Hiranuma, Y. and Hirata, N. and Hirose, E. and
                   Hochheim, S. and Hofman, D. and Hohmann, J. N. and
                   Holcomb, D. G. and Holland, N. A. and Hollows, I. J. and
                   Holmes, Z. J. and Holt, K. and Holz, D. E. and
                   Hong, Z. and Hopkins, P. and Hough, J. and
                   Hourihane, S. and Howell, E. J. and Hoy, C. G. and
                   Hoyland, D. and Hreibi, A. and Hsieh, B. -H. and
                   Hsu, Y. and Huang, G. -Z. and Huang, H. -Y. and
                   Huang, P. and Huang, Y. -C. and Huang, Y. -J. and
                   Huang, Y. and H{\"u}bner, M. T. and Huddart, A. D. and
                   Hughey, B. and Hui, D. C. Y. and Hui, V. and Husa, S. and
                   Huttner, S. H. and Huxford, R. and {Huynh-Dinh}, T. and
                   Ide, S. and Idzkowski, B. and Iess, A. and
                   Ikenoue, B. and Imam, S. and Inayoshi, K. and
                   Ingram, C. and Inoue, Y. and Ioka, K. and Isi, M. and
                   Isleif, K. and Ito, K. and Itoh, Y. and Iyer, B. R. and
                   Izumi, K. and Jaberianhamedan, V. and Jacqmin, T. and
                   Jadhav, S. J. and Jadhav, S. P. and James, A. L. and
                   Jan, A. Z. and Jani, K. and Janquart, J. and
                   Janssens, K. and Janthalur, N. N. and Jaranowski, P. and
                   Jariwala, D. and Jaume, R. and Jenkins, A. C. and
                   Jenner, K. and Jeon, C. and Jeunon, M. and Jia, W. and
                   Jin, H. -B. and Johns, G. R. and Jones, A. W. and
                   Jones, D. I. and Jones, J. D. and Jones, P. and
                   Jones, R. and Jonker, R. J. G. and Ju, L. and
                   Jung, P. and Jung, K. and Junker, J. and Juste, V. and
                   Kaihotsu, K. and Kajita, T. and Kakizaki, M. and
                   Kalaghatgi, C. V. and Kalogera, V. and Kamai, B. and
                   Kamiizumi, M. and Kanda, N. and Kandhasamy, S. and
                   Kang, G. and Kanner, J. B. and Kao, Y. and
                   Kapadia, S. J. and Kapasi, D. P. and Karat, S. and
                   Karathanasis, C. and Karki, S. and Kashyap, R. and
                   Kasprzack, M. and Kastaun, W. and Katsanevas, S. and
                   Katsavounidis, E. and Katzman, W. and Kaur, T. and
                   Kawabe, K. and Kawaguchi, K. and Kawai, N. and
                   Kawasaki, T. and K{\'e}f{\'e}lian, F. and Keitel, D. and
                   Key, J. S. and Khadka, S. and Khalili, F. Y. and
                   Khan, S. and Khazanov, E. A. and Khetan, N. and
                   Khursheed, M. and Kijbunchoo, N. and Kim, C. and
                   Kim, J. C. and Kim, J. and Kim, K. and Kim, W. S. and
                   Kim, Y. -M. and Kimball, C. and Kimura, N. and
                   {Kinley-Hanlon}, M. and Kirchhoff, R. and
                   Kissel, J. S. and Kita, N. and Kitazawa, H. and
                   Kleybolte, L. and Klimenko, S. and Knee, A. M. and
                   Knowles, T. D. and Knyazev, E. and Koch, P. and
                   Koekoek, G. and Kojima, Y. and Kokeyama, K. and
                   Koley, S. and Kolitsidou, P. and Kolstein, M. and
                   Komori, K. and Kondrashov, V. and Kong, A. K. H. and
                   Kontos, A. and Koper, N. and Korobko, M. and
                   Kotake, K. and Kovalam, M. and Kozak, D. B. and
                   Kozakai, C. and Kozu, R. and Kringel, V. and
                   Krishnendu, N. V. and Kr{\'o}lak, A. and Kuehn, G. and
                   Kuei, F. and Kuijer, P. and Kulkarni, S. and
                   Kumar, A. and Kumar, P. and Kumar, Rahul and
                   Kumar, Rakesh and Kume, J. and Kuns, K. and Kuo, C. and
                   Kuo, H. -S. and Kuromiya, Y. and Kuroyanagi, S. and
                   Kusayanagi, K. and Kuwahara, S. and Kwak, K. and
                   Lagabbe, P. and Laghi, D. and Lalande, E. and
                   Lam, T. L. and Lamberts, A. and Landry, M. and
                   Landry, P. and Lane, B. B. and Lang, R. N. and
                   Lange, J. and Lantz, B. and La Rosa, I. and
                   {Lartaux-Vollard}, A. and Lasky, P. D. and Laxen, M. and
                   Lazzarini, A. and Lazzaro, C. and Leaci, P. and
                   Leavey, S. and Lecoeuche, Y. K. and Lee, H. K. and
                   Lee, H. M. and Lee, H. W. and Lee, J. and Lee, K. and
                   Lee, R. and Lehmann, J. and Lema{\^i}tre, A. and
                   Leonardi, M. and Leroy, N. and Letendre, N. and
                   Levesque, C. and Levin, Y. and Leviton, J. N. and
                   Leyde, K. and Li, A. K. Y. and Li, B. and Li, J. and
                   Li, K. L. and Li, T. G. F. and Li, X. and Lin, C. -Y. and
                   Lin, F. -K. and Lin, F. -L. and Lin, H. L. and
                   Lin, L. C. -C. and Linde, F. and Linker, S. D. and
                   Linley, J. N. and Littenberg, T. B. and Liu, G. C. and
                   Liu, J. and Liu, K. and Liu, X. and Llamas, F. and
                   {Llorens-Monteagudo}, M. and Lo, R. K. L. and
                   Lockwood, A. and Loh, M. and London, L. T. and
                   Longo, A. and Lopez, D. and Portilla, M. Lopez and
                   Lorenzini, M. and Loriette, V. and Lormand, M. and
                   Losurdo, G. and Lott, T. P. and Lough, J. D. and
                   Lousto, C. O. and Lovelace, G. and Lucaccioni, J. F. and
                   L{\"u}ck, H. and Lumaca, D. and Lundgren, A. P. and
                   Luo, L. -W. and Lynam, J. E. and Macas, R. and
                   Macinnis, M. and MacLeod, D. M. and
                   MacMillan, I. A. O. and Macquet, A. and
                   Hernandez, I. Maga{\~n}a and Magazz{\`u}, C. and
                   Magee, R. M. and Maggiore, R. and Magnozzi, M. and
                   Mahesh, S. and Majorana, E. and Makarem, C. and
                   Maksimovic, I. and Maliakal, S. and Malik, A. and
                   Man, N. and Mandic, V. and Mangano, V. and
                   Mango, J. L. and Mansell, G. L. and Manske, M. and
                   Mantovani, M. and Mapelli, M. and Marchesoni, F. and
                   Marchio, M. and Marion, F. and Mark, Z. and
                   M{\'a}rka, S. and M{\'a}rka, Z. and Markakis, C. and
                   Markosyan, A. S. and Markowitz, A. and Maros, E. and
                   Marquina, A. and Marsat, S. and Martelli, F. and
                   Martin, I. W. and Martin, R. M. and Martinez, M. and
                   Martinez, V. A. and Martinez, V. and Martinovic, K. and
                   Martynov, D. V. and Marx, E. J. and Masalehdan, H. and
                   Mason, K. and Massera, E. and Masserot, A. and
                   Massinger, T. J. and {Masso-Reid}, M. and
                   Mastrogiovanni, S. and Matas, A. and
                   {Mateu-Lucena}, M. and Matichard, F. and
                   Matiushechkina, M. and Mavalvala, N. and
                   McCann, J. J. and McCarthy, R. and McClelland, D. E. and
                   McClincy, P. K. and McCormick, S. and McCuller, L. and
                   McGhee, G. I. and McGuire, S. C. and McIsaac, C. and
                   McIver, J. and McRae, T. and McWilliams, S. T. and
                   Meacher, D. and Mehmet, M. and Mehta, A. K. and
                   Meijer, Q. and Melatos, A. and Melchor, D. A. and
                   Mendell, G. and {Menendez-Vazquez}, A. and
                   Menoni, C. S. and Mercer, R. A. and Mereni, L. and
                   Merfeld, K. and Merilh, E. L. and Merritt, J. D. and
                   Merzougui, M. and Meshkov, S. and Messenger, C. and
                   Messick, C. and Meyers, P. M. and Meylahn, F. and
                   Mhaske, A. and Miani, A. and Miao, H. and
                   Michaloliakos, I. and Michel, C. and Michimura, Y. and
                   Middleton, H. and Milano, L. and Miller, A. L. and
                   Miller, A. and Miller, B. and Miller, S. and
                   Millhouse, M. and Mills, J. C. and Milotti, E. and
                   Minazzoli, O. and Minenkov, Y. and Mio, N. and
                   Mir, {\relax Ll}. M. and {Miravet-Ten{\'e}s}, M. and
                   Mishra, C. and Mishra, T. and Mistry, T. and
                   Mitra, S. and Mitrofanov, V. P. and Mitselmakher, G. and
                   Mittleman, R. and Miyakawa, O. and Miyamoto, A. and
                   Miyazaki, Y. and Miyo, K. and Miyoki, S. and
                   Mo, Geoffrey and Modafferi, L. M. and Moguel, E. and
                   Mogushi, K. and Mohapatra, S. R. P. and Mohite, S. R. and
                   Molina, I. and {Molina-Ruiz}, M. and Mondin, M. and
                   Montani, M. and Moore, C. J. and Moraru, D. and
                   Morawski, F. and More, A. and Moreno, C. and
                   Moreno, G. and Mori, Y. and Morisaki, S. and
                   Moriwaki, Y. and Morr{\'a}s, G. and Mours, B. and
                   {Mow-Lowry}, C. M. and Mozzon, S. and Muciaccia, F. and
                   Mukherjee, Arunava and Mukherjee, D. and
                   Mukherjee, Soma and Mukherjee, Subroto and
                   Mukherjee, Suvodip and Mukund, N. and Mullavey, A. and
                   Munch, J. and Mu{\~n}iz, E. A. and Murray, P. G. and
                   Musenich, R. and Muusse, S. and Nadji, S. L. and
                   Nagano, K. and Nagano, S. and Nagar, A. and
                   Nakamura, K. and Nakano, H. and Nakano, M. and
                   Nakashima, R. and Nakayama, Y. and Napolano, V. and
                   Nardecchia, I. and Narikawa, T. and Naticchioni, L. and
                   Nayak, B. and Nayak, R. K. and Negishi, R. and
                   Neil, B. F. and Neilson, J. and Nelemans, G. and
                   Nelson, T. J. N. and Nery, M. and Neubauer, P. and
                   Neunzert, A. and Ng, K. Y. and Ng, S. W. S. and
                   Nguyen, C. and Nguyen, P. and Nguyen, T. and
                   Quynh, L. Nguyen and Ni, W. -T. and Nichols, S. A. and
                   Nishizawa, A. and Nissanke, S. and Nitoglia, E. and
                   Nocera, F. and Norman, M. and North, C. and
                   Nozaki, S. and Siles, J. F. Nu{\~n}o and
                   Nuttall, L. K. and Oberling, J. and O'Brien, B. D. and
                   Obuchi, Y. and O'Dell, J. and Oelker, E. and
                   Ogaki, W. and Oganesyan, G. and Oh, J. J. and Oh, K. and
                   Oh, S. H. and Ohashi, M. and Ohishi, N. and
                   Ohkawa, M. and Ohme, F. and Ohta, H. and Okada, M. A. and
                   Okutani, Y. and Okutomi, K. and Olivetto, C. and
                   Oohara, K. and Ooi, C. and Oram, R. and O'Reilly, B. and
                   Ormiston, R. G. and Ormsby, N. D. and Ortega, L. F. and
                   O'Shaughnessy, R. and O'Shea, E. and Oshino, S. and
                   Ossokine, S. and Osthelder, C. and Otabe, S. and
                   Ottaway, D. J. and Overmier, H. and Pace, A. E. and
                   Pagano, G. and Page, M. A. and Pagliaroli, G. and
                   Pai, A. and Pai, S. A. and Palamos, J. R. and
                   Palashov, O. and Palomba, C. and Pan, H. and Pan, K. and
                   Panda, P. K. and Pang, H. and Pang, P. T. H. and
                   Pankow, C. and Pannarale, F. and Pant, B. C. and
                   Panther, F. H. and Paoletti, F. and Paoli, A. and
                   Paolone, A. and Parisi, A. and Park, H. and Park, J. and
                   Parker, W. and Pascucci, D. and Pasqualetti, A. and
                   Passaquieti, R. and Passuello, D. and Patel, M. and
                   Pathak, M. and Patricelli, B. and Patron, A. S. and
                   Paul, S. and Payne, E. and Pedraza, M. and
                   Pegoraro, M. and Pele, A. and
                   Arellano, F. E. Pe{\~n}a and Penn, S. and Perego, A. and
                   Pereira, A. and Pereira, T. and Perez, C. J. and
                   P{\'e}rigois, C. and Perkins, C. C. and Perreca, A. and
                   Perri{\`e}s, S. and Petermann, J. and Petterson, D. and
                   Pfeiffer, H. P. and Pham, K. A. and Phukon, K. S. and
                   Piccinni, O. J. and Pichot, M. and Piendibene, M. and
                   Piergiovanni, F. and Pierini, L. and Pierro, V. and
                   Pillant, G. and Pillas, M. and Pilo, F. and
                   Pinard, L. and Pinto, I. M. and Pinto, M. and
                   Piotrzkowski, B. and Piotrzkowski, K. and Pirello, M. and
                   Pitkin, M. D. and Placidi, E. and Planas, L. and
                   Plastino, W. and Pluchar, C. and Poggiani, R. and
                   Polini, E. and Pong, D. Y. T. and Ponrathnam, S. and
                   Popolizio, P. and Porter, E. K. and Poulton, R. and
                   Powell, J. and Pracchia, M. and Pradier, T. and
                   Prajapati, A. K. and Prasai, K. and Prasanna, R. and
                   Pratten, G. and Principe, M. and Prodi, G. A. and
                   Prokhorov, L. and Prosposito, P. and Prudenzi, L. and
                   Puecher, A. and Punturo, M. and Puosi, F. and
                   Puppo, P. and P{\"u}rrer, M. and Qi, H. and
                   Quetschke, V. and {Quitzow-James}, R. and Raab, F. J. and
                   Raaijmakers, G. and Radkins, H. and Radulesco, N. and
                   Raffai, P. and Rail, S. X. and Raja, S. and Rajan, C. and
                   Ramirez, K. E. and Ramirez, T. D. and
                   {Ramos-Buades}, A. and Rana, J. and Rapagnani, P. and
                   Rapol, U. D. and Ray, A. and Raymond, V. and Raza, N. and
                   Razzano, M. and Read, J. and Rees, L. A. and
                   Regimbau, T. and Rei, L. and Reid, S. and Reid, S. W. and
                   Reitze, D. H. and Relton, P. and Renzini, A. and
                   Rettegno, P. and Reza, A. and Rezac, M. and Ricci, F. and
                   Richards, D. and Richardson, J. W. and Richardson, L. and
                   Riemenschneider, G. and Riles, K. and Rinaldi, S. and
                   Rink, K. and Rizzo, M. and Robertson, N. A. and
                   Robie, R. and Robinet, F. and Rocchi, A. and
                   Rodriguez, S. and Rolland, L. and Rollins, J. G. and
                   Romanelli, M. and Romano, R. and Romel, C. L. and
                   {Romero-Rodr{\'i}guez}, A. and {Romero-Shaw}, I. M. and
                   Romie, J. H. and Ronchini, S. and Rosa, L. and
                   Rose, C. A. and Rosi{\'n}ska, D. and Ross, M. P. and
                   Rowan, S. and Rowlinson, S. J. and Roy, S. and
                   Roy, Santosh and Roy, Soumen and Rozza, D. and
                   Ruggi, P. and Ryan, K. and Sachdev, S. and
                   Sadecki, T. and Sadiq, J. and Sago, N. and Saito, S. and
                   Saito, Y. and Sakai, K. and Sakai, Y. and
                   Sakellariadou, M. and Sakuno, Y. and Salafia, O. S. and
                   Salconi, L. and Saleem, M. and Salemi, F. and
                   Samajdar, A. and Sanchez, E. J. and Sanchez, J. H. and
                   Sanchez, L. E. and {Sanchis-Gual}, N. and
                   Sanders, J. R. and Sanuy, A. and Saravanan, T. R. and
                   Sarin, N. and Sassolas, B. and Satari, H. and
                   Sathyaprakash, B. S. and Sato, S. and Sato, T. and
                   Sauter, O. and Savage, R. L. and Sawada, T. and
                   Sawant, D. and Sawant, H. L. and Sayah, S. and
                   Schaetzl, D. and Scheel, M. and Scheuer, J. and
                   Schiworski, M. and Schmidt, P. and Schmidt, S. and
                   Schnabel, R. and Schneewind, M. and
                   Schofield, R. M. S. and Sch{\"o}nbeck, A. and
                   Schulte, B. W. and Schutz, B. F. and Schwartz, E. and
                   Scott, J. and Scott, S. M. and {Seglar-Arroyo}, M. and
                   Sekiguchi, T. and Sekiguchi, Y. and Sellers, D. and
                   Sengupta, A. S. and Sentenac, D. and Seo, E. G. and
                   Sequino, V. and Sergeev, A. and Setyawati, Y. and
                   Shaffer, T. and Shahriar, M. S. and Shams, B. and
                   Shao, L. and Sharma, A. and Sharma, P. and
                   Shawhan, P. and Shcheblanov, N. S. and Shibagaki, S. and
                   Shikauchi, M. and Shimizu, R. and Shimoda, T. and
                   Shimode, K. and Shinkai, H. and Shishido, T. and
                   Shoda, A. and Shoemaker, D. H. and Shoemaker, D. M. and
                   Shyamsundar, S. and Sieniawska, M. and Sigg, D. and
                   Singer, L. P. and Singh, D. and Singh, N. and
                   Singha, A. and Sintes, A. M. and Sipala, V. and
                   Skliris, V. and Slagmolen, B. J. J. and
                   {Slaven-Blair}, T. J. and Smetana, J. and
                   Smith, J. R. and Smith, R. J. E. and Soldateschi, J. and
                   Somala, S. N. and Somiya, K. and Son, E. J. and
                   Soni, K. and Soni, S. and Sordini, V. and
                   Sorrentino, F. and Sorrentino, N. and Sotani, H. and
                   Soulard, R. and Souradeep, T. and Sowell, E. and
                   Spagnuolo, V. and Spencer, A. P. and Spera, M. and
                   Srinivasan, R. and Srivastava, A. K. and
                   Srivastava, V. and Staats, K. and Stachie, C. and
                   Steer, D. A. and Steinhoff, J. and Steinlechner, J. and
                   Steinlechner, S. and Stevenson, S. P. and
                   Stops, D. J. and Stover, M. and Strain, K. A. and
                   Strang, L. C. and Stratta, G. and Strunk, A. and
                   Sturani, R. and Stuver, A. L. and Sudhagar, S. and
                   Sudhir, V. and Sugimoto, R. and Suh, H. G. and
                   Sullivan, A. G. and Summerscales, T. Z. and Sun, H. and
                   Sun, L. and Sunil, S. and Sur, A. and Suresh, J. and
                   Sutton, P. J. and Suzuki, Takamasa and
                   Suzuki, Toshikazu and Swinkels, B. L. and
                   Szczepa{\'n}czyk, M. J. and Szewczyk, P. and
                   Tacca, M. and Tagoshi, H. and Tait, S. C. and
                   Takahashi, H. and Takahashi, R. and Takamori, A. and
                   Takano, S. and Takeda, H. and Takeda, M. and
                   Talbot, C. J. and Talbot, C. and Tanaka, H. and
                   Tanaka, Kazuyuki and Tanaka, Kenta and Tanaka, Taiki and
                   Tanaka, Takahiro and Tanasijczuk, A. J. and
                   Tanioka, S. and Tanner, D. B. and Tao, D. and Tao, L. and
                   Mart{\'i}n, E. N. Tapia San and Taranto, C. and
                   Tasson, J. D. and Telada, S. and Tenorio, R. and
                   Terhune, J. E. and Terkowski, L. and
                   Thirugnanasambandam, M. P. and Thomas, L. and
                   Thomas, M. and Thomas, P. and Thompson, J. E. and
                   Thondapu, S. R. and Thorne, K. A. and Thrane, E. and
                   Tiwari, Shubhanshu and Tiwari, Srishti and Tiwari, V. and
                   Toivonen, A. M. and Toland, K. and Tolley, A. E. and
                   Tomaru, T. and Tomigami, Y. and Tomura, T. and
                   Tonelli, M. and {Torres-Forn{\'e}}, A. and
                   Torrie, C. I. and E Melo, I. Tosta and
                   T{\"o}yr{\"a}, D. and Trapananti, A. and Travasso, F. and
                   Traylor, G. and Trevor, M. and Tringali, M. C. and
                   Tripathee, A. and Troiano, L. and Trovato, A. and
                   Trozzo, L. and Trudeau, R. J. and Tsai, D. S. and
                   Tsai, D. and Tsang, K. W. and Tsang, T. and
                   Tsao, J. -S. and Tse, M. and Tso, R. and Tsubono, K. and
                   Tsuchida, S. and Tsukada, L. and Tsuna, D. and
                   Tsutsui, T. and Tsuzuki, T. and Turbang, K. and
                   Turconi, M. and Tuyenbayev, D. and Ubhi, A. S. and
                   Uchikata, N. and Uchiyama, T. and Udall, R. P. and
                   Ueda, A. and Uehara, T. and Ueno, K. and Ueshima, G. and
                   Unnikrishnan, C. S. and Uraguchi, F. and Urban, A. L. and
                   Ushiba, T. and Utina, A. and Vahlbruch, H. and
                   Vajente, G. and Vajpeyi, A. and Valdes, G. and
                   Valentini, M. and Valsan, V. and {van Bakel}, N. and
                   {van Beuzekom}, M. and {van den Brand}, J. F. J. and
                   {van den Broeck}, C. and {Vander-Hyde}, D. C. and
                   {van der Schaaf}, L. and {van Heijningen}, J. V. and
                   Vanosky, J. and {van Putten}, M. H. P. M. and
                   {van Remortel}, N. and Vardaro, M. and Vargas, A. F. and
                   Varma, V. and Vas{\'u}th, M. and Vecchio, A. and
                   Vedovato, G. and Veitch, J. and Veitch, P. J. and
                   Venneberg, J. and Venugopalan, G. and Verkindt, D. and
                   Verma, P. and Verma, Y. and Veske, D. and Vetrano, F. and
                   Vicer{\'e}, A. and Vidyant, S. and Viets, A. D. and
                   Vijaykumar, A. and {Villa-Ortega}, V. and
                   Vinet, J. -Y. and Virtuoso, A. and Vitale, S. and
                   Vo, T. and Vocca, H. and {von Reis}, E. R. G. and
                   {von Wrangel}, J. S. A. and Vorvick, C. and
                   Vyatchanin, S. P. and Wade, L. E. and Wade, M. and
                   Wagner, K. J. and Walet, R. C. and Walker, M. and
                   Wallace, G. S. and Wallace, L. and Walsh, S. and
                   Wang, J. and Wang, J. Z. and Wang, W. H. and
                   Ward, R. L. and Warner, J. and Was, M. and
                   Washimi, T. and Washington, N. Y. and Watchi, J. and
                   Weaver, B. and Webster, S. A. and Weinert, M. and
                   Weinstein, A. J. and Weiss, R. and Weller, C. M. and
                   Wellmann, F. and Wen, L. and We{\ss}els, P. and
                   Wette, K. and Whelan, J. T. and White, D. D. and
                   Whiting, B. F. and Whittle, C. and Wilken, D. and
                   Williams, D. and Williams, M. J. and
                   Williamson, A. R. and Willis, J. L. and Willke, B. and
                   Wilson, D. J. and Winkler, W. and Wipf, C. C. and
                   Wlodarczyk, T. and Woan, G. and Woehler, J. and
                   Wofford, J. K. and Wong, I. C. F. and Wu, C. and
                   Wu, D. S. and Wu, H. and Wu, S. and Wysocki, D. M. and
                   Xiao, L. and Xu, W. -R. and Yamada, T. and
                   Yamamoto, H. and Yamamoto, Kazuhiro and
                   Yamamoto, Kohei and Yamamoto, T. and Yamashita, K. and
                   Yamazaki, R. and Yang, F. W. and Yang, L. and
                   Yang, Y. and Yang, Yang and Yang, Z. and Yap, M. J. and
                   Yeeles, D. W. and Yelikar, A. B. and Ying, M. and
                   Yokogawa, K. and Yokoyama, J. and Yokozawa, T. and
                   Yoo, J. and Yoshioka, T. and Yu, Hang and Yu, Haocun and
                   Yuzurihara, H. and Zadro{\.z}ny, A. and Zanolin, M. and
                   Zeidler, S. and Zelenova, T. and Zendri, J. -P. and
                   Zevin, M. and Zhan, M. and Zhang, H. and Zhang, J. and
                   Zhang, L. and Zhang, T. and Zhang, Y. and Zhao, C. and
                   Zhao, G. and Zhao, Y. and Zhao, Yue and Zheng, Y. and
                   Zhou, R. and Zhou, Z. and Zhu, X. J. and Zhu, Z. -H. and
                   Zimmerman, A. B. and Zlochower, Y. and Zucker, M. E. and
                   Zweizig, J. and {LIGO Scientific Collaboration} and
                   {VIRGO Collaboration} and {KAGRA Collaboration}},
  journal =       {Phys. Rev. X},
  month =         jan,
  pages =         {011048},
  publisher =     {APS},
  title =         {Population of {{Merging Compact Binaries Inferred
                   Using Gravitational Waves}} through {{GWTC-3}}},
  volume =        {13},
  year =          {2023},
  doi =           {10.1103/PhysRevX.13.011048},
}

@article{schneider2023a,
  author =        {Schneider, Fabian R. N. and Podsiadlowski, Philipp and
                   Laplace, Eva},
  journal =       {ApJ},
  month =         jun,
  number =        {L9},
  pages =         {L9},
  title =         {Bimodal Black Hole Mass Distribution and Chirp Masses
                   of Binary Black Hole Mergers},
  volume =        {950},
  year =          {2023},
  doi =           {10.3847/2041-8213/acd77a},
}

@misc{willcox2025a,
  author =        {Willcox, Reinhold and Schneider, Fabian R. N. and
                   Laplace, Eva and Podsiadlowski, Philipp and
                   Maltsev, Kiril and Mandel, Ilya and Marchant, Pablo and
                   Sana, Hugues and Li, Tjonnie G. F. and
                   Hertog, Thomas},
  month =         oct,
  publisher =     {arXiv},
  title =         {Good Things Always Come in 3s: Trimodality in the
                   Binary Black-Hole Chirp-Mass Distribution Supports
                   Bimodal Black-Hole Formation},
  year =          {2025},
}

@article{miller-jones2021a,
  author =        {{Miller-Jones}, James C. A. and Bahramian, Arash and
                   Orosz, Jerome A. and Mandel, Ilya and Gou, Lijun and
                   Maccarone, Thomas J. and Neijssel, Coenraad J. and
                   Zhao, Xueshan and Zi{\'o}{\l}kowski, Janusz and
                   Reid, Mark J. and Uttley, Phil and Zheng, Xueying and
                   Byun, Do-Young and Dodson, Richard and
                   Grinberg, Victoria and Jung, Taehyun and
                   Kim, Jeong-Sook and Marcote, Benito and Markoff, Sera and
                   Rioja, Mar{\'\i}a J. and Rushton, Anthony P. and
                   Russell, David M. and Sivakoff, Gregory R. and
                   Tetarenko, Alexandra J. and Tudose, Valeriu and
                   Wilms, Joern},
  journal =       {Science},
  month =         mar,
  number =        {6533},
  pages =         {1046--1049},
  title =         {Cygnus {{X-1}} Contains a 21-Solar Mass Black
                   Hole---{{Implications}} for Massive Star Winds},
  volume =        {371},
  year =          {2021},
  doi =           {10.1126/science.abb3363},
}

@article{binder2021a,
  author =        {Binder, Breanna A. and Sy, Janelle M. and
                   Eracleous, Michael and Christodoulou, Dimitris M. and
                   Bhattacharya, Sayantan and Cappallo, Rigel and
                   Laycock, Silas and Plucinsky, Paul P. and
                   Williams, Benjamin F.},
  journal =       {ApJ},
  month =         mar,
  pages =         {74},
  title =         {The {{Wolf-Rayet}} + {{Black Hole Binary NGC}} 300
                   {{X-1}}: {{What}} Is the {{Mass}} of the {{Black
                   Hole}}?},
  volume =        {910},
  year =          {2021},
  doi =           {10.3847/1538-4357/abe6a9},
  issn =          {0004-637X},
}

@article{casares2014b,
  author =        {Casares, J. and Jonker, P. G.},
  journal =       {Space\textasciitilde Sci.\textasciitilde Rev.},
  month =         sep,
  number =        {1-4},
  pages =         {223--252},
  title =         {Mass Measurements of Stellar and Intermediate-Mass
                   Black Holes},
  volume =        {183},
  year =          {2014},
  doi =           {10.1007/s11214-013-0030-6},
}

@article{giesers2019a,
  author =        {Giesers, Benjamin and Kamann, Sebastian and
                   Dreizler, Stefan and Husser, Tim-Oliver and
                   Askar, Abbas and G{\"o}ttgens, Fabian and
                   Brinchmann, Jarle and Latour, Marilyn and
                   Weilbacher, Peter M. and Wendt, Martin and
                   Roth, Martin M.},
  journal =       {A\&A},
  month =         dec,
  number =        {A3},
  pages =         {A3},
  title =         {A Stellar Census in Globular Clusters with {{MUSE}}:
                   {{Binaries}} in {{NGC}} 3201},
  volume =        {632},
  year =          {2019},
  doi =           {10.1051/0004-6361/201936203},
}

@article{el-badry2023a,
  author =        {{El-Badry}, Kareem and Rix, Hans-Walter and
                   Cendes, Yvette and Rodriguez, Antonio C. and
                   Conroy, Charlie and Quataert, Eliot and
                   Hawkins, Keith and Zari, Eleonora and Hobson, Melissa and
                   Breivik, Katelyn and Rau, Arne and Berger, Edo and
                   Shahaf, Sahar and Seeburger, Rhys and
                   Burdge, Kevin B. and Latham, David W. and
                   Buchhave, Lars A. and Bieryla, Allyson and
                   Bashi, Dolev and Mazeh, Tsevi and Faigler, Simchon},
  journal =       {MNRAS},
  month =         may,
  number =        {3},
  pages =         {4323--4348},
  title =         {A Red Giant Orbiting a Black Hole},
  volume =        {521},
  year =          {2023},
  doi =           {10.1093/mnras/stad799},
}

@article{panuzzo2024a,
  author =        {{Gaia Collaboration} and Panuzzo, P. and Mazeh, T. and
                   Arenou, F. and Holl, B. and Caffau, E. and
                   Jorissen, A. and Babusiaux, C. and Gavras, P. and
                   Sahlmann, J. and Bastian, U. and Wyrzykowski, {\L}. and
                   Eyer, L. and Leclerc, N. and Bauchet, N. and
                   Bombrun, A. and Mowlavi, N. and Seabroke, G. M. and
                   Teyssier, D. and Balbinot, E. and Helmi, A. and
                   Brown, A. G. A. and Vallenari, A. and Prusti, T. and
                   {de Bruijne}, J. H. J. and Barbier, A. and
                   Biermann, M. and Creevey, O. L. and Ducourant, C. and
                   Evans, D. W. and Guerra, R. and Hutton, A. and
                   Jordi, C. and Klioner, S. A. and Lammers, U. and
                   Lindegren, L. and Luri, X. and Mignard, F. and
                   Nicolas, C. and Randich, S. and Sartoretti, P. and
                   Smiljanic, R. and Tanga, P. and Walton, N. A. and
                   Aerts, C. and {Bailer-Jones}, C. A. L. and
                   Cropper, M. and Drimmel, R. and Jansen, F. and
                   Katz, D. and Lattanzi, M. G. and Soubiran, C. and
                   Th{\'e}venin, F. and {van Leeuwen}, F. and Andrae, R. and
                   Audard, M. and Bakker, J. and Blomme, R. and
                   Casta{\~n}eda, J. and De Angeli, F. and Fabricius, C. and
                   Fouesneau, M. and Fr{\'e}mat, Y. and Galluccio, L. and
                   Guerrier, A. and Heiter, U. and Masana, E. and
                   Messineo, R. and Nienartowicz, K. and Pailler, F. and
                   Riclet, F. and Roux, W. and Sordo, R. and
                   {Gracia-Abril}, G. and Portell, J. and Altmann, M. and
                   Benson, K. and Berthier, J. and Burgess, P. W. and
                   Busonero, D. and Busso, G. and Cacciari, C. and
                   C{\'a}novas, H. and Carrasco, J. M. and Carry, B. and
                   Cellino, A. and Cheek, N. and Clementini, G. and
                   Damerdji, Y. and Davidson, M. and {de Teodoro}, P. and
                   Delchambre, L. and Dell'Oro, A. and Fraile Garcia, E. and
                   Garabato, D. and {Garc{\'\i}a-Lario}, P. and
                   Haigron, R. and Hambly, N. C. and Harrison, D. L. and
                   Hatzidimitriou, D. and Hern{\'a}ndez, J. and
                   Hestroffer, D. and Hodgkin, S. T. and Jamal, S. and
                   {Jevardat de Fombelle}, G. and Jordan, S. and
                   {Krone-Martins}, A. and Lanzafame, A. C. and
                   L{\"o}ffler, W. and Lorca, A. and Marchal, O. and
                   Marrese, P. M. and Moitinho, A. and Muinonen, K. and
                   Nu{\~n}ez Campos, M. and {Oreshina-Slezak}, I. and
                   Osborne, P. and Pancino, E. and Pauwels, T. and
                   {Recio-Blanco}, A. and Riello, M. and Rimoldini, L. and
                   Robin, A. C. and Roegiers, T. and Sarro, L. M. and
                   Schultheis, M. and Smith, M. and Sozzetti, A. and
                   Utrilla, E. and {van Leeuwen}, M. and Weingrill, K. and
                   Abbas, U. and {\'A}brah{\'a}m, P. and
                   Abreu Aramburu, A. and Ahmed, S. and Altavilla, G. and
                   {\'A}lvarez, M. A. and Anders, F. and Anderson, R. I. and
                   Anglada Varela, E. and Antoja, T. and Baig, S. and
                   Baines, D. and Baker, S. G. and
                   {Balaguer-N{\'u}{\~n}ez}, L. and Balog, Z. and
                   Barache, C. and Barros, M. and Barstow, M. A. and
                   Bartolom{\'e}, S. and Bashi, D. and Bassilana, J. -L. and
                   Baudeau, N. and Becciani, U. and Bedin, L. R. and
                   {Bellas-Velidis}, I. and Bellazzini, M. and
                   Beordo, W. and Bernet, M. and Bertolotto, C. and
                   Bertone, S. and Bianchi, L. and Binnenfeld, A. and
                   {Blanco-Cuaresma}, S. and {Bland-Hawthorn}, J. and
                   Blazere, A. and Boch, T. and Bossini, D. and
                   Bouquillon, S. and Bragaglia, A. and Braine, J. and
                   Bratsolis, E. and Breedt, E. and Bressan, A. and
                   Brouillet, N. and Brugaletta, E. and Bucciarelli, B. and
                   Butkevich, A. G. and Buzzi, R. and Camut, A. and
                   Cancelliere, R. and {Cantat-Gaudin}, T. and
                   Capilla Guilarte, D. and Carballo, R. and
                   Carlucci, T. and Carnerero, M. I. and Carretero, J. and
                   Carton, S. and Casamiquela, L. and Casey, A. and
                   Castellani, M. and {Castro-Ginard}, A. and Ceraj, L. and
                   Cesare, V. and Charlot, P. and Chaudet, C. and
                   Chemin, L. and Chiavassa, A. and Chornay, N. and
                   Chosson, D. and Cooper, W. J. and Cornez, T. and
                   Cowell, S. and Crosta, M. and Crowley, C. and
                   Cruz Reyes, M. and Dafonte, C. and Dal Ponte, M. and
                   David, M. and {de Laverny}, P. and De Luise, F. and
                   De March, R. and {de Torres}, A. and
                   {del Peloso}, E. F. and Delbo, M. and Delgado, A. and
                   Delisle, J. -B. and Demouchy, C. and Denis, E. and
                   Dharmawardena, T. E. and Di Giacomo, F. and
                   Diener, C. and Distefano, E. and Dolding, C. and
                   Dsilva, K. and Enke, H. and Fabre, C. and
                   Fabrizio, M. and Faigler, S. and Fatovi{\'c}, M. and
                   Fedorets, G. and {Fern{\'a}ndez-Hern{\'a}ndez}, J. and
                   Fernique, P. and Figueras, F. and Fouron, C. and
                   Fragkoudi, F. and Gai, M. and Galinier, M. and
                   {Garcia-Serrano}, A. and {Garc{\'\i}a-Torres}, M. and
                   Garofalo, A. and Gerlach, E. and Geyer, R. and
                   Giacobbe, P. and Gilmore, G. and Girona, S. and
                   Giuffrida, G. and Gomboc, A. and Gomez, A. and
                   {Gonz{\'a}lez-Santamar{\'\i}a}, I. and Gosset, E. and
                   Granvik, M. and Gregori Barrera, V. and
                   {Guti{\'e}rrez-S{\'a}nchez}, R. and Haywood, M. and
                   Helmer, A. and Hidalgo, S. L. and Hilger, T. and
                   Hobbs, D. and Hottier, C. and Huckle, H. E. and
                   {Jim{\'e}nez-Arranz}, {\'O}. and
                   Juaristi Campillo, J. and Kaczmarek, Z. and
                   Kervella, P. and Khanna, S. and Kontizas, M. and
                   Kordopatis, G. and Korn, A. J. and
                   K{\'o}sp{\'a}l, {\'A}. and {Kostrzewa-Rutkowska}, Z. and
                   Kruszy{\'n}ska, K. and Kun, M. and Lambert, S. and
                   Lanza, A. F. and Lebreton, Y. and Lebzelter, T. and
                   Leccia, S. and Lecoutre, G. and Liao, S. and
                   Liberato, L. and Licata, E. and Livanou, E. and
                   Lobel, A. and {L{\'o}pez-Miralles}, J. and Loup, C. and
                   Madar{\'a}sz, M. and Mahy, L. and Mann, R. G. and
                   Manteiga, M. and Marcellino, C. P. and
                   Marchant, J. M. and Marconi, M. and
                   Mar{\'\i}n Pina, D. and Marinoni, S. and
                   Marshall, D. J. and Mart{\'\i}n Lozano, J. and
                   Martin Polo, L. and {Mart{\'\i}n-Fleitas}, J. M. and
                   Marton, G. and Mascarenhas, D. and Masip, A. and
                   {Mastrobuono-Battisti}, A. and McMillan, P. J. and
                   Meichsner, J. and Merc, J. and Messina, S. and
                   Millar, N. R. and Mints, A. and Mohamed, D. and
                   Molina, D. and Molinaro, R. and Moln{\'a}r, L. and
                   Mongui{\'o}, M. and Montegriffo, P. and Monti, L. and
                   Mora, A. and Morbidelli, R. and Morris, D. and
                   Mudimadugula, R. and Muraveva, T. and Musella, I. and
                   Nagy, Z. and Nardetto, N. and Navarrete, C. and
                   Oh, S. and Ordenovic, C. and Orenstein, O. and
                   Pagani, C. and Pagano, I. and Palaversa, L. and
                   Palicio, P. A. and {Pallas-Quintela}, L. and
                   Pawlak, M. and Penttil{\"a}, A. and Pesciullesi, P. and
                   Pinamonti, M. and Plachy, E. and Planquart, L. and
                   Plum, G. and Poggio, E. and Pourbaix, D. and
                   {Price-Whelan}, A. M. and Pulone, L. and Rabin, V. and
                   Rainer, M. and Raiteri, C. M. and Ramos, P. and
                   {Ramos-Lerate}, M. and Ratajczak, M. and
                   Re Fiorentin, P. and Regibo, S. and Reyl{\'e}, C. and
                   Ripepi, V. and Riva, A. and Rix, H. -W. and Rixon, G. and
                   Robert, G. and Robichon, N. and Robin, C. and
                   {Romero-G{\'o}mez}, M. and Rowell, N. and
                   Ruz Mieres, D. and Rybicki, K. A. and Sadowski, G. and
                   Sagrist{\`a} Sell{\'e}s, A. and Sanna, N. and
                   Santove{\~n}a, R. and Sarasso, M. and
                   Sarmiento, M. H. and Sarrate Riera, C. and
                   Sciacca, E. and S{\'e}gransan, D. and Semczuk, M. and
                   Shahaf, S. and Siebert, A. and Slezak, E. and
                   Smart, R. L. and Snaith, O. N. and Solano, E. and
                   Solitro, F. and Souami, D. and Souchay, J. and
                   Spitoni, E. and Spoto, F. and Squillante, L. A. and
                   Steele, I. A. and Steidelm{\"u}ller, H. and
                   Surdej, J. and Szabados, L. and Taris, F. and
                   Taylor, M. B. and Teixeira, R. and
                   {Tepper-Garcia}, T. and Thuillot, W. and Tolomei, L. and
                   Tonello, N. and Torra, F. and Torralba Elipe, G. and
                   Trabucchi, M. and Trentin, E. and Tsantaki, M. and
                   Turon, C. and Ulla, A. and Unger, N. and
                   Valtchanov, I. and Vanel, O. and Vecchiato, A. and
                   Vicente, D. and Villar, E. and Weiler, M. and
                   Zhao, H. and Zorec, J. and Zucker, S. and
                   {\v Z}upi{\'c}, A. and Zwitter, T.},
  journal =       {A\&A},
  month =         jun,
  number =        {L2},
  pages =         {L2},
  title =         {Discovery of a Dormant 33 Solar-Mass Black Hole in
                   Pre-Release {{Gaia}} Astrometry},
  volume =        {686},
  year =          {2024},
  doi =           {10.1051/0004-6361/202449763},
}

@article{wang2024b,
  author =        {Wang, Song and Zhao, Xinlin and Feng, Fabo and
                   Ge, Hongwei and Shao, Yong and Cui, Yingzhen and
                   Gao, Shijie and Zhang, Lifu and Wang, Pei and Li, Xue and
                   Bai, Zhongrui and Yuan, Hailong and Huang, Yang and
                   Yuan, Haibo and Zhang, Zhixiang and Yi, Tuan and
                   Xiang, Maosheng and Li, Zhenwei and Li, Tanda and
                   Zhang, Junbo and Zhang, Meng and Han, Henggeng and
                   Fan, Dongwei and Li, Xiangdong and Chen, Xuefei and
                   Liu, Zhengwei and Meng, Xiangcun and Liu, Qingzhong and
                   Zhang, Haotong and Gu, Wei-Min and Liu, Jifeng},
  journal =       {Nat. Astron.},
  month =         dec,
  pages =         {1583--1591},
  title =         {A Potential Mass-Gap Black Hole in a Wide Binary with
                   a Circular Orbit},
  volume =        {8},
  year =          {2024},
  doi =           {10.1038/s41550-024-02359-9},
  issn =          {2397-3366},
}

@article{wyrzykowski2020a,
  author =        {Wyrzykowski, {\L}ukasz and Mandel, Ilya},
  journal =       {A\&A},
  month =         apr,
  number =        {A20},
  pages =         {A20},
  title =         {Constraining the Masses of Microlensing Black Holes
                   and the Mass Gap with {{Gaia DR2}}},
  volume =        {636},
  year =          {2020},
  doi =           {10.1051/0004-6361/201935842},
}

@article{sahu2022a,
  author =        {Sahu, Kailash C. and Anderson, Jay and
                   Casertano, Stefano and Bond, Howard E. and
                   Udalski, Andrzej and Dominik, Martin and
                   Calamida, Annalisa and Bellini, Andrea and
                   Brown, Thomas M. and Rejkuba, Marina and Bajaj, Varun and
                   Kains, No{\'e} and Ferguson, Henry C. and
                   Fryer, Chris L. and Yock, Philip and
                   Mr{\'o}z, Przemek and Koz{\l}owski, Szymon and
                   Pietrukowicz, Pawe{\l} and Poleski, Radek and
                   Skowron, Jan and Soszy{\'n}ski, Igor and
                   Szyma{\'n}ski, Micha{\l} K. and Ulaczyk, Krzysztof and
                   Wyrzykowski, {\L}ukasz and Barry, Richard K. and
                   Bennett, David P. and Bond, Ian A. and Hirao, Yuki and
                   Silva, Stela Ishitani and Kondo, Iona and
                   Koshimoto, Naoki and Ranc, Cl{\'e}ment and
                   Rattenbury, Nicholas J. and Sumi, Takahiro and
                   Suzuki, Daisuke and Tristram, Paul J. and
                   Vandorou, Aikaterini and Beaulieu, Jean-Philippe and
                   Marquette, Jean-Baptiste and Cole, Andrew and
                   Fouqu{\'e}, Pascal and Hill, Kym and Dieters, Stefan and
                   Coutures, Christian and {Dominis-Prester}, Dijana and
                   Bennett, Clara and Bachelet, Etienne and
                   Menzies, John and Albrow, Michael and Pollard, Karen and
                   Gould, Andrew and Yee, Jennifer C. and Allen, William and
                   Almeida, Leonardo A. and Christie, Grant and
                   Drummond, John and {Gal-Yam}, Avishay and
                   Gorbikov, Evgeny and Jablonski, Francisco and
                   Lee, Chung-Uk and Maoz, Dan and Manulis, Ilan and
                   McCormick, Jennie and Natusch, Tim and
                   Pogge, Richard W. and Shvartzvald, Yossi and
                   J{\o}rgensen, Uffe G. and Alsubai, Khalid A. and
                   Andersen, Michael I. and Bozza, Valerio and
                   Novati, Sebastiano Calchi and Burgdorf, Martin and
                   Hinse, Tobias C. and Hundertmark, Markus and
                   Husser, Tim-Oliver and Kerins, Eamonn and
                   {Longa-Pe{\~n}a}, Penelope and Mancini, Luigi and
                   Penny, Matthew and Rahvar, Sohrab and Ricci, Davide and
                   Sajadian, Sedighe and Skottfelt, Jesper and
                   Snodgrass, Colin and Southworth, John and
                   {Tregloan-Reed}, Jeremy and Wambsganss, Joachim and
                   Wertz, Olivier and Tsapras, Yiannis and
                   Street, Rachel A. and Bramich, D. M. and Horne, Keith and
                   Steele, Iain A. and {RoboNet Collaboration}},
  journal =       {ApJ},
  month =         jul,
  pages =         {83},
  publisher =     {IOP},
  title =         {An {{Isolated Stellar}}-Mass {{Black Hole Detected}}
                   through {{Astrometric Microlensing}}},
  volume =        {933},
  year =          {2022},
  doi =           {10.3847/1538-4357/ac739e},
  issn =          {0004-637X},
}

@article{sahu2025a,
  author =        {Sahu, Kailash C. and Anderson, Jay and
                   Casertano, Stefano and Bond, Howard E. and
                   Dominik, Martin and Calamida, Annalisa and
                   Bellini, Andrea and Brown, Thomas M. and
                   Ferguson, Henry C. and Rejkuba, Marina},
  journal =       {ApJ},
  month =         apr,
  pages =         {104},
  title =         {{{OGLE-2011-BLG-0462}}: {{An Isolated Stellar}}-Mass
                   {{Black Hole Confirmed Using New HST Astrometry}} and
                   {{Updated Photometry}}},
  volume =        {983},
  year =          {2025},
  doi =           {10.3847/1538-4357/adbe6e},
  issn =          {0004-637X},
}

@misc{bambi2025a,
  author =        {Bambi, Cosimo},
  month =         jul,
  publisher =     {arXiv},
  title =         {Stellar-{{Mass Black Holes}}},
  year =          {2025},
}

@article{rinaldi2024a,
  author =        {Rinaldi, Stefano and Del Pozzo, Walter and
                   Mapelli, Michela and {Lorenzo-Medina}, Ana and
                   Dent, Thomas},
  journal =       {A\&A},
  month =         apr,
  pages =         {A204},
  title =         {Evidence of Evolution of the Black Hole Mass Function
                   with Redshift},
  volume =        {684},
  year =          {2024},
  doi =           {10.1051/0004-6361/202348161},
  issn =          {0004-6361},
}

@article{farr2011a,
  author =        {Farr, W. M. and Sravan, N. and Cantrell, A. and
                   Kreidberg, L. and Bailyn, C. D. and Mandel, I. and
                   Kalogera, V.},
  journal =       {ApJ},
  month =         nov,
  number =        {103},
  pages =         {103},
  title =         {The Mass Distribution of Stellar-Mass Black Holes},
  volume =        {741},
  year =          {2011},
  doi =           {10.1088/0004-637X/741/2/103},
}

@article{kreidberg2012a,
  author =        {Kreidberg, Laura and Bailyn, Charles D. and
                   Farr, Will M. and Kalogera, Vicky},
  journal =       {ApJ},
  month =         sep,
  number =        {36},
  pages =         {36},
  title =         {Mass Measurements of Black Holes in X-Ray Transients:
                   {{Is}} There a Mass Gap?},
  volume =        {757},
  year =          {2012},
  doi =           {10.1088/0004-637X/757/1/36},
}

@article{bailyn1998a,
  author =        {Bailyn, Charles D. and Jain, Raj K. and Coppi, Paolo and
                   Orosz, Jerome A.},
  journal =       {ApJ},
  month =         may,
  number =        {1},
  pages =         {367--374},
  title =         {The Mass Distribution of Stellar Black Holes},
  volume =        {499},
  year =          {1998},
  doi =           {10.1086/305614},
}

@article{ozel2010a,
  author =        {{\"O}zel, F. and Psaltis, D. and Narayan, R. and
                   McClintock, J. E.},
  journal =       {ApJ},
  month =         dec,
  pages =         {1918--1927},
  title =         {The Black Hole Mass Distribution in the Galaxy},
  volume =        {725},
  year =          {2010},
  doi =           {10.1088/0004-637X/725/2/1918},
}

@article{abac2024a,
  author =        {Abac, A. G. and Abbott, R. and Abouelfettouh, I. and
                   Acernese, F. and Ackley, K. and Adhicary, S. and
                   Adhikari, N. and Adhikari, R. X. and Adkins, V. K. and
                   Agarwal, D. and Agathos, M. and Aghaei Abchouyeh, M. and
                   Aguiar, O. D. and Aguilar, I. and Aiello, L. and
                   Ain, A. and Ajith, P. and Ak{\c c}ay, S. and
                   Akutsu, T. and Albanesi, S. and Alfaidi, R. A. and
                   {Al-Jodah}, A. and All{\'e}n{\'e}, C. and Allocca, A. and
                   {Al-Shammari}, S. and Altin, P. A. and
                   {Alvarez-Lopez}, S. and Amato, A. and {Amez-Droz}, L. and
                   Amorosi, A. and Amra, C. and Ananyeva, A. and
                   Anderson, S. B. and Anderson, W. G. and Andia, M. and
                   Ando, M. and Andrade, T. and Andres, N. and
                   {Andr{\'e}s-Carcasona}, M. and Andri{\'c}, T. and
                   Anglin, J. and Ansoldi, S. and Antelis, J. M. and
                   Antier, S. and Aoumi, M. and Appavuravther, E. Z. and
                   Appert, S. and Apple, S. K. and Arai, K. and
                   Araya, A. and Araya, M. C. and Areeda, J. S. and
                   Argianas, L. and Aritomi, N. and Armato, F. and
                   Arnaud, N. and Arogeti, M. and Aronson, S. M. and
                   Arun, K. G. and Ashton, G. and Aso, Y. and
                   Assiduo, M. and {Assis de Souza Melo}, S. and
                   Aston, S. M. and Astone, P. and Attadio, F. and
                   Aubin, F. and Aultoneal, K. and Avallone, G. and
                   Azrad, D. and Babak, S. and Badaracco, F. and
                   Badger, C. and Bae, S. and Bagnasco, S. and Bagui, E. and
                   Baier, J. G. and Baiotti, L. and Bajpai, R. and
                   Baka, T. and Ball, M. and Ballardin, G. and
                   Ballmer, S. W. and Banagiri, S. and Banerjee, B. and
                   Bankar, D. and Baral, P. and Barayoga, J. C. and
                   Barish, B. C. and Barker, D. and Barneo, P. and
                   Barone, F. and Barr, B. and Barsotti, L. and
                   Barsuglia, M. and Barta, D. and Bartoletti, A. M. and
                   Barton, M. A. and Bartos, I. and Basak, S. and
                   Basalaev, A. and Bassiri, R. and Basti, A. and
                   Bates, D. E. and Bawaj, M. and Baxi, P. and
                   Bayley, J. C. and Baylor, A. C. and Baynard, P. A. and
                   Bazzan, M. and Bedakihale, V. M. and Beirnaert, F. and
                   Bejger, M. and Belardinelli, D. and Bell, A. S. and
                   Benedetto, V. and Benoit, W. and Bentara, I. and
                   Bentley, J. D. and Ben Yaala, M. and Bera, S. and
                   Berbel, M. and Bergamin, F. and Berger, B. K. and
                   Bernuzzi, S. and Beroiz, M. and Berry, C. P. L. and
                   Bersanetti, D. and Bertolini, A. and Betzwieser, J. and
                   Beveridge, D. and Bevins, N. and Bhandare, R. and
                   Bhardwaj, U. and Bhatt, R. and Bhattacharjee, D. and
                   Bhaumik, S. and Bhowmick, S. and Bianchi, A. and
                   Bilenko, I. A. and Billingsley, G. and Binetti, A. and
                   Bini, S. and Birnholtz, O. and Biscoveanu, S. and
                   Bisht, A. and Bitossi, M. and Bizouard, M. -A. and
                   Blackburn, J. K. and Blagg, L. A. and Blair, C. D. and
                   Blair, D. G. and Bobba, F. and Bode, N. and
                   Boileau, G. and Boldrini, M. and Bolingbroke, G. N. and
                   Bolliand, A. and Bonavena, L. D. and Bondarescu, R. and
                   Bondu, F. and Bonilla, E. and Bonilla, M. S. and
                   Bonino, A. and Bonnand, R. and Booker, P. and
                   Borchers, A. and Boschi, V. and Bose, S. and
                   Bossilkov, V. and Boudart, V. and Boudon, A. and
                   Bozzi, A. and Bradaschia, C. and Brady, P. R. and
                   Braglia, M. and Branch, A. and Branchesi, M. and
                   Brandt, J. and Braun, I. and Breschi, M. and
                   Briant, T. and Brillet, A. and Brinkmann, M. and
                   Brockill, P. and Brockmueller, E. and Brooks, A. F. and
                   Brown, B. C. and Brown, D. D. and Brozzetti, M. L. and
                   Brunett, S. and Bruno, G. and Bruntz, R. and
                   Bryant, J. and Bucci, F. and Buchanan, J. and
                   Bulashenko, O. and Bulik, T. and Bulten, H. J. and
                   Buonanno, A. and Burtnyk, K. and Buscicchio, R. and
                   Buskulic, D. and Buy, C. and Byer, R. L. and
                   Cabourn Davies, G. S. and Cabras, G. and Cabrita, R. and
                   {C{\'a}ceres-Barbosa}, V. and Cadonati, L. and
                   Cagnoli, G. and Cahillane, C. and
                   Calder{\'o}n Bustillo, J. and Callister, T. A. and
                   Calloni, E. and Camp, J. B. and Canepa, M. and
                   Caneva Santoro, G. and Cannon, K. C. and Cao, H. and
                   Capistran, L. A. and Capocasa, E. and Capote, E. and
                   Carapella, G. and Carbognani, F. and Carlassara, M. and
                   Carlin, J. B. and Carpinelli, M. and Carrillo, G. and
                   Carter, J. J. and Carullo, G. and Casanueva Diaz, J. and
                   Casentini, C. and {Castro-Lucas}, S. Y. and
                   Caudill, S. and Cavagli{\`a}, M. and Cavalieri, R. and
                   Cella, G. and {Cerd{\'a}-Dur{\'a}n}, P. and
                   Cesarini, E. and Chaibi, W. and Chakraborty, P. and
                   Chalathadka Subrahmanya, S. and Chan, J. C. L. and
                   Chan, M. and Chandra, K. and Chang, R. -J. and
                   Chao, S. and Char, P. and Charlton, E. L. and
                   Charlton, P. and {Chassande-Mottin}, E. and
                   Chatterjee, C. and Chatterjee, Debarati and
                   Chatterjee, Deep and Chattopadhyay, D. and
                   Chaturvedi, M. and Chaty, S. and Chatziioannou, K. and
                   Chen, A. and Chen, A. H. -Y. and Chen, D. and
                   Chen, H. and Chen, H. Y. and Chen, J. and Chen, K. H. and
                   Chen, Y. and Chen, Yanbei and Chen, Yitian and
                   Cheng, H. P. and Chessa, P. and Cheung, H. T. and
                   Cheung, S. Y. and Chiadini, F. and Chiarini, G. and
                   Chierici, R. and Chincarini, A. and Chiofalo, M. L. and
                   Chiummo, A. and Chou, C. and Choudhary, S. and
                   Christensen, N. and Chua, S. S. Y. and Chugh, P. and
                   Ciani, G. and Ciecielag, P. and Cie{\'s}lar, M. and
                   Cifaldi, M. and Ciolfi, R. and Clara, F. and
                   Clark, J. A. and Clarke, J. and Clarke, T. A. and
                   Clearwater, P. and Clesse, S. and Coccia, E. and
                   Codazzo, E. and Cohadon, P. -F. and Colace, S. and
                   Colleoni, M. and Collette, C. G. and Collins, J. and
                   Colloms, S. and Colombo, A. and Colpi, M. and
                   Compton, C. M. and Connolly, G. and Conti, L. and
                   Corbitt, T. R. and {Cordero-Carri{\'o}n}, I. and
                   Corezzi, S. and Cornish, N. J. and Corsi, A. and
                   Cortese, S. and Costa, C. A. and Cottingham, R. and
                   Coughlin, M. W. and Couineaux, A. and Coulon, J. -P. and
                   Countryman, S. T. and Coupechoux, J. -F. and
                   Couvares, P. and Coward, D. M. and Cowart, M. J. and
                   Coyne, R. and Craig, K. and Creed, R. and
                   Creighton, J. D. E. and Creighton, T. D. and
                   Cremonese, P. and Criswell, A. W. and
                   {Crockett-Gray}, J. C. G. and Crook, S. and
                   Crouch, R. and Csizmazia, J. and Cudell, J. R. and
                   Cullen, T. J. and Cumming, A. and Cuoco, E. and
                   Cusinato, M. and Dabadie, P. and Dal Canton, T. and
                   Dall'Osso, S. and Dal Pra, S. and D{\'a}lya, G. and
                   D'Angelo, B. and Danilishin, S. and D'Antonio, S. and
                   Danzmann, K. and Darroch, K. E. and Dartez, L. P. and
                   Dasgupta, A. and Datta, S. and Dattilo, V. and
                   Daumas, A. and Davari, N. and Dave, I. and
                   Davenport, A. and Davier, M. and Davies, T. F. and
                   Davis, D. and Davis, L. and Davis, M. C. and
                   Davis, P. J. and Dax, M. and {de Bolle}, J. and
                   Deenadayalan, M. and Degallaix, J. and
                   {de Laurentis}, M. and Del{\'e}glise, S. and
                   {de Lillo}, F. and Dell'Aquila, D. and Del Pozzo, W. and
                   De Marco, F. and {de Matteis}, F. and D'Emilio, V. and
                   Demos, N. and Dent, T. and Depasse, A. and
                   Depergola, N. and {de Pietri}, R. and De Rosa, R. and
                   {de Rossi}, C. and Desalvo, R. and {de Simone}, R. and
                   Dhani, A. and Diab, R. and D{\'i}az, M. C. and
                   {di Cesare}, M. and Dideron, G. and Didio, N. A. and
                   Dietrich, T. and {di Fiore}, L. and {di Fronzo}, C. and
                   {di Giovanni}, M. and {di Girolamo}, T. and
                   Diksha, D. and {di Michele}, A. and Ding, J. and
                   {di Pace}, S. and {di Palma}, I. and {di Renzo}, F. and
                   {Divyajyoti} and Dmitriev, A. and Doctor, Z. and
                   Dohmen, E. and Doleva, P. P. and Dominguez, D. and
                   D'Onofrio, L. and Donovan, F. and Dooley, K. L. and
                   Dooney, T. and Doravari, S. and Dorosh, O. and
                   Drago, M. and Driggers, J. C. and Ducoin, J. -G. and
                   Dunn, L. and Dupletsa, U. and D'Urso, D. and
                   Duval, H. and Duverne, P. -A. and Dwyer, S. E. and
                   Eassa, C. and Ebersold, M. and Eckhardt, T. and
                   Eddolls, G. and Edelman, B. and Edo, T. B. and
                   Edy, O. and Effler, A. and Eichholz, J. and
                   Einsle, H. and Eisenmann, M. and Eisenstein, R. A. and
                   Ejlli, A. and Eleveld, R. M. and Emma, M. and
                   Endo, K. and Engl, A. J. and Enloe, E. and Errico, L. and
                   Essick, R. C. and Estell{\'e}s, H. and Estevez, D. and
                   Etzel, T. and Evans, M. and Evstafyeva, T. and
                   Ewing, B. E. and Ezquiaga, J. M. and Fabrizi, F. and
                   Faedi, F. and Fafone, V. and Fairhurst, S. and
                   Farah, A. M. and Farr, B. and Farr, W. M. and
                   Favaro, G. and Favata, M. and Fays, M. and Fazio, M. and
                   Feicht, J. and Fejer, M. M. and Felicetti, R. . and
                   Fenyvesi, E. and Ferguson, D. L. and Ferraiuolo, S. and
                   Ferrante, I. and Ferreira, T. A. and Fidecaro, F. and
                   Figura, P. and Fiori, A. and Fiori, I. and
                   Fishbach, M. and Fisher, R. P. and Fittipaldi, R. and
                   Fiumara, V. and Flaminio, R. and Fleischer, S. M. and
                   Fleming, L. S. and Floden, E. and Foley, E. M. and
                   Fong, H. and Font, J. A. and Fornal, B. and
                   Forsyth, P. W. F. and Franceschetti, K. and
                   Franchini, N. and Frasca, S. and Frasconi, F. and
                   Frattale Mascioli, A. and Frei, Z. and Freise, A. and
                   Freitas, O. and Frey, R. and Frischhertz, W. and
                   Fritschel, P. and Frolov, V. V. and Fronz{\'e}, G. G. and
                   {Fuentes-Garcia}, M. and Fujii, S. and Fujimori, T. and
                   Fulda, P. and Fyffe, M. and Gadre, B. and Gair, J. R. and
                   Galaudage, S. and Galdi, V. and Gallagher, H. and
                   Gallardo, S. and Gallego, B. and Gamba, R. and
                   Gamboa, A. and Ganapathy, D. and Ganguly, A. and
                   Garaventa, B. and {Garc{\'i}a-Bellido}, J. and
                   Garc{\'i}a N{\'u}{\~n}ez, C. and
                   {Garc{\'i}a-Quir{\'o}s}, C. and Gardner, J. W. and
                   Gardner, K. A. and Gargiulo, J. and Garron, A. and
                   Garufi, F. and Gasbarra, C. and Gateley, B. and
                   Gayathri, V. and Gemme, G. and Gennai, A. and
                   Gennari, V. and George, J. and George, R. and
                   Gerberding, O. and Gergely, L. and Ghonge, S. and
                   Ghosh, Archisman and Ghosh, Sayantan and Ghosh, Shaon and
                   Ghosh, Shrobana and Ghosh, Suprovo and
                   Ghosh, Tathagata and Giacoppo, L. and Giaime, J. A. and
                   Giardina, K. D. and Gibson, D. R. and Gibson, D. T. and
                   Gier, C. and Giri, P. and Gissi, F. and
                   Gkaitatzis, S. and Glanzer, J. and Glotin, F. and
                   Godfrey, J. and Godwin, P. and Goebbels, N. L. and
                   Goetz, E. and Golomb, J. and Gomez Lopez, S. and
                   Goncharov, B. and Gong, Y. and Gonz{\'a}lez, G. and
                   Goodarzi, P. and Goode, S. and {Goodwin-Jones}, A. W. and
                   Gosselin, M. and G{\"o}ttel, A. S. and Gouaty, R. and
                   Gould, D. W. and Govorkova, K. and Goyal, S. and
                   Grace, B. and Grado, A. and Graham, V. and
                   Granados, A. E. and Granata, M. and Granata, V. and
                   Gras, S. and Grassia, P. and Gray, A. and Gray, C. and
                   Gray, R. and Greco, G. and Green, A. C. and
                   Green, S. M. and Green, S. R. and Gretarsson, A. M. and
                   Gretarsson, E. M. and Griffith, D. and
                   Griffiths, W. L. and Griggs, H. L. and Grignani, G. and
                   Grimaldi, A. and Grimaud, C. and Grote, H. and
                   Guerra, D. and Guetta, D. and Guidi, G. M. and
                   Guimaraes, A. R. and Gulati, H. K. and Gulminelli, F. and
                   Gunny, A. M. and Guo, H. and Guo, W. and Guo, Y. and
                   Gupta, Anchal and Gupta, Anuradha and Gupta, Ish and
                   Gupta, N. C. and Gupta, P. and Gupta, S. K. and
                   Gupta, T. and Gupte, N. and Gurs, J. and
                   Gutierrez, N. and Guzman, F. and H, H. -Y. and
                   Haba, D. and Haberland, M. and Haino, S. and
                   Hall, E. D. and Hamilton, E. Z. and Hammond, G. and
                   Han, W. -B. and Haney, M. and Hanks, J. and Hanna, C. and
                   Hannam, M. D. and Hannuksela, O. A. and
                   Hanselman, A. G. and Hansen, H. and Hanson, J. and
                   Harada, R. and Hardison, A. R. and Haris, K. and
                   Harmark, T. and Harms, J. and Harry, G. M. and
                   Harry, I. W. and Hart, J. and Haskell, B. and
                   Haster, C. -J. and Hathaway, J. S. and Haughian, K. and
                   Hayakawa, H. and Hayama, K. and Hayes, R. and
                   Heffernan, A. and Heidmann, A. and Heintze, M. C. and
                   Heinze, J. and Heinzel, J. and Heitmann, H. and
                   Hellman, F. and Hello, P. and
                   {Helmling-Cornell}, A. F. and Hemming, G. and
                   {Henderson-Sapir}, O. and Hendry, M. and Heng, I. S. and
                   Hennes, E. and Henshaw, C. and Hertog, T. and
                   Heurs, M. and Hewitt, A. L. and Heyns, J. and
                   Higginbotham, S. and Hild, S. and Hill, S. and
                   Himemoto, Y. and Hirata, N. and Hirose, C. and
                   Hoang, S. and Hochheim, S. and Hofman, D. and
                   Holland, N. A. and {Holley-Bockelmann}, K. and
                   Holmes, Z. J. and Holz, D. E. and Honet, L. and
                   Hong, C. and Hornung, J. and Hoshino, S. and
                   Hough, J. and Hourihane, S. and Howell, E. J. and
                   Hoy, C. G. and Hrishikesh, C. A. and Hsieh, H. -F. and
                   Hsiung, C. and Hsu, H. C. and Hsu, W. -F. and Hu, P. and
                   Hu, Q. and Huang, H. Y. and Huang, Y. -J. and
                   Huddart, A. D. and Hughey, B. and Hui, D. C. Y. and
                   Hui, V. and Husa, S. and Huxford, R. and
                   {Huynh-Dinh}, T. and Iampieri, L. and Iandolo, G. A. and
                   Ianni, M. and Iess, A. and Imafuku, H. and
                   Inayoshi, K. and Inoue, Y. and Iorio, G. and
                   Iqbal, M. H. and Irwin, J. and Ishikawa, R. and
                   Isi, M. and Ismail, M. A. and Itoh, Y. and
                   Iwanaga, H. and Iwaya, M. and Iyer, B. R. and
                   Jaberianhamedan, V. and Jacquet, C. and
                   Jacquet, P. -E. and Jadhav, S. J. and Jadhav, S. P. and
                   Jain, T. and James, A. L. and James, P. A. and
                   Jamshidi, R. and Janquart, J. and Janssens, K. and
                   Janthalur, N. N. and Jaraba, S. and Jaranowski, P. and
                   Jaume, R. and Javed, W. and Jennings, A. and Jia, W. and
                   Jiang, J. and Kubisz, J. and Johanson, C. and
                   Johns, G. R. and Johnson, N. A. and
                   {Johnson-McDaniel}, N. K. and Johnston, M. C. and
                   Johnston, R. and Johny, N. and Jones, D. H. and
                   Jones, D. I. and Jones, R. and Jose, S. and Joshi, P. and
                   Ju, L. and Jung, K. and Junker, J. and Juste, V. and
                   Kajita, T. and Kaku, I. and Kalaghatgi, C. and
                   Kalogera, V. and Kamiizumi, M. and Kanda, N. and
                   Kandhasamy, S. and Kang, G. and Kanner, J. B. and
                   Kapadia, S. J. and Kapasi, D. P. and Karat, S. and
                   Karathanasis, C. and Kashyap, R. and Kasprzack, M. and
                   Kastaun, W. and Kato, T. and Katsavounidis, E. and
                   Katzman, W. and Kaushik, R. and Kawabe, K. and
                   Kawamoto, R. and Kazemi, A. and Kedia, A. and
                   Keitel, D. and {Kelley-Derzon}, J. and Kennington, J. and
                   Kesharwani, R. and Key, J. S. and Khadela, R. and
                   Khadka, S. and Khalili, F. Y. and Khan, F. and
                   Khan, I. and Khanam, T. and Khursheed, M. and
                   Khusid, N. M. and Kiendrebeogo, W. and Kijbunchoo, N. and
                   Kim, C. and Kim, J. C. and Kim, K. and Kim, M. H. and
                   Kim, S. and Kim, Y. -M. and Kimball, C. and
                   {Kinley-Hanlon}, M. and Kinnear, M. and Kissel, J. S. and
                   Klimenko, S. and Knee, A. M. and Knust, N. and
                   Kobayashi, K. and Koch, P. and Koehlenbeck, S. M. and
                   Koekoek, G. and Kohri, K. and Kokeyama, K. and
                   Koley, S. and Kolitsidou, P. and Kolstein, M. and
                   Komori, K. and Kong, A. K. H. and Kontos, A. and
                   Korobko, M. and Kossak, R. V. and Kou, X. and
                   Koushik, A. and Kouvatsos, N. and Kovalam, M. and
                   Kozak, D. B. and Kranzhoff, S. L. and Kringel, V. and
                   Krishnendu, N. V. and Kr{\'o}lak, A. and Kruska, K. and
                   Kuehn, G. and Kuijer, P. and Kulkarni, S. and
                   Kulur Ramamohan, A. and Kumar, A. and Kumar, Praveen and
                   Kumar, Prayush and Kumar, Rahul and Kumar, Rakesh and
                   Kume, J. and Kuns, K. and Kuntimaddi, N. and
                   Kuroyanagi, S. and Kurth, N. J. and Kuwahara, S. and
                   Kwak, K. and Kwan, K. and Kwok, J. and Lacaille, G. and
                   Lagabbe, P. and Laghi, D. and Lai, S. and
                   Laity, A. H. and Lakkis, M. H. and Lalande, E. and
                   Lalleman, M. and Lalremruati, P. C. and Landry, M. and
                   Landry, P. and Lane, B. B. and Lang, R. N. and
                   Lange, J. and Lantz, B. and La Rana, A. and
                   La Rosa, I. and {Lartaux-Vollard}, A. and
                   Lasky, P. D. and Lawrence, J. and Lawrence, M. N. and
                   Laxen, M. and Lazzarini, A. and Lazzaro, C. and
                   Leaci, P. and Lecoeuche, Y. K. and Lee, H. M. and
                   Lee, H. W. and Lee, K. and Lee, R. -K. and Lee, R. and
                   Lee, S. and Lee, Y. and Legred, I. N. and Lehmann, J. and
                   Lehner, L. and Le Jean, M. and Lema{\^i}tre, A. and
                   Lenti, M. and Leonardi, M. and Lequime, M. and
                   Leroy, N. and Lesovsky, M. and Letendre, N. and
                   Lethuillier, M. and Levin, S. E. and Levin, Y. and
                   Leyde, K. and Li, A. K. Y. and Li, K. L. and
                   Li, T. G. F. and Li, X. and Li, Z. and Lihos, A. and
                   Lin, C. -Y. and Lin, C. -Y. and Lin, E. T. and
                   Lin, F. and Lin, H. and Lin, L. C. -C. and
                   Lin, Y. -C. and Linde, F. and Linker, S. D. and
                   Littenberg, T. B. and Liu, A. and Liu, G. C. and
                   Liu, Jian and Llamas Villarreal, F. and
                   {Llobera-Querol}, J. and Lo, R. K. L. and
                   Locquet, J. -P. and London, L. T. and Longo, A. and
                   Lopez, D. and Lopez Portilla, M. and Lorenzini, M. and
                   {Lorenzo-Medina}, A. and Loriette, V. and Lormand, M. and
                   Losurdo, G. and Lott, T. P. and Lough, J. D. and
                   Loughlin, H. A. and Lousto, C. O. and Lowry, M. J. and
                   Lu, N. and L{\"u}ck, H. and Lumaca, D. and
                   Lundgren, A. P. and Lussier, A. W. and Ma, L. -T. and
                   Ma, S. and Ma'Arif, M. and Macas, R. and Macedo, A. and
                   Macinnis, M. and Maciy, R. R. and MacLeod, D. M. and
                   MacMillan, I. A. O. and Macquet, A. and Macri, D. and
                   Maeda, K. and Maenaut, S. and
                   Maga{\~n}a Hernandez, I. and Magare, S. S. and
                   Magazz{\`u}, C. and Magee, R. M. and Maggio, E. and
                   Maggiore, R. and Magnozzi, M. and Mahesh, M. and
                   Mahesh, S. and Maini, M. and Majhi, S. and
                   Majorana, E. and Makarem, C. N. and Makelele, E. and
                   {Malaquias-Reis}, J. A. and Mali, U. and Maliakal, S. and
                   Malik, A. and Man, N. and Mandic, V. and Mangano, V. and
                   Mannix, B. and Mansell, G. L. and Mansingh, G. and
                   Manske, M. and Mantovani, M. and Mapelli, M. and
                   Marchesoni, F. and Mar{\'i}n Pina, D. and Marion, F. and
                   M{\'a}rka, S. and M{\'a}rka, Z. and Markosyan, A. S. and
                   Markowitz, A. and Maros, E. and Marsat, S. and
                   Martelli, F. and Martin, I. W. and Martin, R. M. and
                   Martinez, B. B. and Martinez, M. and Martinez, V. and
                   Martini, A. and Martinovic, K. and Martins, J. C. and
                   Martynov, D. V. and Marx, E. J. and Massaro, L. and
                   Masserot, A. and {Masso-Reid}, M. and
                   Mastrodicasa, M. and Mastrogiovanni, S. and
                   Matcovich, T. and Matiushechkina, M. and
                   Matsuyama, M. and Mavalvala, N. and Maxwell, N. and
                   McCarrol, G. and McCarthy, R. and McClelland, D. E. and
                   McCormick, S. and McCuller, L. and McEachin, S. and
                   McElhenny, C. and McGhee, G. I. and McGinn, J. and
                   McGowan, K. B. M. and McIver, J. and McLeod, A. and
                   McRae, T. and Meacher, D. and Meijer, Q. and
                   Melatos, A. and Mellaerts, S. and
                   {Menendez-Vazquez}, A. and Menoni, C. S. and Mera, F. and
                   Mercer, R. A. and Mereni, L. and Merfeld, K. and
                   Merilh, E. L. and M{\'e}rou, J. R. and Merritt, J. D. and
                   Merzougui, M. and Messenger, C. and Messick, C. and
                   {Meyer-Conde}, M. and Meylahn, F. and Mhaske, A. and
                   Miani, A. and Miao, H. and Michaloliakos, I. and
                   Michel, C. and Michimura, Y. and Middleton, H. and
                   Miller, A. L. and Miller, S. and Millhouse, M. and
                   Milotti, E. and Milotti, V. and Minenkov, Y. and
                   Mio, N. and Mir, {\relax Ll}. M. and Mirasola, L. and
                   {Miravet-Ten{\'e}s}, M. and Miritescu, C. -A. and
                   Mishra, A. K. and Mishra, A. and Mishra, C. and
                   Mishra, T. and Mitchell, A. L. and Mitchell, J. G. and
                   Mitra, S. and Mitrofanov, V. P. and Mittleman, R. and
                   Miyakawa, O. and Miyamoto, S. and Miyoki, S. and
                   Mo, G. and Mobilia, L. and Mohapatra, S. R. P. and
                   Mohite, S. R. and {Molina-Ruiz}, M. and Mondal, C. and
                   Mondin, M. and Montani, M. and Moore, C. J. and
                   Moraru, D. and More, A. and More, S. and Moreno, G. and
                   Morgan, C. and Morisaki, S. and Moriwaki, Y. and
                   Morras, G. and Moscatello, A. and Mourier, P. and
                   Mours, B. and {Mow-Lowry}, C. M. and Muciaccia, F. and
                   Mukherjee, Arunava and Mukherjee, D. and
                   Mukherjee, Samanwaya and Mukherjee, Soma and
                   Mukherjee, Subroto and Mukherjee, Suvodip and
                   Mukund, N. and Mullavey, A. and Munch, J. and
                   Mundi, J. and Mungioli, C. L. and Munn Oberg, W. R. and
                   Murakami, Y. and Murakoshi, M. and Murray, P. G. and
                   Muusse, S. and Nabari, D. and Nadji, S. L. and
                   Nagar, A. and Nagarajan, N. and Nagler, K. N. and
                   Nakagaki, K. and Nakamura, K. and Nakano, H. and
                   Nakano, M. and Nandi, D. and Napolano, V. and
                   Narayan, P. and Nardecchia, I. and Narikawa, T. and
                   Narola, H. and Naticchioni, L. and Nayak, R. K. and
                   Neilson, J. and Nelson, A. and Nelson, T. J. N. and
                   Nery, M. and Neunzert, A. and Ng, S. and
                   Nguyen Quynh, L. and Nichols, S. A. and
                   Nielsen, A. B. and Nieradka, G. and Niko, A. and
                   Nishino, Y. and Nishizawa, A. and Nissanke, S. and
                   Nitoglia, E. and Niu, W. and Nocera, F. and
                   Norman, M. and North, C. and Novak, J. and
                   Nu{\~n}o Siles, J. F. and Nuttall, L. K. and
                   Obayashi, K. and Oberling, J. and O'Dell, J. and
                   Oertel, M. and Offermans, A. and Oganesyan, G. and
                   Oh, J. J. and Oh, K. and O'Hanlon, T. and Ohashi, M. and
                   Ohkawa, M. and Ohme, F. and Oliveira, A. S. and
                   Oliveri, R. and O'Neal, B. and Oohara, K. and
                   O'Reilly, B. and Ormsby, N. D. and Orselli, M. and
                   O'Shaughnessy, R. and O'Shea, S. and Oshima, Y. and
                   Oshino, S. and Ossokine, S. and Osthelder, C. and
                   Ota, I. and Ottaway, D. J. and Ouzriat, A. and
                   Overmier, H. and Owen, B. J. and Pace, A. E. and
                   Pagano, R. and Page, M. A. and Pai, A. and Pal, A. and
                   Pal, S. and Palaia, M. A. and P{\'a}lfi, M. and
                   Palma, P. P. and Palomba, C. and Palud, P. and
                   Pan, H. and Pan, J. and Pan, K. C. and Panai, R. and
                   Panda, P. K. and Pandey, S. and Panebianco, L. and
                   Pang, P. T. H. and Pannarale, F. and Pannone, K. A. and
                   Pant, B. C. and Panther, F. H. and Paoletti, F. and
                   Paolone, A. and Papalexakis, E. E. and Papalini, L. and
                   Papigkiotis, G. and Paquis, A. and Parisi, A. and
                   Park, B. -J. and Park, J. and Parker, W. and
                   Pascale, G. and Pascucci, D. and Pasqualetti, A. and
                   Passaquieti, R. and Passenger, L. and Passuello, D. and
                   Patane, O. and Pathak, D. and Pathak, M. and
                   Patra, A. and Patricelli, B. and Patron, A. S. and
                   Paul, K. and Paul, S. and Payne, E. and Pearce, T. and
                   Pedraza, M. and Pegna, R. and Pele, A. and
                   Pe{\~n}a Arellano, F. E. and Penn, S. and
                   Penuliar, M. D. and Perego, A. and Pereira, Z. and
                   Perez, J. J. and P{\'e}rigois, C. and Perna, G. and
                   Perreca, A. and Perret, J. and Perri{\`e}s, S. and
                   Perry, J. W. and Pesios, D. and Petracca, S. and
                   Petrillo, C. and Pfeiffer, H. P. and Pham, H. and
                   Pham, K. A. and Phukon, K. S. and Phurailatpam, H. and
                   Piarulli, M. and Piccari, L. and Piccinni, O. J. and
                   Pichot, M. and Piendibene, M. and Piergiovanni, F. and
                   Pierini, L. and Pierra, G. and Pierro, V. and
                   Pietrzak, M. and Pillas, M. and Pilo, F. and
                   Pinard, L. and Pinto, I. M. and Pinto, M. and
                   Piotrzkowski, B. J. and Pirello, M. and Pitkin, M. D. and
                   Placidi, A. and Placidi, E. and Planas, M. L. and
                   Plastino, W. and Poggiani, R. and Polini, E. and
                   Pompili, L. and Poon, J. and Porcelli, E. and
                   Porter, E. K. and Posnansky, C. and Poulton, R. and
                   Powell, J. and Pracchia, M. and Pradhan, B. K. and
                   Pradier, T. and Prajapati, A. K. and Prasai, K. and
                   Prasanna, R. and Prasia, P. and Pratten, G. and
                   Principe, G. and Principe, M. and Prodi, G. A. and
                   Prokhorov, L. and Prosposito, P. and Puecher, A. and
                   Pullin, J. and Punturo, M. and Puppo, P. and
                   P{\"u}rrer, M. and Qi, H. and Qin, J. and
                   Qu{\'e}m{\'e}ner, G. and Quetschke, V. and
                   Quigley, C. and Quinonez, P. J. and Raab, F. J. and
                   Raabith, S. S. and Raaijmakers, G. and Raja, S. and
                   Rajan, C. and Rajbhandari, B. and Ramirez, K. E. and
                   Ramis Vidal, F. A. and {Ramos-Buades}, A. and
                   Rana, D. and Ranjan, S. and Ransom, K. and
                   Rapagnani, P. and Ratto, B. and Rawat, S. and Ray, A. and
                   Raymond, V. and Razzano, M. and Read, J. and
                   Recaman Payo, M. and Regimbau, T. and Rei, L. and
                   Reid, S. and Reitze, D. H. and Relton, P. and
                   Renzini, A. I. and Rettegno, P. and Revenu, B. and
                   Reyes, R. and Rezaei, A. S. and Ricci, F. and
                   Ricci, M. and Ricciardone, A. and Richardson, J. W. and
                   Richardson, M. and Rijal, A. and Riles, K. and
                   Riley, H. K. and Rinaldi, S. and Rittmeyer, J. and
                   Robertson, C. and Robinet, F. and Robinson, M. and
                   Rocchi, A. and Rolland, L. and Rollins, J. G. and
                   Romano, A. E. and Romano, R. and Romero, A. and
                   {Romero-Shaw}, I. M. and Romie, J. H. and
                   Ronchini, S. and Roocke, T. J. and Rosa, L. and
                   Rosauer, T. J. and Rose, C. A. and Rosi{\'n}ska, D. and
                   Ross, M. P. and Rossello, M. and Rowan, S. and
                   Roy, S. K. and Roy, S. and Rozza, D. and Ruggi, P. and
                   Ruhama, N. and Ruiz Morales, E. and {Ruiz-Rocha}, K. and
                   Sachdev, S. and Sadecki, T. and Sadiq, J. and
                   Saffarieh, P. and Sah, M. R. and Saha, S. S. and
                   Saha, S. and Sainrat, T. and Sajith Menon, S. and
                   Sakai, K. and Sakellariadou, M. and Sakon, S. and
                   Salafia, O. S. and {Salces-Carcoba}, F. and
                   Salconi, L. and Saleem, M. and Salemi, F. and
                   Sall{\'e}, M. and Salvador, S. and Sanchez, A. and
                   Sanchez, E. J. and Sanchez, J. H. and Sanchez, L. E. and
                   {Sanchis-Gual}, N. and Sanders, J. R. and
                   S{\"a}nger, E. M. and Santoliquido, F. and
                   Saravanan, T. R. and Sarin, N. and Sasaoka, S. and
                   Sasli, A. and Sassi, P. and Sassolas, B. and
                   Satari, H. and Sathyaprakash, B. S. and Sato, R. and
                   Sato, Y. and Sauter, O. and Savage, R. L. and
                   Sawada, T. and Sawant, H. L. and Sayah, S. and
                   Scacco, V. and Schaetzl, D. and Scheel, M. and
                   Schiebelbein, A. and Schiworski, M. G. and
                   Schmidt, P. and Schmidt, S. and Schnabel, R. and
                   Schneewind, M. and Schofield, R. M. S. and
                   Schouteden, K. and Schulte, B. W. and Schutz, B. F. and
                   Schwartz, E. and Scialpi, M. and Scott, J. and
                   Scott, S. M. and Seetharamu, T. C. and
                   {Seglar-Arroyo}, M. and Sekiguchi, Y. and Sellers, D. and
                   Sengupta, A. S. and Sentenac, D. and Seo, E. G. and
                   Seo, J. W. and Sequino, V. and Serra, M. and
                   Servignat, G. and Sevrin, A. and Shaffer, T. and
                   Shah, U. S. and Shaikh, M. A. and Shao, L. and
                   Sharma, A. K. and Sharma, P. and
                   {Sharma-Chaudhary}, S. and Shaw, M. R. and
                   Shawhan, P. and Shcheblanov, N. S. and Sheridan, E. and
                   Shikano, Y. and Shikauchi, M. and Shimode, K. and
                   Shinkai, H. and Shiota, J. and Shoemaker, D. H. and
                   Shoemaker, D. M. and Short, R. W. and Shyamsundar, S. and
                   Sider, A. and Siegel, H. and Sieniawska, M. and
                   Sigg, D. and Silenzi, L. and Simmonds, M. and
                   Singer, L. P. and Singh, A. and Singh, D. and
                   Singh, M. K. and Singh, S. and Singha, A. and
                   Sintes, A. M. and Sipala, V. and Skliris, V. and
                   Slagmolen, B. J. J. and {Slaven-Blair}, T. J. and
                   Smetana, J. and Smith, J. R. and Smith, L. and
                   Smith, R. J. E. and Smith, W. J. and Soldateschi, J. and
                   Somiya, K. and Song, I. and Soni, K. and Soni, S. and
                   Sordini, V. and Sorrentino, F. and Sorrentino, N. and
                   Sotani, H. and Soulard, R. and Southgate, A. and
                   Spagnuolo, V. and Spencer, A. P. and Spera, M. and
                   Spinicelli, P. and Spoon, J. B. and Sprague, C. A. and
                   Srivastava, A. K. and Stachurski, F. and Steer, D. A. and
                   Steinlechner, J. and Steinlechner, S. and
                   Stergioulas, N. and Stevens, P. and Stevenson, S. and
                   Stpierre, M. and Stratta, G. and Strong, M. D. and
                   Strunk, A. and Sturani, R. and Stuver, A. L. and
                   Suchenek, M. and Sudhagar, S. and Sueltmann, N. and
                   Suleiman, L. and Sullivan, K. D. and Sun, L. and
                   Sunil, S. and Suresh, J. and Sutton, P. J. and
                   Suzuki, T. and Suzuki, Y. and Swinkels, B. L. and
                   Syx, A. and Szczepa{\'n}czyk, M. J. and Szewczyk, P. and
                   Tacca, M. and Tagoshi, H. and Tait, S. C. and
                   Takahashi, H. and Takahashi, R. and Takamori, A. and
                   Takase, T. and Takatani, K. and Takeda, H. and
                   Takeshita, K. and Talbot, C. and Tamaki, M. and
                   Tamanini, N. and Tanabe, D. and Tanaka, K. and
                   Tanaka, S. J. and Tanaka, T. and Tang, D. and
                   Tanioka, S. and Tanner, D. B. and Tao, L. and
                   Tapia, R. D. and Tapia San Mart{\'i}n, E. N. and
                   Tarafder, R. and Taranto, C. and Taruya, A. and
                   Tasson, J. D. and Teloi, M. and Tenorio, R. and
                   Themann, H. and Theodoropoulos, A. and
                   Thirugnanasambandam, M. P. and Thomas, L. M. and
                   Thomas, M. and Thomas, P. and Thompson, J. E. and
                   Thondapu, S. R. and Thorne, K. A. and Thrane, E. and
                   Tissino, J. and Tiwari, A. and Tiwari, P. and
                   Tiwari, S. and Tiwari, V. and Todd, M. R. and
                   Toivonen, A. M. and Toland, K. and Tolley, A. E. and
                   Tomaru, T. and Tomita, K. and Tomura, T. and Tong, H. and
                   {Tong-Yu}, C. and Toriyama, A. and Toropov, N. and
                   {Torres-Forn{\'e}}, A. and Torrie, C. I. and
                   Toscani, M. and Tosta E Melo, I. and Tournefier, E. and
                   Trapananti, A. and Travasso, F. and Traylor, G. and
                   Trevor, M. and Tringali, M. C. and Tripathee, A. and
                   Troian, G. and Troiano, L. and Trovato, A. and
                   Trozzo, L. and Trudeau, R. J. and Tsang, T. T. L. and
                   Tso, R. and Tsuchida, S. and Tsukada, L. and
                   Tsutsui, T. and Turbang, K. and Turconi, M. and
                   Turski, C. and Ubach, H. and Uchikata, N. and
                   Uchiyama, T. and Udall, R. P. and Uehara, T. and
                   Uematsu, M. and Ueno, K. and Ueno, S. and Undheim, V. and
                   Ushiba, T. and Vacatello, M. and Vahlbruch, H. and
                   Vaidya, N. and Vajente, G. and Vajpeyi, A. and
                   Valdes, G. and Valencia, J. and Valentini, M. and
                   {Vallejo-Pe{\~n}a}, S. A. and Vallero, S. and
                   Valsan, V. and {van Bakel}, N. and {van Beuzekom}, M. and
                   {van Dael}, M. and {van den Brand}, J. F. J. and
                   {van den Broeck}, C. and {Vander-Hyde}, D. C. and
                   {van der Sluys}, M. and {van de Walle}, A. and
                   {van Dongen}, J. and Vandra, K. and
                   {van Haevermaet}, H. and {van Heijningen}, J. V. and
                   {van Hove}, P. and Vankeuren, M. and Vanosky, J. and
                   {van Putten}, M. H. P. M. and {van Ranst}, Z. and
                   {van Remortel}, N. and Vardaro, M. and Vargas, A. F. and
                   Varghese, J. J. and Varma, V. and Vas{\'u}th, M. and
                   Vecchio, A. and Vedovato, G. and Veitch, J. and
                   Veitch, P. J. and Venikoudis, S. and Venneberg, J. and
                   Verdier, P. and Verkindt, D. and Verma, B. and
                   Verma, P. and Verma, Y. and Vermeulen, S. M. and
                   Vetrano, F. and Veutro, A. and Vibhute, A. M. and
                   Vicer{\'e}, A. and Vidyant, S. and Viets, A. D. and
                   Vijaykumar, A. and Vilkha, A. and {Villa-Ortega}, V. and
                   Vincent, E. T. and Vinet, J. -Y. and Viret, S. and
                   Virtuoso, A. and Vitale, S. and Vives, A. and
                   Vocca, H. and Voigt, D. and {von Reis}, E. R. G. and
                   {von Wrangel}, J. S. A. and Vyatchanin, S. P. and
                   Wade, L. E. and Wade, M. and Wagner, K. J. and
                   Wajid, A. and Walker, M. and Wallace, G. S. and
                   Wallace, L. and Wang, H. and Wang, J. Z. and
                   Wang, W. H. and Wang, Z. and Waratkar, G. and
                   Warner, J. and Was, M. and Washimi, T. and
                   Washington, N. Y. and Watarai, D. and Wayt, K. E. and
                   Weaver, B. R. and Weaver, B. and Weaving, C. R. and
                   Webster, S. A. and Weinert, M. and Weinstein, A. J. and
                   Weiss, R. and Wellmann, F. and Wen, L. and
                   We{\ss}els, P. and Wette, K. and Whelan, J. T. and
                   Whiting, B. F. and Whittle, C. and Wildberger, J. B. and
                   Wilk, O. S. and Wilken, D. and Wilkin, A. T. and
                   Willadsen, D. J. and Willetts, K. and Williams, D. and
                   Williams, M. J. and Williams, N. S. and Willis, J. L. and
                   Willke, B. and Wils, M. and Winterflood, J. and
                   Wipf, C. C. and Woan, G. and Woehler, J. and
                   Wofford, J. K. and Wolfe, N. E. and Wong, H. T. and
                   Wong, H. W. Y. and Wong, I. C. F. and Wright, J. L. and
                   Wright, M. and Wu, C. and Wu, D. S. and Wu, H. and
                   Wuchner, E. and Wysocki, D. M. and Xu, V. A. and
                   Xu, Y. and Yadav, N. and Yamamoto, H. and
                   Yamamoto, K. and Yamamoto, T. S. and Yamamoto, T. and
                   Yamamura, S. and Yamazaki, R. and Yan, S. and Yan, T. and
                   Yang, F. W. and Yang, F. and Yang, K. Z. and Yang, Y. and
                   Yarbrough, Z. and Yasui, H. and Yeh, S. -W. and
                   Yelikar, A. B. and Yin, X. and Yokoyama, J. and
                   Yokozawa, T. and Yoo, J. and Yu, H. and Yuan, S. and
                   Yuzurihara, H. and Zadro{\.z}ny, A. and Zanolin, M. and
                   Zeeshan, M. and Zelenova, T. and Zendri, J. -P. and
                   Zeoli, M. and Zerrad, M. and Zevin, M. and
                   Zhang, A. C. and Zhang, L. and Zhang, R. and
                   Zhang, T. and Zhang, Y. and Zhao, C. and Zhao, Yue and
                   Zhao, Yuhang and Zheng, Y. and Zhong, H. and Zhou, R. and
                   Zhu, X. -J. and Zhu, Z. -H. and Zimmerman, A. B. and
                   Zucker, M. E. and Zweizig, J. and
                   {Ligo Scientific Collaboration} and
                   {VIRGO Collaboration} and {Kagra Collaboration}},
  journal =       {ApJ},
  month =         aug,
  pages =         {L34},
  publisher =     {IOP},
  title =         {Observation of {{Gravitational Waves}} from the
                   {{Coalescence}} of a 2.5--4.5 {{M}} {$\odot$}
                   {{Compact Object}} and a {{Neutron Star}}},
  volume =        {970},
  year =          {2024},
  doi =           {10.3847/2041-8213/ad5beb},
  issn =          {0004-637X},
}

@article{barr2024a,
  author =        {Barr, Ewan D. and Dutta, Arunima and
                   Freire, Paulo C. C. and Cadelano, Mario and
                   Gautam, Tasha and Kramer, Michael and
                   Pallanca, Cristina and Ransom, Scott M. and
                   Ridolfi, Alessandro and Stappers, Benjamin W. and
                   Tauris, Thomas M. and Venkatraman Krishnan, Vivek and
                   Wex, Norbert and Bailes, Matthew and Behrend, Jan and
                   Buchner, Sarah and Burgay, Marta and Chen, Weiwei and
                   Champion, David J. and Chen, C. -H. Rosie and
                   Corongiu, Alessandro and Geyer, Marisa and Men, Y. P. and
                   Padmanabh, Prajwal Voraganti and Possenti, Andrea},
  journal =       {Science},
  month =         jan,
  pages =         {275--279},
  title =         {A Pulsar in a Binary with a Compact Object in the
                   Mass Gap between Neutron Stars and Black Holes},
  volume =        {383},
  year =          {2024},
  doi =           {10.1126/science.adg3005},
  issn =          {0036-8075},
}

@article{thompson2019a,
  author =        {Thompson, Todd A. and Kochanek, Christopher S. and
                   Stanek, Krzysztof Z. and Badenes, Carles and
                   Post, Richard S. and Jayasinghe, Tharindu and
                   Latham, David W. and Bieryla, Allyson and
                   Esquerdo, Gilbert A. and Berlind, Perry and
                   Calkins, Michael L. and Tayar, Jamie and
                   Lindegren, Lennart and Johnson, Jennifer A. and
                   Holoien, Thomas W. -S. and Auchettl, Katie and
                   Covey, Kevin},
  journal =       {Science},
  month =         nov,
  number =        {6465},
  pages =         {637--640},
  title =         {A Noninteracting Low-Mass Black Hole-Giant Star
                   Binary System},
  volume =        {366},
  year =          {2019},
  doi =           {10.1126/science.aau4005},
}

@misc{kochanek2025a,
  author =        {Kochanek, C. S. and Stanek, K. Z. and Thompson, T. A. and
                   Jayasinghe, T.},
  month =         sep,
  publisher =     {arXiv},
  title =         {The {{HST Ultraviolet Spectrum}} of {{V723 Mon}}:
                   {{Additional Evidence}} of a {{Stellar Companion}}},
  year =          {2025},
  doi =           {10.48550/arXiv.2509.10608},
}

\begin{appendix}

\section{\label{appendix:masses}Observational data}
We derive the individual masses of the 40 black holes of our statistical analysis from the parameters listed in the following tables:
\begin{itemize}
    \item Table~\ref{tab:XRB} for X-ray binaries,
    \item Table~\ref{tab:Dormant} for inert black holes,
    \item Table~\ref{tab:gravL} for the gravitational lens.
\end{itemize}
In each table, alongside the input parameters, we report our computed mass distributions using the Monte Carlo error propagation method described in Sect.~\ref{sec:bh-mass-distr}. For each distribution, we quote the mode and the $68\%$ highest density interval (HDI). In many cases where our result differs from the literature value, we can attribute the discrepancy to the choice of summary statistics. For instance, \cite{casares2027a} reports medians where we quote modes. Below, we summarized specific constraints and adjustments we applied to individual sources.

\subsection{Inclination constraints on X-ray binaries}

\subsubsection{Low-mass X-ray binaries (LMXBs)}
To model the mass probability density function of LMXBs, we impose a no-eclipse constraint on the orbital inclination, \ie
\begin{equation}
    \cos i < \frac{R_2}{a}=\frac{0.49 q^{2/3}}{0.6 q ^{2/3}+ \ln(1+q^{1/3})},
\end{equation}
where $q$ is the mass ratio between the black hole and its companion object. This excludes high inclinations that would produce X-ray eclipses in the systems where they haven't been observed. We apply this constraint to all LMXBs except Swift J1357.2-0933, whose system presents a nearly edge-on inclination  (i=$87.4_{-5.6}^{+2.6}$), and whose inner disc is expected to be permanently eclipsed by the outer disc rim \citep{corral-santana2013a,casares2027a}.
\subsubsection{High-mass X-ray binaries (HMXBs)}
HMXBs pose a challenge, as their massive donor stars typically underfill the Roche lobe,  preventing direct inference of the mass ratio $q$ from the projected rotational velocity of the companion star $v\sin i$ and the radial velocity semi-amplitude $K_2$ as done for LMXBs. Instead, we rely solely on the companion mass $M_*$ in Eq.~(\ref{eq:mass-spectroscopy}) to determine $M_\mathrm{BH}$. Implementing a no-eclipse prior function would demand additional assumptions about the stellar radius, but tests on LMC X-3, LMC X-1, and Cyg X-1, as performed in \cite{casares2027a}, have a negligible impact on the derived masses. For simplicity - and following the approach of \citet{farr2011a} - we therefore omit no-eclipse prior functions when calculating the masses of HMXBs.

\subsection{Individual comments on inert black holes}

\subsubsection{KY TrA, NOph 93, HD130298} 
In the case of the X-ray binaries KyTra, NOph93 \citep{casares2027a}, and the inert black hole HD 130298 \citep{mahy2022a}, our Monte Carlo draws produce unphysical tails extending beyond $10^6\,\msun$. These extreme values arise from poorly constrained low inclination, allowing for draws with very small $i$. $\sin i $ appears in the denominator of Eqs.~(\ref{eq:mass-spectroscopy}) and~(\ref{eq:mass-astrometry}), where the other parameters are relatively well constrained. To prevent these outliers from skewing the analysis, we truncate the mass distributions at $100\,\msun$.  
    
\subsubsection{NGC 3201-1, NGC 3201-2}
\cite{giesers2018a} and \cite{giesers2019a} only report the minimum mass of the dark companion of NGC 3201-1, NGC 3201-2, \ie for $\sin i = 1$. We generate the full posterior by assuming an isotropic inclination angle $i\sim \mathcal{I}(0^\circ, 90^\circ)$. Here, we also truncate the resulting distribution at $100\,\msun$. This results in PDFs of $4.9_{-1.0}^{+3.9}\,\msun$ for NGC 3201-1 and $8.1_{-1.3}^{+6.9}\,\msun$ for NGC 3201-2, which, as one expects, are larger and broader than the published minima.

\subsubsection{VFTS 243}
\citet{shenar2022a} computes the mass of VFTS 243, assuming an isotropic inclination angle distribution $i\sim \mathcal{I}(0^\circ, 90^\circ)$ to get the mass $10.1_{-2.0}^{+2.0}$. We aditionally adopt the light-curve constraint $i\sim \mathcal{I}(40^\circ, 90^\circ)$, yielding a  smaller mass $9.0_{-0.9}^{+1.9}\,\msun$.

\subsubsection{Gaia BH 1, Gaia BH 3}
Using Eq.~(\ref{eq:mass-astrometry}) and direct error propagation of period and semi-major axis (assuming uncorrelated errors) inflates uncertainties by a factor of $\sim 3$ compared to the literature in Gaia BH 1 \citep{el-badry2023b}, and BH3\citep{ panuzzo2024a}. To account for the parameter correlation, we instead directly use the provided mass functions $f_A$ for our BH mass calculations.

\onecolumn
\begin{landscape}
    \begin{table*}[p]
    \centering
    \caption{Orbital parameters of the black hole X-ray binaries used in this work. The first 25 black holes have a low-mass companion, while the last 6 have a high-mass companion.}
    \begingroup
    \renewcommand{\arraystretch}{1.2}
    \scriptsize
    \begin{tabular}{llllllllll}
    \hline \hline 
         System &$P/\mathrm{days}$ &$K_2\,/\,\mathrm{km\,s^{-1}}$ &$i/^\circ$ &$q$ &$M_*/\msun$ &$M_\mathrm{BH, ref}/\msun$&$M_\mathrm{BH, calc}/\msun$&[Fe/H]& Refs\\\hline
         A 0620-00 & $\mathcal{N}(0.32301415,7.0 \cdot 10^{-8})$ & $\mathcal{N}(435.4,0.5)$ & $\mathcal{N}(52.6,2.5)$ & $\mathcal{N}(0.067,0.01)$ & $\mathcal{SN}(0.42,0.08,0.09)$ & $\mathrm{6.2}_{-0.6}^{+0.7}$ & $\mathrm{6.2}_{-0.6}^{+0.6}$ & $+0.14\pm0.20^\mathrm{(comp)}$ & 1,2 \\
GS 2023+338G & $\mathcal{N}(6.47117,2.0 \cdot 10^{-6})$ & $\mathcal{N}(208.4,0.5)$ & $\mathcal{I}(52.0,70.0)$ & $\mathcal{N}(0.067,0.005)$ & $\mathcal{SN}(0.92,0.34,1.84)$ & $\mathrm{10.2}_{-1.4}^{+2.3}$ & $\mathrm{8.6}_{-0.3}^{+2.8}$ & $+0.23\pm0.19^\mathrm{(comp)}$ & 1,3 \\
GS 1354-64 & $\mathcal{N}(2.54451,8.0 \cdot 10^{-5})$ & $\mathcal{N}(279.0,4.7)$ & $\mathcal{I}(50.0,90.0)$ & $\mathcal{N}(0.12,0.035)$ & $\mathcal{SN}(1.18,0.44,0.6)$ & $\mathrm{9.8}_{-1.8}^{+3.4}$ & $\mathrm{8.4}_{-1.2}^{+3.2}$ & -- & 1 \\
4U 1543-475 & $\mathcal{N}(1.1163,0.0005)$ & $\mathcal{N}(124.0,4.0)$ & $\mathcal{I}(18.0,28.0)$ & $\mathcal{U}(0.0,0.5)$ & $\mathcal{SN}(1.35,1.04,1.8)$ & $\mathrm{5.4}_{-1.9}^{+3.5}$ & $\mathrm{4.4}_{-1.8}^{+2.8}$ & $-0.11\pm0.10^\mathrm{(comp)}$ & 1,4 \\
V4641 Sgr & $\mathcal{N}(2.8173,1.0 \cdot 10^{-5})$ & $\mathcal{N}(211.3,1.0)$ & $\mathcal{N}(72.3,1.4)$ & $\mathcal{N}(0.45,0.04)$ & $\mathcal{SN}(2.96,0.42,0.46)$ & $\mathrm{6.7}_{-0.4}^{+0.5}$ & $\mathrm{6.8}_{-0.4}^{+0.4}$ & -- & 1 \\
GRS 1915+105 & $\mathcal{N}(33.85,0.16)$ & $\mathcal{N}(126.0,1.0)$ & $\mathcal{N}(64.0,4.0)$ & $\mathcal{N}(0.042,0.024)$ & $\mathcal{SN}(0.45,0.24,0.29)$ & $\mathrm{10.5}_{-1.0}^{+1.4}$ & $\mathrm{10.2}_{-0.9}^{+1.4}$ & -- & 1 \\
XTE J1650-500 & $\mathcal{N}(0.3205,0.0007)$ & $\mathcal{N}(435.0,30.0)$ & $\mathcal{N}(53.2,4.8)$ & $\mathcal{U}(0.0,0.09)$ & $\mathcal{SN}(0.25,0.18,0.23)$ & $\mathrm{5.8}_{-1.4}^{+1.9}$ & $\mathrm{5.3}_{-1.3}^{+2.0}$ & -- & 1 \\
GRO J1655-40 & $\mathcal{N}(2.62168,0.00014)$ & $\mathcal{N}(215.5,2.5)$ & $\mathcal{N}(68.7,1.5)$ & $\mathcal{N}(0.26,0.04)$ & $\mathcal{SN}(1.39,0.29,0.32)$ & $\mathrm{5.3}_{-0.4}^{+0.4}$ & $\mathrm{5.3}_{-0.4}^{+0.4}$ & $-0.10\pm0.10^\mathrm{(comp)}$ & 1,5 \\
GRS 1124-683 & $\mathcal{N}(0.43260249,9.0 \cdot 10^{-8})$ & $\mathcal{N}(406.8,2.7)$ & $\mathcal{SN}(43.2,2.7,2.1)$ & $\mathcal{N}(0.079,0.007)$ & $\mathcal{SN}(0.86,0.13,0.16)$ & $\mathrm{10.9}_{-1.3}^{+1.6}$ & $\mathrm{10.8}_{-1.3}^{+1.7}$ & -- & 1 \\
GS 2000+25 & $\mathcal{N}(0.3440915,9.0 \cdot 10^{-7})$ & $\mathcal{N}(519.9,5.1)$ & $\mathcal{N}(67.5,5.7)$ & $\mathcal{N}(0.042,0.012)$ & $\mathcal{SN}(0.29,0.09,0.1)$ & $\mathrm{6.9}_{-0.7}^{+1.1}$ & $\mathrm{6.6}_{-0.6}^{+1.0}$ & -- & 1 \\
N Oph 77 & $\mathcal{N}(0.5213,0.0013)$ & $\mathcal{N}(441.0,6.0)$ & $\mathcal{N}(61.6,4.4)$ & $\mathcal{U}(0.0,0.053)$ & $\mathcal{SN}(0.19,0.13,0.14)$ & $\mathrm{7.2}_{-0.8}^{+1.1}$ & $\mathrm{6.9}_{-0.8}^{+1.1}$ & -- & 1 \\
XTE J1118+480 & $\mathcal{N}(0.16993404,5.0 \cdot 10^{-8})$ & $\mathcal{N}(708.8,1.4)$ & $\mathcal{N}(71.8,1.8)$ & $\mathcal{N}(0.024,0.009)$ & $\mathcal{SN}(0.18,0.07,0.07)$ & $\mathrm{7.7}_{-0.3}^{+0.3}$ & $\mathrm{7.6}_{-0.2}^{+0.3}$ & $+0.18\pm0.17^\mathrm{(comp)}$ & 1,6 \\
XTE J1550-564 & $\mathcal{N}(1.5420333,2.0 \cdot 10^{-6})$ & $\mathcal{N}(363.1,6.0)$ & $\mathcal{N}(74.7,3.8)$ & $\mathcal{N}(0.033,0.006)$ & $\mathcal{SN}(0.3,0.06,0.07)$ & $\mathrm{9.1}_{-0.6}^{+0.7}$ & $\mathrm{9.0}_{-0.6}^{+0.7}$ & -- & 1 \\
NVel 93 & $\mathcal{N}(0.285206,1.0 \cdot 10^{-6})$ & $\mathcal{N}(475.4,5.9)$ & $\mathcal{N}(59.7,4.3)$ & $\mathcal{N}(0.055,0.01)$ & $\mathcal{SN}(0.3,0.07,0.08)$ & $\mathrm{5.5}_{-0.7}^{+0.9}$ & $\mathrm{5.3}_{-0.6}^{+0.9}$ & $-0.17\pm0.20^{\mathrm{(comp)}}$ & 1,4 \\
XTE J1859+226 & $\mathcal{N}(0.276,0.003)$ & $\mathcal{N}(562.0,40.0)$ & $\mathcal{N}(66.6,4.3)$ & $\mathcal{N}(0.07,0.01)$ & $\mathcal{SN}(0.53,0.13,0.17)$ & $\mathrm{7.6}_{-1.6}^{+2.0}$ & $\mathrm{7.3}_{-1.7}^{+1.8}$ & -- & 1 \\
GRO J0422+32 & $\mathcal{N}(0.21216,2.0 \cdot 10^{-7})$ & $\mathcal{N}(372.0,10.0)$ & $\mathcal{N}(55.6,4.1)$ & $\mathcal{SN}(0.116,0.071,0.079)$ & $\mathcal{SN}(0.31,0.19,0.26)$ & $\mathrm{2.6}_{-0.5}^{+0.6}$ & $\mathrm{2.4}_{-0.5}^{+0.6}$ & -- & 1 \\
Swift J1753.5-0127 & $\mathcal{N}(0.136,0.001)$ & $\mathcal{N}(820.0,36.0)$ & $\mathcal{N}(79.0,5.0)$ & $\mathcal{N}(0.023,0.006)$ & $\mathcal{SN}(0.2,0.06,0.07)$ & $\mathrm{8.7}_{-1.2}^{+1.3}$ & $\mathrm{8.6}_{-1.1}^{+1.4}$ & -- & 1 \\
Swift J1357.2-0933 & $\mathcal{N}(0.106969,2.0 \cdot 10^{-5})$ & $\mathcal{N}(967.0,49.0)$ & $\mathcal{SN}(87.4,5.6,2.6)$ & $\mathcal{SN}(0.04,0.05,0.003)$ & $\mathcal{SN}(0.44,0.08,0.09)$ & $\mathrm{11.0}_{-1.6}^{+1.8}$ & $\mathrm{10.6}_{-1.6}^{+1.7}$ & -- & 1 \\
MAXI J1305-704 & $\mathcal{N}(0.394,0.004)$ & $\mathcal{N}(554.0,8.0)$ & $\mathcal{SN}(72.0,8.0,5.0)$ & $\mathcal{N}(0.05,0.02)$ & $\mathcal{SN}(0.46,0.2,0.22)$ & $\mathrm{9.1}_{-0.9}^{+1.4}$ & $\mathrm{8.8}_{-0.9}^{+1.5}$ & -- & 1 \\
MAXI J1820+070 & $\mathcal{N}(0.68549,1.0 \cdot 10^{-5})$ & $\mathcal{N}(417.7,3.9)$ & $\mathcal{I}(66.2,90.0)$ & $\mathcal{N}(0.072,0.012)$ & $\mathcal{SN}(0.49,0.1,0.1)$ & $\mathrm{6.8}_{-0.5}^{+0.6}$ & $\mathrm{6.6}_{-0.4}^{+0.7}$ & $-0.55\pm0.17^{\mathrm{(env)}}$ & 1,7 \\
4U 1658-48 & $\mathcal{N}(1.7587,0.0005)$ & $\mathcal{N}(219.0,3.0)$ & $\mathcal{I}(37.0,90.0)$ & $\mathcal{N}(0.18,0.05)$ & $\mathcal{SN}(0.81,0.33,0.67)$ & $\mathrm{4.3}_{-1.2}^{+3.4}$ & $\mathrm{3.3}_{-0.7}^{+2.4}$ & $-0.12\pm0.10^{\mathrm{(comp)}}$ & 1,4 \\
MAXI J1659-152 & $\mathcal{N}(0.10058,2.0 \cdot 10^{-5})$ & $\mathcal{N}(750.0,80.0)$ & $\mathcal{I}(70.0,80.0)$ & $\mathcal{U}(0.02,0.07)$ & $\mathcal{SN}(0.23,0.1,0.14)$ & $\mathrm{5.3}_{-1.5}^{+1.9}$ & $\mathrm{5.0}_{-1.6}^{+1.8}$ & -- & 1 \\
KY TrA & $\mathcal{N}(0.26,0.01)$ & $\mathcal{N}(501.0,52.0)$ & $\mathcal{N}(57.0,13.0)$ & $\mathcal{U}(0.01,0.31)$ & $\mathcal{SN}(0.45,0.28,0.32)$ & $\mathrm{5.8}_{-1.9}^{+3.1}$ & $\mathrm{6.1}_{-2.7}^{+5.4}$ & -- & 1 \\
NOph 93 & $\mathcal{N}(0.278,0.008)$ & $\mathcal{N}(521.0,52.0)$ & $\mathcal{N}(61.0,15.0)$ & $\mathcal{U}(0.02,0.25)$ & $\mathcal{SN}(0.5,0.27,0.32)$ & $\mathrm{6.4}_{-1.9}^{+3.2}$ & $\mathrm{6.8}_{-2.9}^{+5.4}$ & -- & 1 \\
Swift J1727.8-1613 & $\mathcal{N}(0.45017,4.0 \cdot 10^{-5})$ & $\mathcal{N}(390.0,4.0)$ & $\mathcal{I}(42.0,74.0)$ & $\mathcal{N}(0.12,0.03)$ & $\mathcal{SN}(0.59,0.18,0.2)$ & $\mathrm{4.9}_{-0.8}^{+1.7}$ & $\mathrm{4.2}_{-0.5}^{+2.6}$ & -- & 1 \\\hline
         M33 X-7 & $\mathcal{N}(3.453014,2.0 \cdot 10^{-5})$ & $\mathcal{N}(108.9,5.7)$ & $\mathcal{N}(65.0,0.0)$ & -- & $\mathcal{SN}(38.0,10.0,22.0)$ & $11.4_{-1.7}^{+3.3}$ & $\mathrm{12.2}_{-2.4}^{+3.3}$ & $\simeq-0.30^{\mathrm{(atm)}}$ & 1,8,9,10 \\
LMC X-3 & $\mathcal{N}(1.7048089,1.0 \cdot 10^{-6})$ & $\mathcal{N}(241.1,6.2)$ & $\mathcal{N}(69.2,0.7)$ & $\mathcal{N}(0.52,0.05)$ & $\mathcal{SN}(3.6,0.6,0.6)$ & $\mathrm{7.0}_{-0.6}^{+0.6}$ & $\mathrm{6.9}_{-0.7}^{+0.7}$ & $\simeq-0.28^{\mathrm{(env)}}$ & 1,11,12,13 \\
LMC X-1 & $\mathcal{N}(3.90917,5.0 \cdot 10^{-5})$ & $\mathcal{N}(71.6,1.1)$ & $\mathcal{N}(36.4,2.0)$ & -- & $\mathcal{N}(31.8,3.7)$ & $\mathrm{10.9}_{-1.5}^{+1.5}$ & $\mathrm{10.9}_{-1.0}^{+1.0}$ & $\simeq-0.28^{\mathrm{(env)}}$ & 1,13,14 \\
Cyg X-1$^{(a)}$ & $\mathcal{N}(5.599829,1.0 \cdot 10^{-5})$ & $\mathcal{SN}(75.2,0.41,0.42)$ & $\mathcal{I}(25.0,39.0)$ & -- & $\mathcal{SN}(29.0,3.0,6.0)$ & $\mathrm{12.7}\text{ to } {17.8}$ & $\mathrm{13.8}_{-1.8}^{+3.4}$ & $+0.12^{+0.06}_{-0.05}$ & 1,15,16 \\
NGC 300 X1 & $\mathcal{N}(1.366337,1.0 \cdot 10^{-5})$ & $\mathcal{N}(250.0,40.0)$ & $\mathcal{I}(60.0,75.0)$ & -- & $\mathcal{SN}(26.0,5.0,7.0)$ & $\mathrm{17.0}_{-4.0}^{+4.0}$ & $\mathrm{17.0}_{-3.9}^{+5.1}$ & $\simeq-0.55^{\mathrm{(env)}}$ & 17,18 \\
IC 10 X1 & $\mathcal{N}(1.4554,0.0017)$ & $\mathcal{N}(370.0,20.0)$ & $\mathcal{I}(75.0,90.0)$ & $\mathcal{U}(0.7,1.7)$ & -- & $-$ & $\mathrm{29.2}_{-6.1}^{+18.2}$ & $\simeq-0.66^{\mathrm{(env)}}$ & 20,21 

         \\\hline
    \end{tabular}
    \endgroup
    \tablefoot{$P$ is the orbital period, $K_2$ the radial velocity semi-amplitude, $i$ the inclination, $q$ the mass ratio, $M_*$ the companion mass, and [Fe/H] the metallicity of the system. $M_\mathrm{BH, ref}$ is the black hole from the literature, whereas $M_\mathrm{BH, calc}$ is the mass we calculated. $M_\mathrm{BH, calc}$ indicates the mode of the calculated distribution with the $68\%$ credibility interval.\\
    \tablefoottext{(a)}{We assume an eccentricity $e=0$ for all systems, except for Cyg X-1, where we use $e \sim \mathcal{SN}(0.0189,0.0026,0.0028)$.}\\
    \textbf{Notation.} $\mathcal{N}(\mu, \sigma)$ stands for a normal distribution with mean $\mu$ and standard-deviation $\sigma$. $\mathcal{SN}(\mu, \sigma_1, \sigma_2)$ indicates a split normal distribution with mean $\mu$ and left hand side and right hand side standard-deviations $\sigma_1$ and $\sigma_2$. $\mathcal{U}(a, b)$ and $\mathcal{I}(\alpha, \beta)$ indicate uniform and isotropic distributions between $a$ and $b$, and $\alpha$ and $\beta$ respectively. The metallicity indicates the one of the companion star $^\mathrm{(comp)}$, of the atmosphere of the companion $^\mathrm{(atm)}$, or of the local environment $^\mathrm{(env)}$. For metallicities derived from $Z/Z_\odot$, we convert it using $\mathrm{[Fe/H]} = \log_{10}(Z/Z_\odot)$.
    }
    \tablebib{(1)~\citet{casares2027a}; (2)~\citet{gonzalezhernandez2004a}; (3)~\citet{gonzalezhernandez2011a}; (4)~\citet{mlinar2024a}; (5)~\citet{gonzalezhernandez2008a};(6)~\citet{gonzalezhernandez2008b}; (7)~\citet{mikolajewska2022a}; (8)~\citet{pietsch2006a}; (9)~\citet{orosz2007a}; (10)~\citet{ramachandran2022a}; (11)~\citet{song2010a}; (12)~\citet{orosz2014a}; (13)~\citet{val-baker2007a}; (14)~\citet{orosz2009a}; (15)~\citet{miller-jones2021a}; (16)~\citet{ramachandran2025a}; (17)~\citet{binder2021a}; (18)~\citet{crowther2010b}; (19)~\citet{farr2011a}; (20)~\citet{silverman2008a}}
    \label{tab:XRB}
\end{table*}

\begin{table*}[]
    \centering
    \caption{Orbital parameters of the inert black holes used in this work.}
    \begingroup\renewcommand{\arraystretch}{1.2}
    \scriptsize
    \begin{tabular}{llllllllllll}
    \hline \hline 
         System &$P/\mathrm{days}$ &$K_2\,/\,\mathrm{km\,s^{-1}}$ &$i/^\circ$ &$e$  &$f_A$ & $M_*/\msun$ &$M_\mathrm{BH, ref}/\msun$&$M_\mathrm{BH, calc}/\msun$&$\mathrm{[Fe/H]}$& Refs\\\hline
         VFTS 243 & $\mathcal{N}(10.4031,0.0004)$ & $\mathcal{N}(81.4,1.3)$ & $\mathcal{I}(40.0,90.0)$ & $\mathcal{N}(0.017,0.012)$ & -- & $\mathcal{N}(25.0,2.3)$ & $\mathrm{10.1}_{-2.0}^{+2.0}$ & $\mathrm{9.0}_{-0.9}^{+1.9}$ &$\simeq-0.28^{\mathrm{(env)}}$& 21 \\
Gaia BH1 & $\mathcal{N}(184.8,0.7)$ & $\mathcal{N}(66.7,0.6)$ & $\mathcal{N}(53.4,0.4)$ & $\mathcal{N}(0.447,0.005)$ & $\mathcal{N}(8.0,0.16)$ & $\mathcal{N}(0.93,0.05)$ & $\mathrm{9.6}_{-0.2}^{+0.2}$ & $\mathrm{9.6}_{-0.2}^{+0.2}$ &$-0.20 \pm0.05^\mathrm{(comp)}$& 22 \\
Gaia BH2 & $\mathcal{N}(1276.7,0.6)$ & $\mathcal{N}(25.23,0.04)$ & $\mathcal{N}(34.87,0.34)$ & $\mathcal{N}(0.5176,0.0009)$ & -- & $\mathcal{N}(1.07,0.19)$ & $\mathrm{8.9}_{-0.3}^{+0.3}$ & $\mathrm{9.0}_{-0.4}^{+0.3}$ &$-0.22 \pm0.02^\mathrm{(comp)}$& 23 \\
Gaia BH3 & $\mathcal{N}(4253.1,98.5)$ & -- & $\mathcal{N}(69.42,0.095)$ & $\mathcal{N}(0.7291,0.0048)$ & $\mathcal{N}(31.23,0.81)$ & $\mathcal{N}(0.76,0.05)$ & $\mathrm{32.7}_{-0.8}^{+0.8}$ & $\mathrm{32.7}_{-0.8}^{+0.8}$ &$-2.56 \pm0.12^\mathrm{(comp)}$& 24 \\
HD 130298 & $\mathcal{N}(14.62959,0.00085)$ & $\mathcal{N}(71.78,0.68)$ & $\mathcal{N}(54.0,16.0)$ & $\mathcal{N}(0.457,0.007)$ & -- & $\mathcal{N}(24.2,3.8)$ & $\mathrm{8.8}_{-1.5}^{+3.5}$ & $\mathrm{8.5}_{-1.6}^{+2.9}$ &--& 25 \\
G3425 & $\mathcal{SN}(879.11,2.57,3.22)$ & $\mathcal{SN}(22.91,0.14,0.13)$ & $\mathcal{SN}(89.3,10.08,15.48)$ &$\mathcal{N}(0.05,0.01)$ & -- & $\mathcal{SN}(2.66,0.68,1.18)$ & $\mathrm{3.6}_{-0.5}^{+0.8}$ & $\mathrm{3.6}_{-0.5}^{+0.6}$ &$-0.12 \pm0.02^\mathrm{(comp)}$& 26 \\
NGC 3201-1 & $\mathcal{N}(167.01,0.09)$ & $\mathcal{N}(72.3,1.7)$ & $\mathcal{I}(0.0,90.0)$ & $\mathcal{N}(0.61,0.02)$ & -- & $\mathcal{N}(0.81,0.05)$ & $\mathrm{4.5}_{-0.2}^{+0.2}$ & $\mathrm{4.9}_{-1.0}^{+3.9}$ &$-1.53 \pm0.01^\mathrm{(env)}$& 27,28,29 \\
NGC 3201-2 & $\mathcal{N}(2.2422,0.0001)$ & $\mathcal{N}(305.5,7.7)$ & $\mathcal{I}(0.0,90.0)$ & $\mathcal{N}(0.07,0.04)$ & -- & $\mathcal{N}(0.61,0.05)$ & $\mathrm{7.7}_{-0.5}^{+0.5}$ & $\mathrm{8.1}_{-1.3}^{+6.9}$ &$-1.53 \pm0.01^\mathrm{(env)}$& 28,29
         \\\hline
    \end{tabular}
    \endgroup
    \tablefoot{$P$ the orbital period, $K_2$ the radial velocity semi-amplitude, $i$ the inclination, $q$ the mass ratio, $M_*$ the companion mass, and [Fe/H] the metallicity of the system. $M_\mathrm{BH, ref}$ is the black hole from the literature, whereas $M_\mathrm{BH, calc}$ is the mass we calculated. $M_\mathrm{BH, calc}$ indicates the mode of the calculated distribution with the $68\%$ credibility interval.\\
    \textbf{Notation.} $\mathcal{N}(\mu, \sigma)$ stands for a normal distribution with mean $\mu$ and standard-deviation $\sigma$. $\mathcal{SN}(\mu, \sigma_1, \sigma_2)$ indicates a split normal distribution with mean $\mu$ and left hand side and right hand side standard-deviations $\sigma_1$ and $\sigma_2$. $\mathcal{U}(a, b)$ and $\mathcal{I}(\alpha, \beta)$ indicate uniform and isotropic distributions between $a$ and $b$, and $\alpha$ and $\beta$ respectively. The metallicity indicates the one of the companion star $^\mathrm{(comp)}$, or of the local environment $^\mathrm{(env)}$. For metallicities derived from $Z/Z_\odot$, we convert it using $\mathrm{[Fe/H]} = \log_{10}(Z/Z_\odot)$.}
    \tablebib{
    (21)~\citet{shenar2022a}; (22)~\citet{el-badry2023b}; (23)~\citet{el-badry2023a}, (24)~\citet{panuzzo2024a}; (25)~\citet{mahy2022a}; (26)~\citet{wang2024b}; (27)~\citet{giesers2018a};(28)~\citet{giesers2019a}; (29)~\citet{munoz2013a}.}
    \label{tab:Dormant}
\end{table*}

\begin{table*}[]
    \centering
    \caption{Orbital parameters of the gravitational lensing binary black hole used in this work.}
    \begingroup\renewcommand{\arraystretch}{1.2}
    \footnotesize
    \begin{tabular}{llllll}   
    \hline \hline 
         System &$\theta_E$ &$\pi_E$ & $M_\mathrm{BH, ref}/\msun$&$M_\mathrm{BH, calc}/\msun$& Refs\\\hline
         MOA-11-191/OGLE-11-462 & $\mathcal{N}(5.496,0.024)$ & $\mathcal{N}(0.094,0.012)$ & $\mathrm{7.2}_{-1.8}^{+1.8}$ & $\mathrm{6.9}_{-0.8}^{+1.0}$ & 30,31
         \\\hline
    \end{tabular}  
    \endgroup
    \tablefoot{ $\theta_E$ the Einstein radius, $\pi_E$ . $M_\mathrm{BH, ref}$ is the black hole from the literature, whereas $M_\mathrm{BH, ref}$ is the mass we calculated.$M_\mathrm{BH, calc}$ indicates the mode of the calculated distribution with the $68\%$ credibility interval.\\    
    \textbf{Notation.} $\mathcal{N}(\mu, \sigma)$ stands for a normal distribution with mean $\mu$ and standard-deviation $\sigma$. $\mathcal{SN}(\mu, \sigma_1, \sigma_2)$ indicates a split normal distribution with mean $\mu$ and left hand side and right hand side standard-deviations $\sigma_1$ and $\sigma_2$. $\mathcal{U}(a, b)$ and $\mathcal{I}(\alpha, \beta)$ indicate uniform and isotropic distributions between $a$ and $b$, and $\alpha$ and $\beta$ respectively.}
    \tablebib{(30)~\citet{sahu2022a}; (31)~\citet{sahu2025a}.}
    \label{tab:gravL}
\end{table*}
\end{landscape}

\twocolumn
\section{Posterior distributions for the inferred BH mass distributions}\label{appendix:posteriors}
\begin{figure}
    \centering
    \includegraphics[width=\columnwidth]{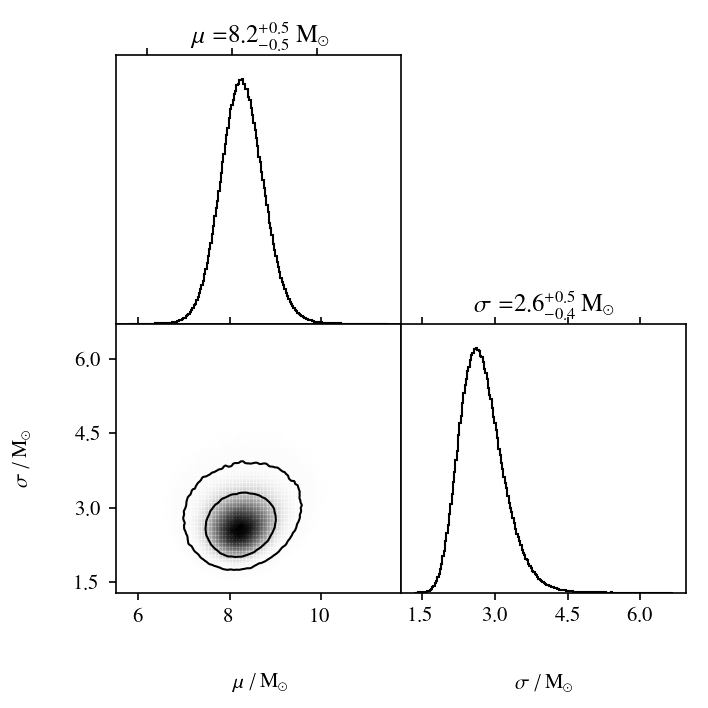}
    \caption{Marginalized posterior probability distribution for the parameters from the Gaussian model. The contours define the $68\%$ and the $95\%$ confidence intervals of the inferred distributions.}
    \label{fig:Gauss-corner}
\end{figure}

\begin{figure}
    \centering
    \includegraphics[width=\columnwidth]{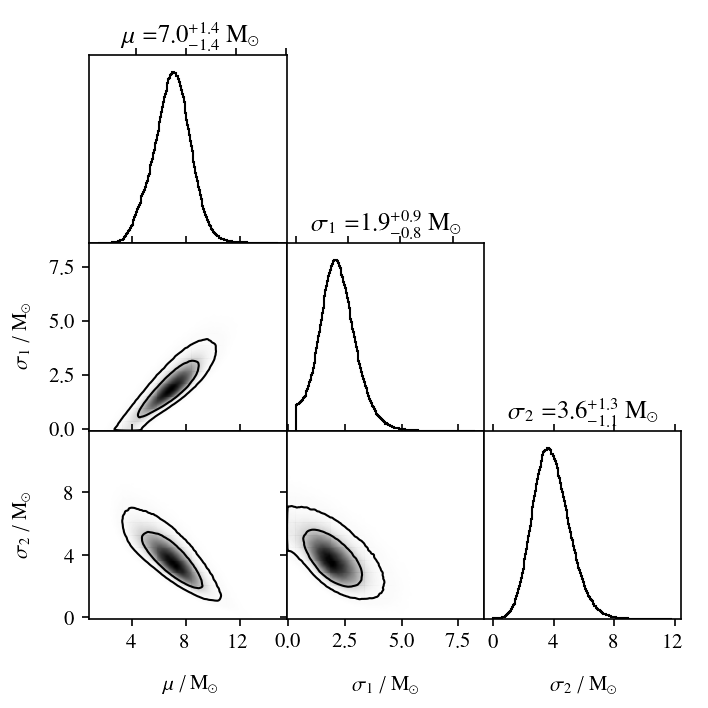}
    \caption{Marginalized posterior probability distributions for the parameters of the
split-normal mass distribution (Eq.~\ref{eq:split-normal}). The contours represent the 68\% and 95\% confidence intervals. }
    \label{fig:SN-corner}
\end{figure}

\begin{figure}
    \centering
    \includegraphics[width=\columnwidth]{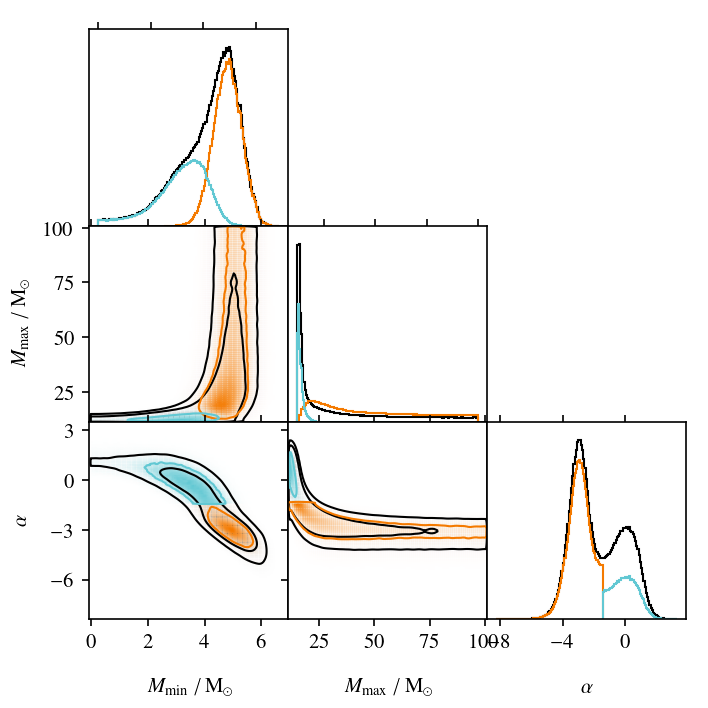}
    \caption{Marginalized posterior probability distributions for the parameters of the power-law population mass distribution (Eq.~\ref{eq:powlaw}). The contours mark the 68\% and 95\% confidence intervals. In blue, we mark the 68\%  contours of Solution 1 and in orange those of Solution 2.}
    \label{fig:PL-corner}
\end{figure}

\begin{figure*}[t]
    \centering
    \includegraphics[width=\linewidth]{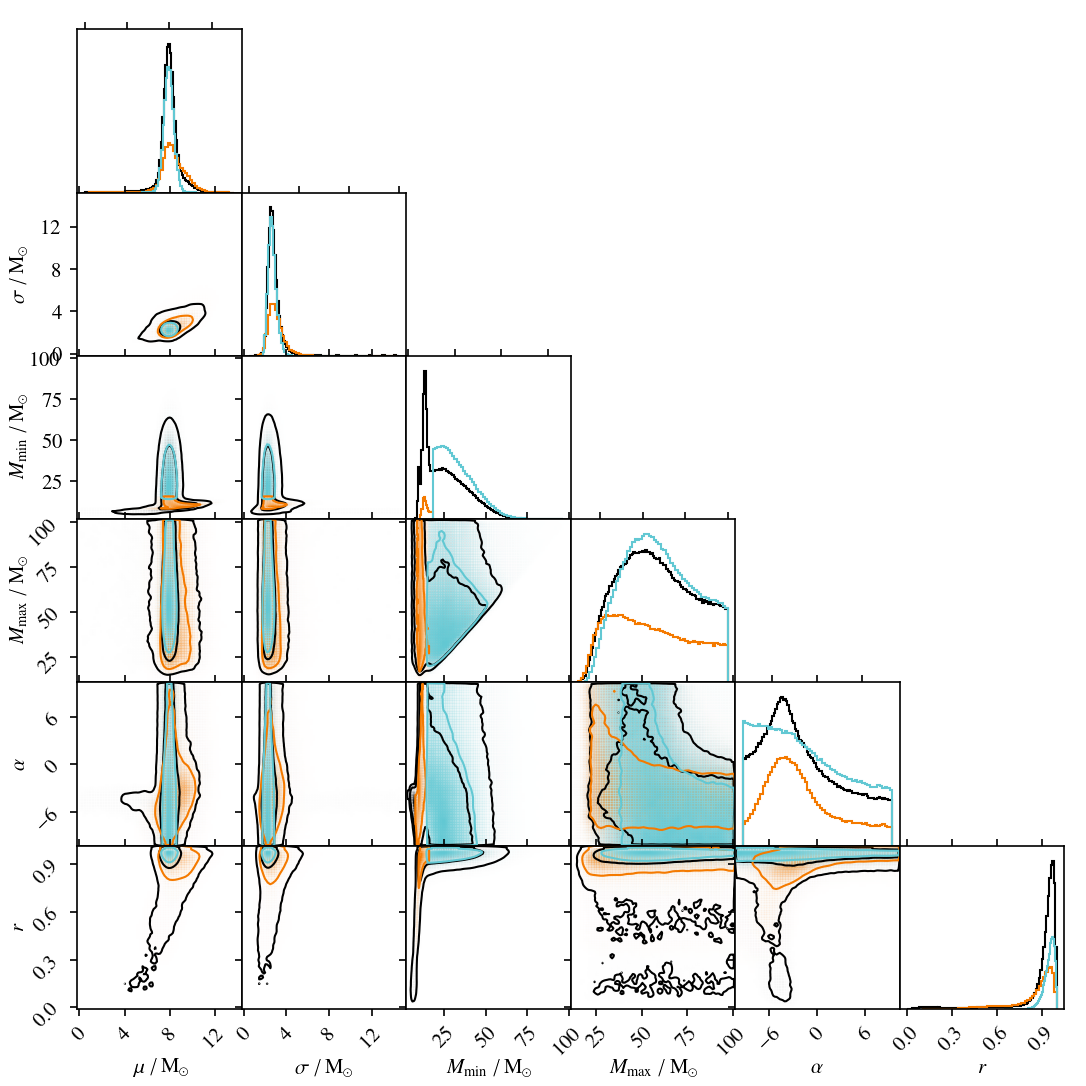}
    \caption{Marginalized posterior probability distributions for the parameters of the
bimodal mass distribution (Eq.~\ref{eq:bimodal}). The contours represent the 68\% and the 95\% confidence intervals of the inferred parameter distributions. In blue, we mark the 68\%  contours of Solution 1 and in orange those of Solution 2.}
    \label{fig:bi-corner}
\end{figure*}

Figures~\ref{fig:Gauss-corner}, \ref{fig:SN-corner}, \ref{fig:PL-corner}, and \ref{fig:bi-corner} present the marginalized posterior distributions of the inferred parameters and their correlations of the different parametric models fitted to the observed BHs. 

\section{Inferred mass distribution including Gaia BH3 \label{appendix:GaiaBH3}}

\begin{figure}
    \centering
    \includegraphics[width=\linewidth]{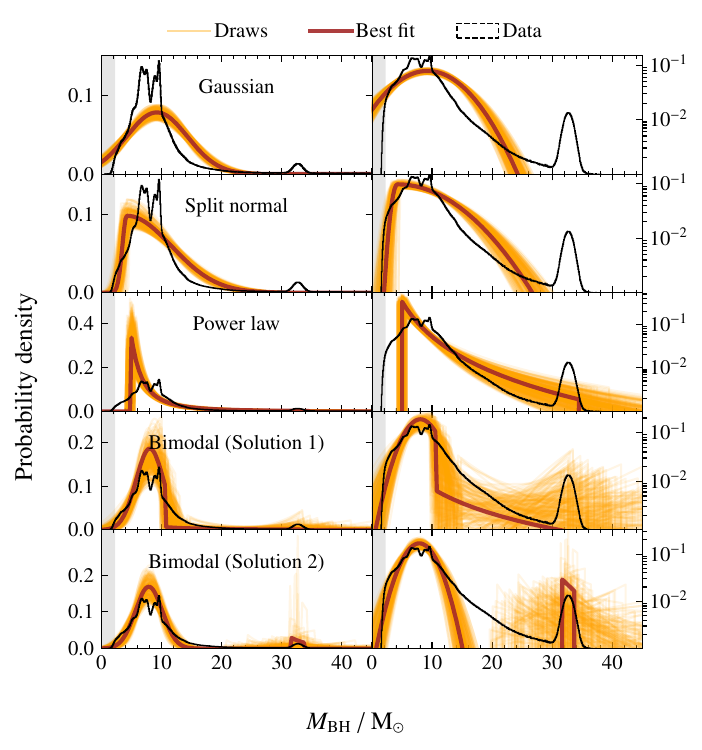}
    \caption{BH mass distribution $p(M\vert\vec{\theta})$ implied by the posterior $p(\vec{\theta}\vert M)$ and the observational data from Sect.~\ref{sec:obs-data} for the models discussed in Sect.~\ref{sec:methods}. The brown line gives the distributions implied by the best-fit parameters, and the orange lines use parameters sampled from the 68\% interval of confidence listed in Table \ref{tab:Res_all}. The black histogram indicates the summed probability density function shown in Fig.~\ref{fig:all-masses}.}
    \label{figA:res}
\end{figure}

We report the model parameters obtained when including Gaia BH 3 in the sample in Table~\ref{tab:Res_all_incl_GBH3} and plot the resulting distributions in Fig.~\ref{figA:res}.

\begin{table}[h]
\caption{Best-fit parameters of the BH mass distribution models.}
\label{tab:Res_all_incl_GBH3}
\begingroup

\renewcommand{\arraystretch}{1.5} 
\small
	\begin{tabular}{l l l l }
		\hline \hline 
		Model & Parameter & Best fit & \\ \hline 
		Gaussian  &$\mu/\msun$ & $9.2_{-0.9}^{+0.9}$&\\
		& $\sigma/\msun$&$5.1_{-0.6}^{+0.8}$&\\\hline

       Split Normal (Eq. \ref{eq:split-normal}) &$\mu/\msun$ &$4.2_{-1.0}^{+1.5}$\\
		& $\sigma_1/\msun$ & $0.7_{-0.7}^{+0.7}$&\\
		& $\sigma_2/\msun$ &$7.4_{-1.1}^{+1.2}$&\\\hline

		Power law (Eq. \ref{eq:powlaw}) &$M_{\min}/\msun$ &$4.9_{-0.8}^{+0.8}$&\\
		& $M_{\max}/\msun$ &$34.7_{-2.4}^{+56.3}$&\\
		& $\alpha$ &$-2.7_{-1.0}^{+0.7}$&\\\hline
		
		Bimodal$^{a}$ (Eq. \ref{eq:bimodal}) &$\mu/\msun$ & $8.1_{-0.9}^{+2.2}$& $7.9_{-0.7}^{+0.8}$ \\
		& $\sigma/\msun$& $2.3_{-0.7}^{+1.7}$& $2.2_{-0.5}^{+0.7}$ \\
		&$M_{\min}/\msun$ & $10.7_{-1.0}^{+4.5}$& $31.5_{-12.4}^{+1.9}$\\
		& $M_{\max}/\msun$& $34.2_{-3.1}^{+47.5}$& $33.9_{-2.4}^{+54.2}$ \\
		& $\alpha$  & $-1.7_{-2.6}^{+10.6}$ & $-9.2_{-0.8}^{+15.5}$\\
		& $r$  & $0.9_{-0.1}^{+0.0}$& $0.9_{-0.1}^{+0.0}$\\\hline
	\end{tabular}
    \endgroup
    \tablefoot{We indicate the mode and the 68\% interval of confidence of the marginalized posterior distributions.\\
    \tablefoottext{a}{Since we find two solutions for $M_\mathrm{min}$, we indicate the parameter values from the marginalized distributions around the two different solutions.}}
    	\end{table}

For the Gaussian model, including Gaia BH3 widens the core substantially, attempting to encompass the full mass range of data (from ${\sim}\, 2\,\msun$ to ${\sim}\, 30\msun$), which in turn smears out the main peak. As a result, the model underestimates the $5\text{--}10\,\msun$ population, where nearly half of the measured BHs reside. Furthermore, the Gaussian overpredicts the intermediate-to-high-mass regime ($15\text{--}20\,\msun$), while assigning non-negligible probability of finding a BH in NS and even unphysical, negative mass ranges within $2\sigma$. This highlights the difficulties of representing the observed mass distribution with a single, symmetric component.

The split-normal seeks to account for asymmetries by shifting the peak toward lower masses ($\mu = 4.2_{-1.0}^{+1.5}\,\msun$) and extending to $\sigma_2=7.4_{-1.1}^{+1.2}\,\msun$. This adjustment leads to an underestimation of the main $5\text{--}10\,\msun$ peak and imposes a sharp cutoff near $4.5\,\msun$, despite the observed PDFs below this value. Similarly to the Gaussian, the split-normal also overpredicts the intermediate masses and fails to account for the highest mass objects ($\gtrsim 30\,\msun$).

The power law, when fitted including Gaia BH3, only finds one solution with a negative slope. This solution shifts its maximum mass from $\sim 20\,\msun$ to $\sim 34\,\msun$ to encompass Gaia BH3. Its slope remains consistent with the inference without Gaia BH3 as well as the sharp low-mass truncation at $M_\mathrm{min}\sim5\,\msun$. 

Finally, the bimodal model also finds two solutions for the onset of the power law. Both solutions infer a higher maximum mass compared to the inference without Gaia BH3. The higher minimum-mass component also shifts upward (to $\sim30\,\msun$ instead of previously $\sim 20\,\msun$), only including Gaia BH3 in the power law component. In both cases, we still find positive and negative slope solutions for the power law component.

In summary, across all models, the inclusion of Gaia BH3 drives significant changes to the parameter values, particularly adapting to the high-mass component (\eg by broadening the core of the distribution) at the expense of the central peak. This behavior suggests that Gaia BH3, potentially with other very massive systems such as IC 10 X1, may not be described with the same distribution as the other BHs.

\section{Inference of the observed BH mass distribution for various subsamples}\label{app:subsamples}
\begin{figure}
    \centering
    \includegraphics[width=\linewidth]{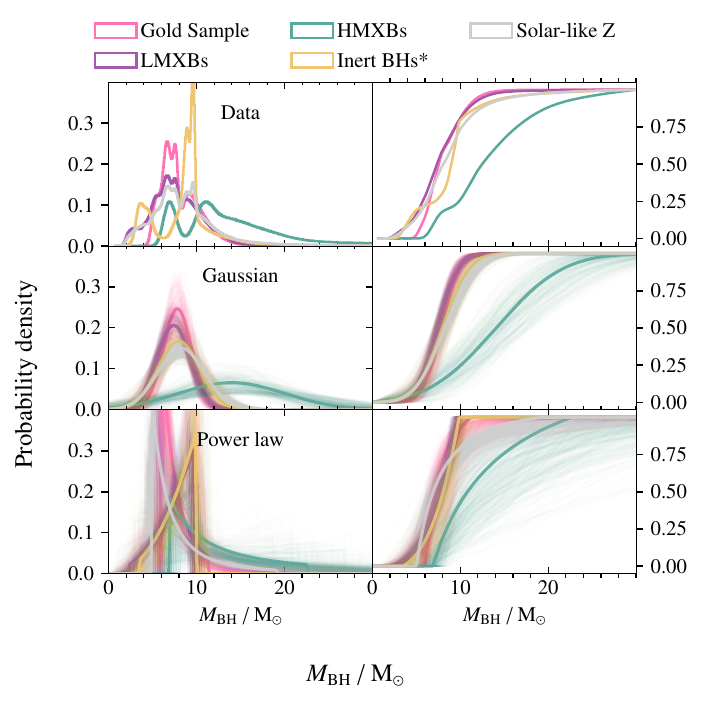}
    \caption{Inference of the observed BH mass distribution for selected subsamples. The upper panel shows the observed PDFs (left) and CDFs (right) for each subsample, while the lower panels indicate the fitted posterior mass distributions for each parametric model.}
    \label{fig:subsamples}
\end{figure}

In our main analysis, we treat BH masses inferred from different observational techniques in the same manner to determine the best-fitting observed mass distribution. However, these groups are likely affected by different observational biases and may sample a different part of the actual BH mass distribution. We now analyze the different groups of observed BH systems to assess whether there are any qualitative differences between their BH mass distributions. In particular, we consider 
\begin{itemize}
    \item the ``golden'' sample defined by \citet{casares2027a}, which comprises ten BH transient X-ray binaries with the most robust mass determinations: A 0620-00, V4641 Sgr, GRS 1915+105, GRO J1655-40, GRS 1124-683, GS 2000+25, XTE J1118+480, XTE J1550-564, MAXI J1305-704, and MAXI J1820+070;
    \item the LMXB sample;
    \item the HMXB sample;
    \item all X-ray binaries, combining low- and high-mass XRBs;
    \item the inert BH sample (excluding Gaia BH3);
    \item the BHs with solar-like metallicity, \ie all BHs except for Gaia BH3 with $\mathrm{[Fe/H] = 2.56\pm 0.12}$ and the two BHs in NGC 3201 with $\mathrm{[Fe/H]}=-1.53 \pm 0.01 $ (see Tables~\ref{tab:XRB}--\ref{tab:gravL}).
\end{itemize}
Furthermore, to assess whether the high-mass component hinges on a few objects with large uncertainties, we consider two additional variations of the main text samples: one removing IC 10 X1, with the largest mass uncertainty, another removing all BHs with uncertainty $\Delta M\geq 5\,\msun$. The latter excludes the LMXBs KY TrA and NOph 93, the HMXBs IC 10 X1 and NGC 300 X1, as well as the inert BH NGC 3201-1 from the inference. We report the best-fit parameters for our parametric models in Table~\ref{tab:subsamples}, and visualize selected fits in Fig.~\ref{fig:subsamples}.

\begin{table*}
    \caption{Inferred parameter values for the different BH subsamples}.\label{tab:subsamples}
    \begingroup
    \renewcommand{\arraystretch}{1.5}
    \scriptsize
    \begin{tabular}{c c |c |c |c | c|c c| c| c | c}
    \hline\hline
        Model & Parameters & Gold Sample & LMXBs & HMXBs & XRBs& \multicolumn{2}{c|}{Inert*}& Solar-like Z & No IC10X1 & $\Delta M < 5\msun$ \\\hline
        Gaussian & $\mu/\msun$ & $7.8_{-0.6}^{+0.7}$ & $7.4_{-0.5}^{+0.5}$ & $14.0_{-3.3}^{+3.9}$ &$8.4_{-0.6}^{+0.6}$& \multicolumn{2}{c|}{$8.0_{-1.2}^{+1.3}$}  & $8.2_{-0.5}^{+0.6}$& $8.0_{-0.5}^{+0.4}$ & $7.9_{-0.4}^{+0.5}$\\
        & $\sigma/\msun$ & $1.6_{-0.4}^{+0.7}$ & $1.9_{-0.4}^{+0.4}$ & $6.0_{-2.5}^{+4.5}$ &$2.9_{-0.5}^{+0.6}$& \multicolumn{2}{c|}{$2.4_{-0.8}^{+1.2}$}& $2.6_{-0.4}^{+0.5}$& $2.2_{-0.3}^{+0.4}$ & $2.2_{-0.3}^{+0.4}$\\\hline
        Split-normal & $\mu/\msun$ & $6.6_{-1.1}^{+2.4}$& $8.9_{-1.9}^{+0.9}$&$8.1_{-2.7}^{+9.1}$ &$6.5_{-1.5}^{+1.5}$& \multicolumn{2}{c|}{$9.6_{-2.8}^{+1.9}$}& $7.1_{-1.6}^{+1.4}$& $8.5_{-1.5}^{+1.3}$ & $8.7_{-1.4}^{+1.3}$\\
        & $\sigma_1/\msun$ & $1.0_{-1.0}^{+1.3}$& $2.4_{-0.9}^{+0.9}$& $3.6_{-3.6}^{+4.7}$&$1.8_{-1.0}^{+0.9}$& \multicolumn{2}{c|}{$2.9_{-1.5}^{+2.4}$}& $1.9_{-0.8}^{+0.9}$& $2.4_{-0.7}^{+0.9}$ & $2.6_{-0.8}^{+0.9}$\\
        & $\sigma_2/\msun$ &$2.2_{-1.6}^{+1.2}$ & $1.0_{-1.0}^{+1.0}$& $8.7_{-5.8}^{+7.4}$ &$4.2_{-1.1}^{+1.5}$&\multicolumn{2}{c|}{$0.0_{-0.0}^{+3.4}$} & $3.7_{-1.1}^{+1.3}$&$2.0_{-1.1}^{+1.0}$&$1.7_{-1.0}^{+1.0}$\\\hline
        Power-law & $M_{\min}/\msun$ & $6.0_{-0.9}^{+0.6}$ & $2.3_{-2.3}^{+2.5}$ & $6.9_{-4.0}^{+2.0}$&$5.4_{-0.8}^{+0.7}$& \multicolumn{2}{c|}{$3.5_{-3.5}^{+0.8}$} & $5.0_{-1.0}^{+0.8}$& $2.6_{-2.6}^{+1.2}$& $2.6_{-2.6}^{+1.0}$\\
		& $M_{\max}/\msun$ & $11.3_{-2.0}^{+77}$ & $9.7_{-1.0}^{+1.9}$ &$23_{-7}^{+64}$ &$28_{-7}^{+69}$& \multicolumn{2}{c|}{$9.9_{-0.7}^{+3.5}$}& $19_{-6}^{+69}$& $10.9_{-1.0}^{+1.4}$ & $10.8_{-1.0}^{+1.3}$\\
		& $\alpha$ & $-4.2_{-3.4}^{+3.6}$ & $1.9_{-3.5}^{+2.8}$ & $-1.8_{-2.4}^{+3.5}$ &$-3.4_{-1.4}^{+1.2}$&\multicolumn{2}{c|}{$2.3_{-4.6}^{+4.8}$ }& $-2.9_{-1.0}^{+1.5}$& $1.5_{-1.7}^{+1.4}$&$1.5_{-1.5}^{+1.5}$\\
        & 2nd solution & yes (17\%) & yes (6\%) & no &no& \multicolumn{2}{c|}{yes (2\%)} & yes(41\%) & no & no\\\hline

        Bimodal &$\mu/\msun$ &$7.8_{-1.4}^{+1.6}$ & $7.4_{-1.0}^{+1.6}$ & $11.9_{-8.8}^{+7.9}$&$8.0_{-1.2}^{+1.2}$&$4.2_{-4.0}^{+3.5}$&$8.0_{-5.4}^{+3.4}\ ^{(a)}$ & $8.0_{-1.0}^{+1.3}$& $8.0_{-1.8}^{+2.2}$& $8.0_{-2.2}^{+2.3}$\\
		& $\sigma/\msun$ & $1.6_{-0.8}^{+2.3}$ & $1.9_{-0.6}^{+1.4}$ & $5.2_{-2.5}^{+9.2}$&$2.4_{-0.8}^{+1.2}$&$2.6_{-1.1}^{+12.2}$ &$2.4_{-1.2}^{+9.4}$& $2.3_{-0.6}^{+1.1}$& $2.2_{-0.8}^{+1.6}$ & $2.2_{-0.9}^{+2.1}$ \\
		&$M_{\min}/\msun$ & $10.0_{-1.7}^{+58}$ & $9.9_{-1.7}^{+56.5}$ &$20.1_{-10.8}^{+42.0}$& $11.6_{-3.5}^{+34.6}$&$8.5_{-1.6}^{+1.1}$& $11_{-3}^{+53}$ &$11_{-3}^{+35}$& $10_{-5}^{+50}$ & $10_{-5}^{+50}$\\
		& $M_{\max}/\msun$& $98_{-68}^{+2}$& $96_{-71}^{+4}$ & $95_{-62}^{+4}$ &$49_{-21}^{+47}$& $10_{-1}^{+77}$&$97_{-72}^{+3}$& $51_{-25}^{+44}$& $13_{-6}^{+87}$ &$13_{-6}^{+88}$\\
		& $\alpha$  & $-9.7_{-0.3}^{+17.5}$ & $-9.6_{-0.4}^{+17.4}$& $-3.8_{-6.2}^{+11.0}$&$-4.0_{-6.0}^{+10.7}$& $-9.4_{-0.6}^{+13.6}$& $-8.9_{-1.1}^{+16.6}$&$-4.6_{-5.4}^{+11.3}$& $-9.1_{-0.9}^{+15.5}$ &$-9.8_{-0.2}^{+16.5}$\\
		& $r$  & $0.79_{-0.05}^{+0.21}$& $0.87_{-0.03}^{+0.13}$ & $0.99_{-0.54}^{+0.01}$&$1.0_{-0.1}^{+0.0}$&  $0.21_{-0.17}^{+0.33}$&$0.88_{-0.16}^{+0.12}$&$1.0_{-0.1}^{+0.0}$& $0.8_{-0.4}^{+0.2}$ & $0.8_{-0.4}^{+0.2}$\\
        & 2nd solution & yes (19\%) & yes (25\%) & no &yes (14\%)& yes (40\%) & yes$^{(a)}$ &yes (7\%) &no&no\\\hline
 \end{tabular}
    \tablefoot{The ``Inert*'' sample refers to the inert BHs excluding Gaia BH3. For the power law and the bimodal models, we indicate whether the inference supports two solutions, \eg a positive or negative exponent $\alpha$ for the power law or different $M_\mathrm{min}$ or $r$ for the bimodal distribution. The percentages indicate how much of the posterior chain supports this second solution that is not shown in the table.\\  
    \tablefoottext{a}{Since we find two substantial solutions for $r$, we indicate the parameter values from the marginalized distributions around the two different solutions.}}
    \endgroup
\end{table*}

\begin{table*}
    \centering
    \caption{Model comparison and fractions of low- (${<}\,5\,\msun$) and high-mass BHs (${>}\,13\,\msun$) for each subsample.}
    \label{tab:subsamples_comp}
    \footnotesize
    \setlength{\tabcolsep}{3pt}
    \renewcommand{\arraystretch}{1.3}
    \begin{tabular}{l|c|cccc|cccc|cccc|cccc}
    \hline\hline
     && \multicolumn{4}{c|}{$\Delta$AICc} & \multicolumn{4}{c|}{$\Delta$BIC} & \multicolumn{4}{c|}{$M_\mathrm{BH}<5\,\msun$} & \multicolumn{4}{c}{$M_\mathrm{BH}>13\,\msun$} \\
     Subsample & $n$ & Gauss & SN & PL & Bi & Gauss & SN & PL & Bi & Data & Gauss & PL & Bi & Data & Gauss & PL & Bi\\\hline
     Gold Sample &10& \textbf{0.0} & 3.8 & 3.1 & 27.5 & \textbf{0.0} & 1.8 & 1.1 & 2.4
       & $0.02$ & $0.0_{-0.0}^{+0.09}$ & $0.003_{-0.003}^{+0.001}$ &$0.0_{-0.0}^{+0.1}$& $0.02$ & $0.0_{-0.0}^{+0.01}$ & $0.0_{-0.0}^{+0.06}$ &$0.0_{-0.0}^{+0.1}$\\
     LMXBs &25& \textbf{0.0} & 2.1 & 1.9 & 1.6 & \textbf{0.0} & 2.7 & 2.5 & 2.4
       & $0.16$ & $0.10_{-0.05}^{+0.07}$ & $0.003_{-0.00}^{+0.18}$ &$0.09_{-0.04}^{+0.09}$& $0.05$ & $0.0_{-0.0}^{+0.0}$ & $0.0_{-0.0}^{+0.0}$&$0.0_{-0.0}^{+0.04}$ \\
     HMXBs &6& \textbf{0.0} &8.6 &8.8&-- & \textbf{0.0} & 0.4 &0.6 &--
       & $0.0$ & $0.0_{-0.0}^{+0.1}$ & $0.004_{-0.004}^{+0.004}$ & $0.0_{-0.0}^{+0.12}$&$0.53$ & $0.66_{-0.18}^{+0.13}$ & $0.44_{-0.18}^{+0.22}$ &$0.59_{-0.17}^{+0.16}$\\
     XRBs &31& \textbf{0.0} & 0.2 & 3.2 & 2.9 & \textbf{0.0} & 1.1 & 4.2 & 5.6
       & $0.13$ & $0.11_{-0.04}^{+0.06}$ & $0.003_{-0.003}^{+0.001}$ &$0.08_{-0.05}^{+0.06}$& $0.14$ & $0.04_{-0.03}^{+0.05}$ & $0.10_{-0.04}^{+0.06}$ &$0.07_{-0.03}^{+0.06}$\\
     Inert* &7& \textbf{0.0} & 3.8 & 2.9 & -- &2.2 & 0.9 & \textbf{0.0} &--
       & $0.18$ & $0.05_{-0.05}^{+0.13}$ & $0.004_{-0.004}^{+0.12}$ &$0.12_{-0.08}^{+0.12}$$^{(a)}$& $0.12$ & $0.0_{-0.0}^{+0.08}$ & $0.0_{-0.0}^{+0.0}$ &$0.0_{-0.0}^{+0.04}$$^{(a)}$\\
     Solar-like Z &37& \textbf{0.0} & 1.3 & 7.4 & 2.3 & \textbf{0.0} & 2.5 & 8.6 & 6.3
       & $0.13$ & $0.11_{-0.04}^{+0.05}$ & $0.003_{-0.003}^{+0.1}$ &$0.09_{-0.04}^{+0.05}$& $0.13$ & $0.02_{-0.02}^{+0.04}$ & $0.12_{-0.05}^{+0.05}$ &$0.05_{-0.02}^{+0.05}$ \\
     No IC10X1 &38& \textbf{0.0} & 2.2 & 4.0 & 9.0 & \textbf{0.0} & 3.5 & 5.3 & 13.2
       & $0.13$ & $0.09_{-0.04}^{+0.05}$ & $0.10_{-0.04}^{+0.06}$ &$0.09_{-0.04}^{+0.06}$& $0.13$ & $0.0_{-0.0}^{+0.02}$ & $0.0_{-0.0}^{+-0.0}$ &$0.02_{-0.02}^{+0.03}$\\
     $\Delta M < 5\,\msun$ &34& \textbf{0.0} & 2.0 & 3.6 & 8.7 & \textbf{0.0} & 3.2 & 4.8 & 12.4
       & $0.14$ & $0.09_{-0.04}^{+0.05}$ & $0.10_{-0.04}^{+0.06}$ &$0.09_{-0.04}^{+0.06}$& $0.08$ & $0.0_{-0.0}^{+0.02}$ & $0.0_{-0.0}^{+-0.0}$ & $0.02_{-0.02}^{+0.03}$\\
    \hline
    \end{tabular}
    \tablefoot{$\Delta$AICc and $\Delta$BIC are given with respect to the best model (shown in bold). $n$ is the size of the subsamples, and ``Gauss'', ``SN'', ``PL'' and ``Bi'' refer to the Gaussian, split-normal, power-law and bimodal BH mass models. The ``Inert*'' sample refers to the inert BHs excluding Gaia BH3. In the HMXBs and Inert* samples, we do not report AICc and BIC of the bimodal distribution as they diverge or break down when the number of parameters $k=6$ approaches the sample size $n$.
    \tablefoottext{a}{Since we find two substantial bimodal solutions for the inert BH sample, we additionally report the fractions found in the low-mass ($0.06_{-0.05}^{+0.24}$) and the high-mass tail ($0.0_{-0.0}^{+0.19}$) of the second solution. See Table~\ref{tab:subsamples}.}}
\end{table*}

\paragraph{Gaussian model.} All subsamples except for the HMXBs share the same mean at ${\approx}\, 8\,\msun$ within a 68\% confidence interval. They even share a similar standard deviation of ${\approx}\,2\,\msun$, identifying the same $5\text{--}10\,\msun$ main component in the BH mass distribution across different subsamples. The HMXBs push $\mu$ and $\sigma$ to higher masses, as the Gaussian with only two free parameters is highly sensitive to the single, unconstrained outlier IC 10 X1 among only six black holes.

\paragraph{Split-normal model.} All subsamples share again a common peak around $7\text{--}9\,\msun$, however, with different asymmetry adjustments. HMXBs, for instance, have a much broader high-mass side ($\sigma_2=8.7_{-5.8}^{+7.4}\,\msun$), accommodating once more IC 10 X1, while the other subsamples only show little asymmetry.

\paragraph{Power-law model.} The best-fit power-law models for the different BH samples differ widely, and no obvious common trend is found. Either the different BH populations have inherently different demographics or the power law is simply not well-suited to describe these data.

\paragraph{Bimodal model.} Interestingly, for each subsample, we find a solution that shares the same Gaussian component within its 68\% confidence interval. Furthermore, it represents most of the BHs of each subsample (79--99\%). In contrast, the power-law component is not constrained and differs across the subsamples.

\paragraph{Model comparison.} 
We compare the AICc and BIC of the models for each sample, and report the BH fraction in the low- and high-mass tails from each subsample in Table~\ref{tab:subsamples_comp}. The Gaussian unanimously emerges as the preferred model, further highlighting the main component of the BH mass distribution at $5\text{--}10\,\msun$ as a robust feature across all subsamples.

In the low-mass tail (${<}\,5\,\msun$), we find $13\text{--}18\%$ of the BH measurements in all subsamples except for the gold sample and the HMXBs. This trend is well-reproduced by the Gaussian and bimodal models within $68\%$ confidence. The power-law, when negative-sloped, fails to represent the BHs in the lower-mass interval. The HXMB sample has no BHs below $5\,\msun$, which is well-reproduced in all models. The gold sample has no significant amount of BHs in the low-mass tail either, potentially suggesting the low-mass tail to be an artifact of measurement uncertainties.

The high-mass tail is mostly populated by HMXBs and inert BHs, and remains largely unconstrained by the low number of observed BHs. If the different subsamples, even when removing the BHs with large uncertainties, display $8\text{--}14\%$ of BHs above $13\,\msun$, the Gaussian distributions systematically underestimate this fraction. Negative-sloped power-law distributions reproduce the high-mass data within $95\%$ confidence. The bimodal distribution, even with more degrees of freedom, underestimates the high-mass tail as well.

To summarize, the dominant $5\text{--}10\,\msun$ component appears as a common feature across different BH subsamples and is well-reproduced by a Gaussian distribution. The Gaussian model also reproduces the low-mass region rather well, but fails to accommodate the high-mass tail. In fact, none of the models successfully reproduces the main component and the two tails simultaneously.

\end{appendix}

\end{document}